\documentclass[12pt]{article}
\usepackage[linesnumbered,ruled]{algorithm2e}
\usepackage{titlesec}
\SetAlCapNameFnt{\small	}
\SetAlCapFnt{\small	}
\usepackage[title]{appendix}
\usepackage{soul}
\usepackage{amsmath,amssymb,amsbsy,amsthm,calc,cases,enumerate,url,framed,fancyhdr,lipsum,tikz,enumitem,natbib,etoolbox,bbm,booktabs,mathtools}
\usepackage{graphicx}
\usepackage[pageanchor,hypertexnames=false,hidelinks]{hyperref}
\usepackage{minitoc}

\usepackage[a4paper,margin=1in]{geometry}

\usepackage[font=small,skip=0pt]{caption}

\newcommand{\unparskip}{\vspace{-\parskip}}

\usepackage{setspace}

\newtheorem{theorem}{Theorem}
\newtheorem{corollary}{Corollary}
\newtheorem{lemma}{Lemma}%
\usepackage{bbm}
\usepackage{stmaryrd}
\newtheorem{proposition}{Proposition}%
\newtheorem{assumption}{Assumption}%

\BeforeBeginEnvironment{proof}{\unparskip}
\BeforeBeginEnvironment{proof_outline}{\unparskip}
  
\BeforeBeginEnvironment{theorem}{\vspace{0.5\parskip}}
\BeforeBeginEnvironment{corollary}{\vspace{0.5\parskip}}
\BeforeBeginEnvironment{lemma}{\vspace{0.5\parskip}}
\BeforeBeginEnvironment{proposition}{\vspace{0.5\parskip}}
  
\BeforeBeginEnvironment{thm}{\vspace{0.5\parskip}}
\BeforeBeginEnvironment{cor}{\vspace{0.5\parskip}}
\BeforeBeginEnvironment{prop}{\vspace{0.5\parskip}}
\BeforeBeginEnvironment{lem}{\vspace{0.5\parskip}}

\newcommand{\iid}{\overset{\mathrm{iid}}{\sim}}

\newcommand{\dotcup}

\usepackage{xr}

\theoremstyle{definition}
\newtheorem{example}{Example}
\theoremstyle{plain}
\newtheorem{definition}{Definition}

\usepackage[normalem]{ulem}

\begin{document}

\title{Recursive moment updates for scalable Bayesian nonparametric regression}

\author{Peter Liu$^{1}$, \ Andrew Yiu$^{2}$ \ and Edwin Fong$^{1,}$\thanks{Corresponding author. Email: chefong@hku.hk}  \\ \\
{\small $^1$Department of Statistics and Actuarial Science, University of Hong Kong}\\
{\small $^2$School of Mathematical Sciences, University of Southampton }\\
}
\date{}

\maketitle

\begin{abstract}
Bayesian nonparametric regression methods have wide applications in statistics and machine learning, but inference can often be expensive or require approximations. This paper introduces a class of recursive moment function updates which are closed form, interpretable, expedient, and fully quantify Bayesian uncertainty through the martingale posterior. The proposed recursive updates can be used off-the-shelf and encompass challenging scenarios such as nonparametric heteroscedasticity, heavy-tailed or skewed noise, and importantly generalize to the multi-output case with missing outcomes without any significant loss in tractability or speed. We show frequentist consistency under weak assumptions in the aforementioned scenarios, and demonstrate empirically that our proposed method is orders of magnitude faster than competing Bayesian nonparametric approaches, with competitive or even superior prediction and estimation accuracy. 

\end{abstract}
\newpage
\section{Introduction}

Bayesian nonparametric regression has wide applications in areas like machine learning, spatial statistics, and biostatistics. Unfortunately, methods such as dependent mixture models \citep{Maceachern2000, Quintana2022} and Gaussian processes (GP, \citet{Williams2006}) can have prohibitively expensive computation due to Markov chain Monte Carlo (MCMC) or large matrix inversions.  Furthermore, these models have many moving parts, and prior and model elicitation often lacks clear guidance.
These issues are further compounded in \textit{multi-output} regression \citep{alvarez2011computationally}, where the outcome is multivariate.

An alternative MCMC-free approach to Bayesian nonparametric inference for mixture models is based on {recursive density updates}, pioneered by the {predictive recursion} method  \citep{Newton1998} and more recently by \cite{Hahn2018}. Advances within the {predictive Bayesian} literature have demonstrated recursive density updates can even be utilized to obtain full posterior uncertainty through the imputation of unseen future variables \citep{Fortini2020, berti2023bayesian, Fong2023}, resulting in the martingale posterior. 

Within the single-output regression setting, \cite{Fong2023, cappello2025recursive} extend the bivariate copula update of \cite{Hahn2018} to conditional density estimation, and \cite{lin2026fast} extend predictive recursion to estimate a covariate-dependent mixing density. However, in many cases in regression, we are often primarily interested in the mean or variance function, as opposed to the entire conditional density. While the previously mentioned recursive updates are flexible, they all require numerical integration on a grid if the moment functions are of interest, which is already computationally demanding in the single-output case, and quickly becomes infeasible as the dimension of the outcome grows.

In this paper, we introduce flexible, tractable, and interpretable recursive updates for the covariate-dependent {moment functions}, with uncertainty quantification through a semiparametric martingale posterior. Our moment updates easily handle heteroscedasticity, heavy-tailed or skewed residuals, and importantly scale very well to the multi-output setting with missing outcomes and covariate-dependent correlations. Our approach achieves competitive or superior statistical performance compared to other recursive approaches and traditional Bayesian nonparametric methods, while being orders of magnitude faster as we do not rely on numerical integration. Hyperparameter tuning is automatic, so our approach can be used off-the-shelf for Bayesian nonparametric regression. Finally, tractability of the moment updates allows thorough study of predictive and frequentist asymptotics.

\section{Methodology for univariate outcomes}

In this section,  we consider single outcome regression with random covariates to illustrate the main idea, before extending to the multi-output case in Section \ref{sec:mo}. Let $Y \in \mathbb{R}$ denote the outcome and  $X \in \mathcal{X}\subseteq \mathbb{R}^p$ denote the covariates where $\mathcal{X}$ is compact. Let  $(Y_i,X_i)_{i = 1,\ldots, n}$  be $n$ independently and identically distributed (i.i.d.) copies of  $(Y,X)$ from an unknown distribution $P^{*}$ with Lebesgue density $p^*(y,x)$. Furthermore, let $p_X^*(x)$ denote the marginal density of $X$, and  let $p^*(y \mid x)$ denote the conditional density of interest which we assume exists for all $x \in \mathcal{X}$.

\subsection{Recursive bivariate copula updates}

We begin by revisiting a recursive approach for conditional density estimation introduced in \cite{Fong2023} using bivariate copula densities. Whilst not the main focus, our recursive moment updates will be derived from this copula update. The starting point is that the Dirichlet process (DP) mixture is tractable when only a single datum is observed \citep{Newton1998}. In the covariate-free DP mixture, \citet{Hahn2018} derived the predictive density update in terms of a Gaussian copula density, and proposed extending the $n = 1$ rule recursively for $n > 1$. 

The same idea can be carried out for the dependent DP mixture model with a Gaussian kernel and base measure. Let $p_i(y\mid x)$ and $u_i = P_i(y\mid x)$ denote the conditional probability density function (PDF) and cumulative distribution function (CDF) after $i$ observations.  \cite{Fong2023} suggests
the recursive update
\begin{equation}\label{eq:condit_copula}
    p_i(y\mid x)
    =
    \left[
        1-\alpha_i(x,X_i)
        +
        \alpha_i(x,X_i) \,
        c_\rho
        \left(u_{i-1}, v_{i-1}
        \right)
    \right]
    p_{i-1}(y\mid x),
\end{equation}
where $v_{i-1} = P_{i-1}(Y_i\mid X_i)$, and  $c_\rho$ is the bivariate Gaussian copula density:
\begin{equation*}
c_\rho(u,v) = \frac{{\phi_2\left( \Phi^{-1}(u), \Phi^{-1}(v); \rho\right)}}{\phi\left(\Phi^{-1}(u)\right)\, \phi\left(\Phi^{-1}(v)\right)}
\end{equation*}
and $\phi_2({y}_1,y_2; \rho)$ is the standard bivariate Gaussian density with correlation $\rho$,  $\phi(y)$ is the standard Gaussian density, $\Phi^{-1}$ is the Gaussian quantile function, $\rho\in(0,1)$ controls the smoothness of the density, and the covariate-dependent learning rate $\alpha_i(x,X_i)$ determines the information borrowed between covariates.

\subsubsection{Covariate-dependent learning rate}\label{sec:cov_lr}

We consider a general form of $\alpha_i(x,x')$ based on the `conditional' approach of \cite{Fong2023}. Let $s(x,x')$ be a \textit{similarity function}, 
 \begin{align}\label{eq:dxx}
     s(x,x') = \frac{E_{\Psi}\left[k(x; \Psi) \, k(x';\Psi)\right]}{E_{\Psi}\left[k(x;\Psi)\right]\,E_{\Psi}\left[k(x';\Psi)\right]},
 \end{align}
where $\Psi \sim \Pi$,  the kernel $k(\,\cdot \, ; \psi)$ is a PDF for all $\psi \in \text{supp}(\Pi)$, and we assume that $s(x,x')$ is well-defined for all $(x,x') \in \mathcal{X} \times \mathcal{X}$.
We then define 
\begin{align*}
    \alpha_i(x,x') = \frac{\alpha_i s(x,x')}{1-\alpha_i + \alpha_i s(x,x')}
\end{align*}
where $0 < \alpha_i < 1$ and $\alpha_i = O(i^{-m})$ for some $0.5 < m \leq 1$. In our implementations, we choose the learning rate $\alpha_i = \left(2 - i^{-1}\right)(i+1)^{-0.9}$, which works well in practice.
Further discussions on the role of the learning rate can be found in \cite{Fortini2020}, \cite{agnoletto2026nonparametric} and \citet[Appendix E]{Fong2023}.

 Importantly, we have $\alpha_i(x,x') \to 0$ if $s(x,x') \to 0$, so the datum $(Y_i,X_i)$ is less relevant for the conditional density at $x$ if  $s(x,X_i)$ is small, and $\alpha_i(x,x') \to 0$ as $i \to \infty$.  Like \cite{Antoniano2014}, our model elicitation involves a kernel $k(x;\psi)$ with a prior for $\psi$. We now illustrate an example of $s(x,x')$.
\begin{example}\label{ex:Gaussian_cop}
Consider the Gaussian kernel $k(x; \psi) = \mathcal{N}(x; \psi, 1 - \tau^{-1})$ and $\Pi = \mathcal{N}(0, \tau^{-1})$ for a fixed $\tau > 1$. 
In this case, we recover $s(x,x') = c_{\rho_x}\left\{P_0(x), P_0(x')\right\}$ where $P_0 = \Phi$ is the normal CDF and $\rho_x = \tau^{-1}$, as considered in \cite{Fong2023}. 
\end{example}

More generally, a closed form $s(x,x')$ can be obtained if $k(\,\cdot\,;\psi)$ is an exponential family with conjugate prior $\Pi$; we outline more examples in Appendix \ref{app:sxx}. Multivariate covariates can be handled by defining $k(x;\psi) = \prod_{j = 1}^p k(x^{(j)};\psi^{(j)})$, and  hyperparameters of $s(x,x')$, e.g. $\rho_x$ for the Gaussian copula, can be tuned in an empirical Bayesian manner using the prequential log-likelihood. In the multivariate case, the bandwidth tuning procedure can also carry out variable selection.

The main property we assert on $s(x,x')$ is that the family $\{k(x;\psi):\psi \in \textnormal{supp}(\Pi)\}$ is complete \citep{Lehmann2006}  as defined in Definition \ref{def:complete}. Completeness will be necessary for consistency of the recursive moment updates, and is satisfied by many examples including Example \ref{ex:Gaussian_cop}. 
Intuitively, this means that $s(x,x')$ does not have any `gaps' which may impede consistency. 
\begin{definition}\label{def:complete}
    The family $\{k(x;\psi):\psi \in \text{supp}(\Pi)\}$ is \textit{complete} if
    for every measurable function $f:\mathbb{R}^p \to \mathbb{R}$, the condition  $\int \, f(x) \, k(x;\psi) \, dx = 0$ for all $\psi \in \text{supp}(\Pi)$
implies $f(x) = 0$ for $k(\,\cdot\,;\psi)$-almost all $x$ for every $\psi \in \text{supp}(\Pi)$.
\end{definition}

\subsection{Recursive moment updates}\label{sec:rec_mom}
In nonparametric regression, the mean function $E[Y \mid X = x]$ is often of primary interest for prediction, or one may be interested in assessing {heteroscedasticity} through the conditional variance $\text{Var}(Y \mid X = x)$.  A significant drawback of the nonparametric copula approach is that the moment functions must be computed numerically on a grid from the conditional density estimates $p(y \mid x)$, which can be very computationally expensive, and quickly becomes completely infeasible in the multi-output setting in the next Section. In addition, the selection of the $y$-grid is non-trivial, especially under strong heteroscedasticity or heavy tails where a very high resolution is needed to cover the range of $y$ values. We highlight that a grid over the mixing parameter is also needed for the predictive recursion approach of \cite{lin2026fast}, which will scale exponentially poorly with outcome dimensionality. 

To that end, we introduce a \textit{semiparametric} recursive update, which can be viewed as a projection of the nonparametric copula update onto a parametric family. Instead of a recursive update for the conditional density, we restrict $p_i(y \mid x)$ to a specific parametric family, and derive tractable recursive updates for the moment functions directly. The moment updates are simple, flexible and interpretable, and are significantly more expedient than the copula update and other Bayesian competitors. Importantly, uncertainty quantification of the moment functions can still be obtained through the martingale posterior, and Section \ref{sec:freq} will illustrate that consistency can still be attained even if the family is misspecified.

\subsubsection{Gaussian family}

 The core idea is to distill the full nonparametric copula update into its first two moments, then project the updated density onto the Gaussian family through moment matching. Let us restrict $p_{i-1}(y \mid x)$ to be Gaussian, that is 
\begin{align*}
    p_{i-1}(y \mid x)
    = \mathcal{N}\left(y; \mu_{i-1}(x), \sigma_{i-1}^2(x)\right).
\end{align*}
Whilst applying (\ref{eq:condit_copula}) to the above yields a non-Gaussian $p_i(y \mid x)$, its updated mean and variance can be computed exactly from conjugacy of the predictive CDF with the Gaussian copula. We then use these updated moments to define the Gaussian predictive density for the next step. To derive the moment updates, note that
$P_{i-1}(y \mid x)
=\Phi\{(y-\mu_{i-1}(x))/\sigma_{i-1}(x)\}$, and Gaussian conditioning then gives
\begin{align}\label{eq:condit_gauss}
   c_\rho\left(u_{i-1}, v_{i-1}\right) p_{i-1}(y \mid x) = \mathcal{N}\left(y\,;\, \mu_{i-1}(x) + \rho{\sigma_{i-1}(x)}Z_i, (1-\rho^2)\,\sigma_{i-1}^2(x)\right),
\end{align}
where $Z_i$ is the standardized residual of the form
\begin{align}\label{eq:Zi}
    Z_i = \frac{Y_i - \mu_{i-1}(X_i)}{\sigma_{i-1}(X_i)}\cdot
\end{align}
Crucially, this gives a tractable recursive update of the $k$-th moment $m_i^{(k)}(x) = \int y^k\,  p_i(y \mid x)\, dy$, as integrating (\ref{eq:condit_copula}) results in the recursion
\begin{align*}
m_i^{(k)}(x) =
\left[1-\alpha_{i}(x,X_{i})\right]m^{(k)}_{i-1}(x) + \alpha_{i}(x,X_{i})\,  \widetilde{m}_i^{(k)}(x)
\end{align*}
where $\widetilde{m}_i^{(k)}$ is the $k$-th moment of the conditional Gaussian given in (\ref{eq:condit_gauss}). Evaluating the above for $k = 1,2$ then gives the tractable mean and variance recursion
\begin{align}
\mu_i(x) &= \mu_{i-1}(x) + \alpha_i(x,X_i)  \, \sigma_{i-1}(x) \,\rho Z_i, \label{eq:mean_gauss}\\
    \sigma_i^2(x) &= \sigma_{i-1}^2(x)\left(1 + \alpha_i(x,X_i)\,\rho^2 \left[\{1-\alpha_i(x,X_i)\}Z_i^2 - 1 \right]\right). \label{eq:var_gauss}
\end{align}
The form of the mean update is interpretable, where the residual $Z_i$ drives the update to the mean function, with $\alpha_i(x,X_i)$ playing a similar weighting role. The scaling of the residual by $\sigma_{i-1}(x)$ then rescales the normalization by $\sigma_{i-1}(X_i)$ at the test point of interest. The variance update is similarly driven by the error term $(Z_i^2 - 1)$. Interestingly, the two updates are heavily coupled, so studying the frequentist consistency will rely on a carefully constructed almost-supermartingale \citep{Robbins1971}. We note that $\rho \in(0,1)$ now governs the influence of a new datum through scaling of the update. Finally, while mean/variance function updates exist in the non-Bayesian literature for estimation, e.g. \citet{stadtmuller1995nonparametric}, our above update is sufficient for full Bayesian uncertainty quantification. 
\begin{example}\label{ex:het}
We demonstrate the Gaussian moment updating scheme following the simulation setup of \cite{Antoniano2014}, with 
$p^*_X = \operatorname{Unif}(-7,7)$ and $p^*(y \mid x) =  \mathcal N\!\left(m(x),s^2(x)\right)$ where
\begin{align*}
    m(x)=\frac{5}{1+\exp(-x)},
    \qquad
    s(x)=\frac14+\exp\!\left(\frac{x-6}{3}\right).
\end{align*}
Figure \ref{fig:het_sim} illustrates the results with $n = 200$ simulated observations and $B = 1000$ posterior samples with a truncation of $N = n + 4000$. 
The copula approach requires a grid of $200$ $y$-values for numerical accuracy. We see that both the predictive densities and the posterior over $\mu(x)$ are  similar, with slightly higher posterior uncertainty for the moment update. The key difference is that the copula updates required 180s to run, whereas the moment updates only required 0.4s (on an Apple M2 Pro CPU).
\begin{figure}[!h]
\begin{center}
\includegraphics[width=\textwidth]{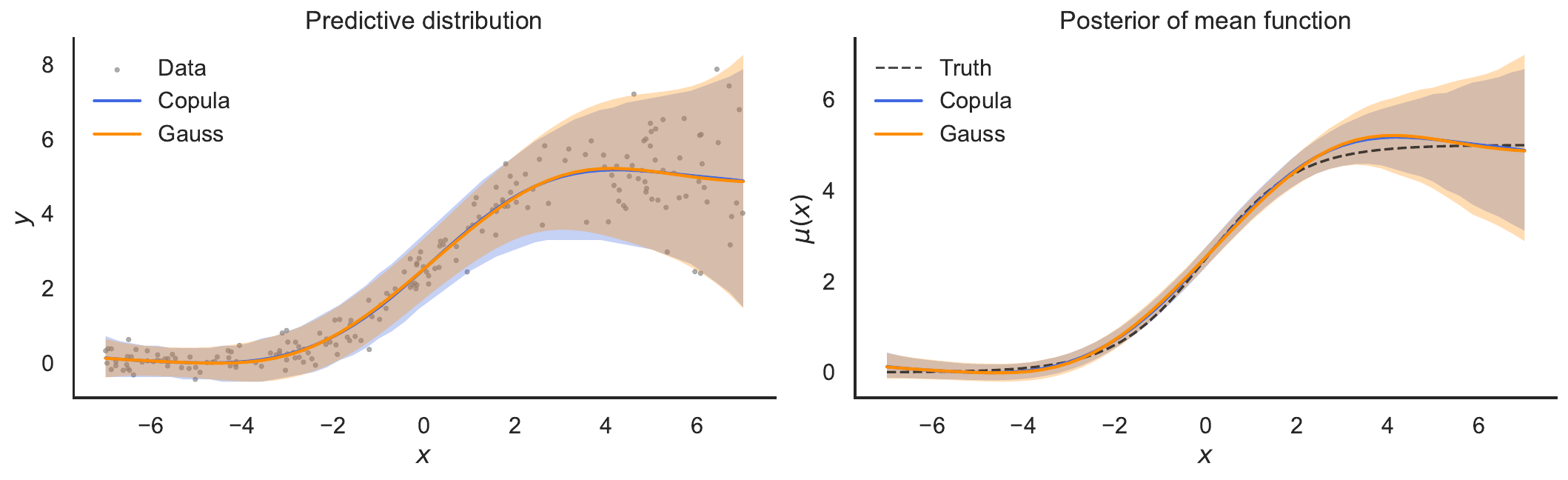}
\caption{Predictive mean with 95\% predictive intervals (Left); Posterior mean with 95\% credible intervals for $\mu(x)$ (Right).} 
\label{fig:het_sim}
\end{center}
\end{figure}
\end{example}

\subsubsection{Predictive resampling for uncertainty}
Equipped with the above, it is natural to define our sequence of conditional predictives as the Gaussian density $p_i(y \mid x) = \mathcal{N}\left(y; \mu_i(x), \sigma_i^2(x)\right)$ with moment updates given by (\ref{eq:mean_gauss}) and (\ref{eq:var_gauss}). Under predictive resampling, it is not too hard to see that $Z_i \iid \mathcal{N}(0,1)$ for $i\geq n+1$, which is similar to the i.i.d. uniform random variables in the copula case. As a result, predictive resampling is computationally very simple, as it just involves applying (\ref{eq:mean_gauss}) and (\ref{eq:var_gauss}) to i.i.d. Gaussian noise. 
Furthermore, the first two raw moments can be shown to be function-valued martingales, and Algorithm \ref{alg:gauss} illustrates how to sample from the martingale posterior, where $P_X$ is a chosen design distribution.
 We briefly discuss the role of $P_X$, which was investigated in \cite{sun2026design}. Since our objects of interest only depend on $p(y \mid x)$, we do not require uncertainty in $P_X$, so we suggest fixing it as the default choice of the uniform distribution over $\mathcal{X}$ if the covariate space is compact, or the Gaussian distribution if $\mathcal{X} = \mathbb{R}^p$. 
In Section \ref{sec:pred_asym}, we will study the convergence of Algorithm \ref{alg:gauss} using function-valued martingale theory \citep{Pisier2016}. In addition to proving convergence of our algorithm, we will show that the martingale posterior has support over smooth mean/variance functions under appropriate assumptions on $s(x,x')$. 
\begin{figure}[!h]
\center

\begin{minipage}{0.9\linewidth}
\begin{algorithm}[H]
{Estimate $p_n(y \mid x)$ from $(Y_{1:n},X_{1:n})$ using Equations (\ref{eq:mean_gauss}) and (\ref{eq:var_gauss})}\\
{Draw $X_{n+1:N} \iid P_X$}\\
\For{$i \gets n+1$ \textnormal{\textbf{to}} $N$}{
{Draw $Z_i \iid \mathcal{N}(0,1)$}\\
{Update $\{\mu_{i-1}(x), \sigma_{i-1}^2(x)\} \mapsto \{\mu_{i}(x), \sigma_{i}^2(x)\}$ using Equations (\ref{eq:mean_gauss}) and (\ref{eq:var_gauss})}}
 {Return $\left\{{\mu_N(x), \sigma_N^2(x)}\right\}$}
\caption{Predictive resampling with recursive moment updates}\label{alg:gauss}
\end{algorithm}
\end{minipage}
\end{figure}

\subsubsection{Heavy-tailed family}\label{sec:heavy}
A key property of our moment matching framework is that it extends naturally to more general choices of $p_i(y \mid x)$ beyond the Gaussian, which may be more appropriate depending on the setting. To illustrate the idea, we investigate the induced moment updates for a heavy-tailed choice of $p_i(y \mid x)$. Our proposed parametric family is a transformed Gaussian random variable which is still `conjugate' to the copula, and induces a tractable robust moment update with only a small computational overhead. This is in stark contrast to other heavy-tailed nonparametric regression methods, e.g. GPs with a Student-t likelihood \citep{jylanki2011robust}, which require approximate inference, although recent generalized Bayes procedures preserve conjugacy \citep{pmlr-v235-altamirano24a, rooijakkers2025multi}. 

\begin{figure}[!h]
\begin{center}
\includegraphics[width=\textwidth]{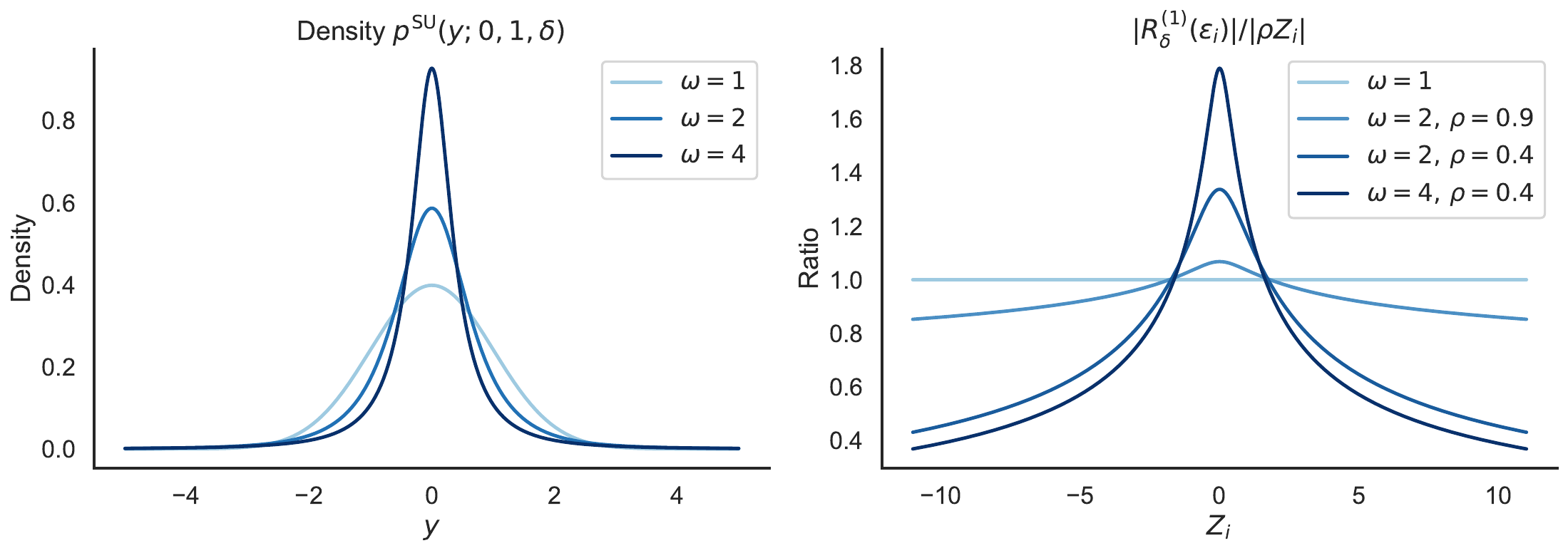}
\caption{Density $p^{\text{SU}}(y; 0,1,\delta)$ (Left); Ratio of mean update $|R^{(1)}_{\delta}(\varepsilon_i)|/|\rho Z_i|$ (Right)}
\label{fig:tails}
\end{center}

\end{figure}

Let $\delta  > 0$ be a parameter controlling the kurtosis, and let $\omega = \exp(1/\delta^2)$ for shorthand. We define the family $p^{\text{SU}}(y; \mu,\sigma^2,\delta)$ as the
random variable generated as 
\begin{align}\label{eq:SU_fam}
    Y &= \mu + \sigma\left[\frac{\sinh\left(\frac{\varepsilon}{\delta}\right)}{\sqrt{\frac{1}{2}(\omega^2 - 1)}}\right], 
\end{align}
where $\varepsilon \sim \mathcal{N}(0,1)$, which is a reparametrized Johnson's SU distribution \citep{johnson1949systems}. One can verify that $Y$ has mean $\mu$ and variance $\sigma^2$ with kurtosis equal to ${(\omega^4  + 2\omega^2 + 3)}/{2}$, so we can attain the range of kurtosis values of $[3,\infty)$ for $\delta \in (0,\infty]$. Figure \ref{fig:tails} (left) illustrates this family for different values of $\delta$, where taking $\delta \to \infty$  (i.e. $\omega \to 1$) recovers the Gaussian. We then have our main result.
\begin{proposition}\label{prop:SU_update}
Let $p_{i-1}(y \mid x) = p^{\textnormal{SU}}\left(y; \mu_{i-1}(x), \sigma_{i-1}^2(x),\delta\right)$ for fixed $\delta > 0$. Then the mean and variance function updates corresponding to (\ref{eq:condit_copula}) take the form
    \begin{align} 
    \mu_i(x) &= \mu_{i-1}(x) + \alpha_i(x,X_i) \, \sigma_{i-1}(x)\, R^{(1)}_{\delta}(\varepsilon_i), \label{eq:mean_SU}\\
\sigma_i^2(x)
&=
\sigma_{i-1}^2(x)
\left[
1
+
\alpha_i(x,X_i) \left\{R^{(2)}_{\delta}(\varepsilon_i) -1
-
\alpha_i(x,X_i)\, R^{(1)}_{\delta}(\varepsilon_i)^2
\right\}\right] \label{eq:var_SU},
    \end{align}
    where the update terms take the form
 \begin{align}\label{eq:R_delt}
  R^{(1)}_{\delta}(\varepsilon_i) = \frac{\omega^{{(1-\rho^2)}/{2}}\,  \sinh\left(\frac{\rho \varepsilon_i}{\delta}\right)}{\sqrt{\frac{1}{2}(\omega^2 - 1)}}, \quad \quad R^{(2)}_{\delta}(\varepsilon_i)  = \frac{\omega^{2(1-\rho^2)}\cosh\left(\frac{2\rho \varepsilon_i}{\delta}\right) - 1}{\omega^2 - 1}
\end{align}
    and $\varepsilon_i= \delta \sinh^{-1}\left(\sqrt{\frac{1}{2}(\omega^2 - 1)}\,\, Z_i\right)$ where $Z_i$ is the standardized residual as in (\ref{eq:Zi}).
    Furthermore, $\delta \to \infty$ recovers the Gaussian update in (\ref{eq:mean_gauss}) and (\ref{eq:var_gauss}).
\end{proposition}
The above update generalizes the Gaussian update: under $p_i(y \mid x) = p^{\textnormal{SU}}\left(y; \mu_{i}(x), \sigma_{i}^2(x),\delta\right)$, we have $\varepsilon_i \iid \mathcal{N}(0,1)$, so predictive resampling in Algorithm \ref{alg:gauss} can easily be extended by replacing the moment updates with (\ref{eq:mean_SU}) and (\ref{eq:var_SU}), and $Z_i$ with $\varepsilon_i$.  A key property of the SU moment update is the increased robustness to outliers. In Figure \ref{fig:tails} (right), we plot the ratio of $|R_{\delta}^{(1)}(\varepsilon_i)|$ to the Gaussian counterpart $\left|\rho Z_i\right|$, which clearly illustrates a reduced sensitivity to outliers for large values of $|Z_i|$. For the variance update, the Gaussian counterpart to $R_{\delta}^{(2)}(\varepsilon_i)$ is $1-\rho^2 + \rho^2 Z_i^2$, where $R_{\delta}^{(2)}$ also has a robustifying effect.
This is formalized below. 
\begin{proposition}\label{prop:SU_robust}
For fixed $\rho \in (0,1)$ and $\delta > 0$ and taking $|Z_i|\to \infty$, (\ref{eq:R_delt}) satisfies $|R_{\delta}^{(1)}(\varepsilon_i)| = O(\,|Z_i|^\rho)$ and $R_{\delta}^{(2)}(\varepsilon_i) = \,O(|Z_i|^{2\rho})$, where the constants depend on $(\delta,\rho)$. Furthermore, for any $\delta > 0$ and fixed $Z_i$, we have $\lim_{\rho \to 1} R_{\delta}^{(1)}(\varepsilon_i) = Z_i$ and $\lim_{\rho \to 1} R_{\delta}^{(2)}(\varepsilon_i) = Z_i^2$, which are equivalent to the Gaussian updates at  $\rho = 1$. 
\end{proposition}
Proposition \ref{prop:SU_robust} confirms our observations in Figure \ref{fig:tails} (right), where a smaller value of $\rho$ increases the relative robustness, and $\delta$ comes into the constant. 
In practice, we learn $\delta$ and $\rho$ alongside the hyperparameters of $s(x,x')$ using a prequential log-likelihood score, which only adds a small computational overhead on top of the Gaussian update. Finally, we highlight that the robustness emerged naturally from the copula update under the assumption of a heavy-tailed predictive density. 

\begin{example}\label{ex:kurt}
Consider the same setup as Example \ref{ex:het}  except with 
$$p^*(y \mid x) = 0.95\,\mathcal N\!\left(m(x),s^2(x)\right)
      +0.05\,\mathcal N\!\left(m(x),100s^2(x)\right),$$
      corresponding to a significant Huber contamination. Figure \ref{fig:kurt_sim} illustrates the Gaussian versus the SU moment updates. We see that the Gaussian's mean and variance functions are noticeably influenced by the outliers, whereas the SU update is robust to the outliers (prequentially tuned $\rho = 0.52$). The run-times for the Gauss and SU methods are 0.5s and 0.8s respectively, indicating the small overhead for the robust approach. We also note that the credible intervals for the SU update are narrower for the mean. 
      \begin{figure}[!h]
\begin{center}
\includegraphics[width=\textwidth]{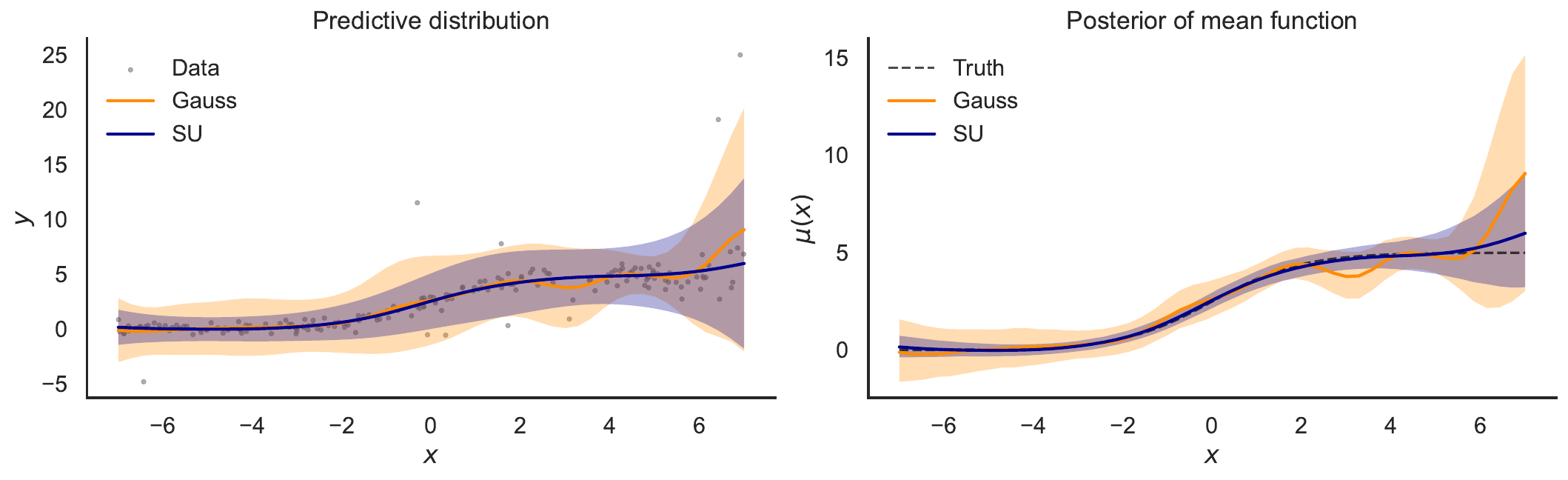}

\caption{Predictive mean with 95\% predictive intervals (Left); Posterior mean with 95\% credible intervals for $\mu(x)$ (Right).} 
\label{fig:kurt_sim}
\end{center}
\end{figure}
\end{example}
\subsubsection{General families}\label{sec:uni_gen}

A natural generalization is to replace the $\sinh$ transformation with other invertible transforms of a Gaussian random variable; we show in Appendix \ref{app:gen_th} that the moment updates take the same form as in Proposition \ref{prop:SU_update} but with a transform-specific $R^{(j)}(\varepsilon_i)$.
    Suitable choices of the transform will preserve tractability, and predictive resampling will still only require i.i.d. Gaussian random variables. 
    In Appendix \ref{app:gen}, we show generalizations for skewed residuals and the platykurtic setting.  

\section{Methodology for multivariate outcomes}\label{sec:mo}

We now extend the previous ideas to the multi-output setting. Let $Y \in \mathbb{R}^q$ and let $Y^{(j)}$ denote the $j$-th component of $Y$. The multi-output setting is where the computational advantage of the moment update is clearest: both the copula update and predictive recursion require numerical integration on a grid and have a computational cost that can potentially scale exponentially with $q$, and methods based on GPs are inflexible, scale poorly, or require complex posterior approximations. We begin with a natural extension of the copula update to the multi-output setting:
\begin{align}\label{eq:condit_copula_mv}
        p_{i}({y} \mid x) =\left[1-\alpha_{i}(x,X_{i}) + \alpha_{i}(x,X_{i}) \, \prod_{j = 1}^q c_{\rho}\left(u_{i-1}^{(j)}, v_{i-1}^{(j)} \right)\right] p_{i-1}({y} \mid x),
\end{align}
where $u_{i-1}^{(j)} = P_{i-1}(y^{(j)} \mid y^{(1:j-1)},x)$ and $v_{i-1}^{(j)} = P_{i-1}(Y_i^{(j)} \mid Y_i^{(1:j-1)},X_i)$. This again follows from the same argument as in \cite{Fong2023} based on the multivariate copula update where the only difference is that $y$ can now also be multivariate. We highlight that an ordering of the dimensions of $y$ must be specified, but this poses no real concern in practice; we discuss this in Section \ref{sec:ordering}.

Following  Section \ref{sec:rec_mom},  we let $p_{i-1}(y \mid x) = \mathcal{N}\left(y; \mu_{i-1}(x), \Sigma_{i-1}(x)\right)$,
where $\mu_{i-1}(x)$ and $\Sigma_{i-1}(x)$ are now the covariate-dependent mean vector and covariance matrix respectively.  Plugging the Gaussian family into (\ref{eq:condit_copula_mv}) yields the multivariate updates: 
\begin{align}
\mu_i(x) &= \mu_{i-1}(x) + \alpha_i(x,X_i) \, L_{i-1}(x) \,\rho Z_i, \label{eq:mean_gauss_mv}\\
   \Sigma_i(x)    &= L_{i-1}(x) \left(I_q + \alpha_i(x,X_i)\rho^2\left[\left\{1-\alpha_i(x,X_i) \right\}Z_iZ_i^\top  - I_q\right] \right) L_{i-1}(x)^\top \label{eq:var_gauss_mv}
\end{align}
where $L_i(x)$ is the Cholesky factor of $\Sigma_{i}(x)$, and the residual is now
\begin{align}\label{eq:mv_Zi}
    Z_i = L_{i-1}^{-1}(X_i)\left[Y_i - \mu_{i-1}(X_i)\right].
\end{align}
The main difference is that the standard deviation function has now been replaced by the Cholesky factor of the covariance matrix. Algorithm \ref{alg:gauss} naturally extends to the multi-output case, where we now have that $Z_i \sim \mathcal{N}(0,I)$ under predictive resampling, and the first two raw moment functions will once again be martingales. 

Our recursive updates are highly flexible: they allow for arbitrary covariate-dependent variance and correlation between outcomes, and the updates are consistent under weak conditions and perform very well in practice. This is generally more challenging to attain with traditional GP approaches, where a latent function model must be explicitly assumed. Standard tractable coregionalization models \citep{alvarez2011computationally} cannot attain covariate-dependent residual variance/correlations, and flexible models like \citet{wilson2012,requeima2019gaussian} can have intractable components and require tuning-heavy approximations. 
\begin{example}
    We follow the same setup as Example \ref{ex:het} (for $n$, $B$, etc.) and extend to the $q = 2$ case, where the first outcome $y^{(1)}$ is as before, and the second outcome $y^{(2)}$ has mean and variance functions $m(-x)$ and $s^2(-x)$ respectively, and the outcome correlation function between $y^{(1)}$ and $y^{(2)}$ is $r(x) = 0.8 \tanh(x/2)$. The recursive moment update required $1.0$s to run, which is slightly longer than the single-output case as expected. Since the copula approach would now require a $200 \times 200$ grid for $y^{(1)} \times y^{(2)}$, this would cost on the order of 200 times the run-time of the single-output case, which would amount to roughly 10 hours. 
\end{example}

\subsection{Ordering}\label{sec:ordering}

The Cholesky preconditioning by $L_{i-1}^{-1}(X_i)$ in (\ref{eq:mv_Zi}) gives an autoregressive interpretation for $Z_i$ which we outline below.
\begin{proposition}\label{prop:Zi}
    Let $p_i(y \mid x) = \mathcal{N}(y; \mu_i(x), \Sigma_i(x))$ with $\mu_i$ and $\Sigma_i$ following the updates (\ref{eq:mean_gauss_mv}) and (\ref{eq:var_gauss_mv}). Then the $j$-th element of the residual takes the form
    \begin{align*}
        Z_i^{(j)} = \frac{Y_i^{(j)} - \tilde{\mu}^{(j)}_{i-1}\left(X_i; \, Y_i^{(1:j-1)}\right)}{\widetilde{\sigma}_{i-1}^{(j)}\left(X_i; \, Y_i^{(1:j-1)}\right)}
    \end{align*}
    where $\tilde{\mu}^{(j)}_{i-1}(x;\, y^{(1:j-1)})$ and $\widetilde{\sigma}_{i-1}^{(j)}(x;\, y^{(1:j-1)})$ are the mean and standard deviation respectively of the conditional density $p_{i-1}(y^{(j)} \mid y^{(1:j-1)},x)$. 
\end{proposition} 
Thus, $Z_i^{(j)}$ is the standardized residual when sequentially predicting $Y_i^{(j)}$ given $Y_i^{(1:j-1)}$.
 In particular, the update for $\mu_i^{(j)}$ only depends on $Y_i^{(1:j)}$, and one can see that the update for $\mu_i^{(j)}$ would be unchanged if only $Y^{(1:j)}$ were observed instead of the full vector $Y^{(1:q)}$, akin to the marginal consistency property observed for the multivariate copula update in \cite{Fong2023}.
Importantly, we highlight that the ordering is not an assumption on the data generating process (DGP): the outcome covariance matrix $\Sigma_i(x)$ can still be fully dense.

 In practice, we do not see a large difference in performance due to ordering, and frequentist consistency is also agnostic to this choice. If there is missing data, it can be helpful to order the outcomes so that the dimensions with the most missing entries come last as we will see later.
Another benefit of the Cholesky form of $L_i(x)$ is that each update only costs $O(q^2)$ as we can update $L_{i-1}$ to $L_{i}$ with a rank-1 Cholesky update, avoiding the potential $O(q^3)$ complexity of a generic matrix square root.

\subsection{Missing outcomes}\label{sec:missing}

Missing outcomes occur very naturally within the multi-output setting, and handling missingness is a popular use case for multi-output regression methods \citep{alvarez2011computationally}. For example, a sensor may only capture a subset of signals at each location, or certain measurements may be much more expensive than others:  multi-output regression allows for the prediction of missing outcomes using the learned correlations, which in our case can be covariate-dependent.

We now derive the moment matching updates for the copula update when marginalizing out the missing components. Let $O_i \subseteq \{1,\ldots,q\}$ denote the set of observed outputs at observation $i$, and let $S_i$ be the corresponding selection matrix of size  $|O_i| \times q$. Let $M_i = \{1,\ldots,q\} \setminus O_i$ denote the missingness pattern. We thus have $Y_i^{(O_i)} = S_i Y_i$ and $Y_i^{(M_i)} = Y_i \setminus Y_i^{(O_i)}$, where we only observe $Y_i^{(O_i)}$. We assume throughout that the outcomes are missing at random (MAR), which implies that
$$p_{i-1}(Y_i^{(M_i)} \mid Y_i^{(O_i)}, O_i, X_i)= p_{i-1}(Y_i^{(M_i)} \mid Y_i^{(O_i)}, X_i).$$
Note that the subscript $(i-1)$ indicates additional conditioning on $(Y_j, X_j)_{j = 1:i-1}$.
Starting again with the fully observed copula update, we consider the update from $p_{i-1}(y \mid x)$ to $p_{i-1}(y \mid x, Y_i^{(O_i)}, X_i)$.
The key is to write
\begin{align*}
    p_{i-1}(y \mid x, Y_i^{(O_i)}, X_i) = E_{Y_i^{(M_i)}}\left[p_{i}\left({y} \mid x\right) \mid Y_i^{(O_i)}, X_i\right]
\end{align*}
where $Y_i^{(M_i)} \sim p_{i-1}(Y_i^{(M_i)} \mid Y_i^{(O_i)}, X_i)$, and $p_i(y \mid x)$ denotes the predictive density updated with (\ref{eq:condit_copula_mv}) given the fully observed outcome $Y_i = (Y_i^{(O_i)}, Y_i^{(M_i)})$. The $k$-th updated moment ${m}_i^{(k)} := \int y^{\otimes k} \, p_{i-1}(y \mid x, Y_i^{(O_i)}, X_i)\, dy$ then satisfies
\begin{align}\label{eq:mk_missing}
{m}_i^{(k)}  =\left[1-\alpha_{i}(x,X_{i})\right]m^{(k)}_{i-1}(x) + \alpha_{i}(x,X_{i})\,  E_{Y_i^{(M_i)}}\left[\widetilde{m}_i^{(k)}(x) \mid Y_i^{(O_i)}, X_i\right]
\end{align}
where $\widetilde{m}_i^{(k)}(x)$ is the moment of the copula kernel as in the single-output case. We now present the general Gaussian moment update under missingness. 

\begin{proposition}\label{prop:mean_var_missing}
Let  $p_{i-1}(y \mid x) = \mathcal{N}(y; \mu_{i-1}(x), \Sigma_{i-1}(x))$. Then (\ref{eq:mk_missing}) corresponds to the mean and variance updates
\begin{align}
    \mu_i(x)
    &=
    \mu_{i-1}(x)
    +
    \alpha_i(x,X_i) \rho
    L_{i-1}(x)\overline{Z}_i,
    \label{eq:mean_gauss_missing}
    \\
\Sigma_i(x)
&=
L_{i-1}(x)
\left[
I
+
\alpha_i(x,X_i)\rho^2
\left\{
\{1-\alpha_i(x,X_i)\}\overline{Z}_i\overline{Z}_i^\top
-
Q_i
\right\}
\right]
L_{i-1}(x)^\top
    \label{eq:cov_gauss_missing}
\end{align}
where $Q_i$ is a projection matrix taking the form
\begin{align}\label{eq:proj_mat}
Q_i
    =
    A_i^\top (A_iA_i^\top)^{-1}A_i, \quad 
    A_i
    =
    S_i L_{i-1}(X_i), 
\end{align}
and $\overline{Z}_i =Q_i Z_i$ is the projected residual with $Z_i$ defined in (\ref{eq:mv_Zi}).
\end{proposition}
The above moment updates are simply projected versions of the fully observed updates (\ref{eq:mean_gauss_mv}) and (\ref{eq:var_gauss_mv}).
Here, $Q_i$ projects the fully observed residual $Z_i$ into the subspace revealed by $Y_i^{(O_i)}$, and the update only depends on $Y_i^{(O_i)}$ as $A_i Z_i  = S_i\left[Y_i-\mu_{i-1}(X_i)\right]$. The projection space is particularly clear when the missingness pattern is \textit{order-compatible} in the sense that $O_i = \{1,\ldots,q_i\}$ for some $q_i \leq q$, as the Cholesky form of $L_{i-1}$ gives $Q_i = \begin{bmatrix}
    I_{q_i} & 0\\
    0 & 0 
\end{bmatrix}$ in this setting. For example, for $q = 2$ where $Y_i^{(2)}$ is missing, we have
\begin{align*}
    \mu_i^{(2)}(x) &= \mu_{i-1}^{(2)}(x) +  \alpha_i(x,X_i) \, L^{(2,1)}_{i-1}(x) \,\rho  Z_i^{(1)},
\end{align*}
where $Z_i^{(1)}$ is as in Proposition \ref{prop:Zi}, which follows as  $E_{Y_i^{(2)}}\left[Z_i^{(2)} \mid Y_i^{(1)}, X_i\right] = 0$. The update for $\mu_i^{(1)}(x)$ is then unchanged.

Our proposed moment update under missingness also satisfies an intuitive property on the preservation of conditional distributions. 
\begin{proposition}\label{prop:missing_invar}
Let $p_i(y \mid x) = \mathcal{N}(y; \mu_i(x), \Sigma_i(x))$ for all $i \geq 0$.
Under the moment updates (\ref{eq:mean_gauss_missing}) and (\ref{eq:cov_gauss_missing}), the conditional densities satisfy
\begin{align*}
    p_i(y^{(M_i)} \mid y^{(O_i)}, x = X_i) =   p_{i-1}(y^{(M_i)} \mid y^{(O_i)}, x= X_i)
\end{align*}
for all $y = \{y^{(M_i)} \cup y^{(O_i)}\}  \in \mathbb{R}^q$. Furthermore, if $O_i$ is order-compatible, then the above holds with $X_i$ replaced with any $x \in \mathcal{X}$.
    \end{proposition}
The above highlights a desirable property of the update: only observing $Y_i^{(O_i)}$ affects the marginal distribution of $y^{(O_i)}$, but not the predictive distribution of $y^{(M_i)}$ given $y^{(O_i)}$. We highlight that the marginal distribution of $y^{(M_i)}$ can still change, but only due to mixing over the updated marginal of $y^{(O_i)}$.  In Section \ref{sec:freq}, we further verify frequentist consistency of the updates under the slightly stricter regime where the outcomes are covariate-dependent MAR, that is where $O_i$ may depend on $X_i$ but not $Y_i^{(O_i)}$. Finally, we briefly discuss predictive resampling -
it is natural to carry out predictive resampling with fully observed outcomes for identifiability, corresponding to drawing $X_i \iid P_X$ and always setting $O_i = \{1,\ldots,q\}$.

\subsection{Generalizations}

Generalizations to use other families beyond the Gaussian in the multi-output setting follow naturally from Sections \ref{sec:heavy} and \ref{sec:uni_gen}, and allow us to capture heavy-tailedness or other properties in the multi-output setting. In the interest of space, we defer further details to Appendix \ref{app:gen_mo}. 

\subsection{Practical considerations}

We briefly discuss some practical considerations which apply to both the single- and multi-output settings in the fully observed case. In Appendix \ref{app:imp}, we outline similar details for the missing-output updates. As the recursive approaches are not exchangeable, we follow \cite{Fong2023} and average over $10$ random permutations of the dataset when estimating $\mu_n(x)$ and $\sigma_n(x)$, averaging on the raw moment scale. For larger $n$, we use 1 or 2 permutations for hyperparameter tuning, but 10 for subsequent fitting. We standardize the outcomes and covariates, and initialize our estimates at $\mu_0(x) = 0$ and $\Sigma_0(x) = I_q$.  All hyperparameters ($\delta$, $\rho$ and $\rho_x$) are tuned by maximizing the prequential log-likelihood $\sum_{i = 1}^n \log p_{i-1}(Y_i \mid X_i)$. 
Fitting the model involves computing $Z_{1:n}$ which costs $O(n^2 (p + q^2))$, with an additional $O(n(p+q^2))$ for each prediction.  Although the cost grows quadratically with $n$, it is significantly faster than the cubic case for exact GPs (and even variational approximations), and orders of magnitude faster than MCMC approaches for dependent DP mixtures. 

\section{Asymptotic theory}\label{sec:asymptotics}
\subsection{Predictive asymptotics}\label{sec:pred_asym}
We now consider the theory underlying predictive resampling in our semiparametric martingale posterior. Since $p_i(y \mid x)$ is restricted to be parametric and is no longer a martingale, we do not satisfy the conditionally identically distributed (c.i.d.) property \citep{Berti2004}. Instead, we will leverage the theory of Banach-valued martingales \citep{Pisier2016, fong2025bayesian} to directly show convergence of $\mu_i(\cdot)$ and $\sigma_i^2(\cdot)$, which then drives convergence of the joint predictive density $p_i(y,x)$. 

We consider the univariate case first. As a reminder, we will assume throughout that $\mathcal{X}$ is compact. For convenience, we will usually omit the dependence on $x$. 
\begin{assumption}[Similarity]\label{as:bounded_sxx}
   The similarity function $s$ from (\ref{eq:dxx}) is well-defined on $\mathcal{X} \times \mathcal{X}$ and satisfies $S = \sup_{x \in \mathcal{X}} s(x,x) < \infty$. 
\end{assumption}
\begin{assumption}[Bandwidth/learning rate]\label{as:lr}
 Let $\rho \in (0,1)$ and $\alpha_i = O(i^{-m})$ where $0.5 < m \leq 1$ with $0 < \alpha_i < 1$ for all $i$.  
\end{assumption}
\begin{assumption}[Initialization]\label{as:init}
Let $\mu_0(x)=0$ and $\sigma_0^2(x) = 1$ for all $x \in \mathcal{X}$.
\end{assumption}
\begin{assumption}[Design]\label{as:design}
    Let the covariate distribution for predictive resampling $P_X$ have compact support $\mathcal{X}$ and Lebesgue density $p_X$. 
\end{assumption}
The first assumption holds if $s(x,x')$ is continuous and $\mathcal{X}$ is compact, and the second is satisfied as we have $\alpha_i = \left(2 - i^{-1}\right)(i+1)^{-0.9}$. Assumption \ref{as:init} can easily be relaxed, but this form is convenient. We then have our main convergence result.
\begin{theorem}\label{th:pred_asymp}
Under Algorithm \ref{alg:gauss} and Assumptions \ref{as:bounded_sxx}-\ref{as:design}, there exist random  $\mu_\infty \in L^2(\mathcal{X})$ and $\sigma_\infty^2 \in L^2(\mathcal{X})$, with $\sigma_\infty^2(x) > 0$ for Lebesgue-almost all $x \in \mathcal{X}$, such that
\begin{align*}
\int_{\mathcal{X}}\left[\mu_\infty(x) - \mu_i(x)\right]^2 dx \to 0, \quad \int_{\mathcal{X}}\left[\sigma_\infty^2(x) - \sigma_i^2(x)\right]^2 dx \to 0 \quad \textnormal{a.s.}
\end{align*}
Furthermore, we have the total variation convergence
\begin{align*}
    \frac{1}{2}\int_{\mathcal{X}} \left[\int \left|p_i(y \mid x) - \mathcal{N}\{y; \mu_\infty(x),\sigma_\infty^2(x)\} \right| dy\,\right] p_X(x) \, dx \to 0 \quad \textnormal{a.s.}
\end{align*}
\end{theorem}

The above theorem validates the convergence of predictive resampling, as the mean and variance functions and the predictive density all converge a.s. The total variation result can be viewed as a functional version of the results in \cite{garelli2024asymptotics}, and moment martingales in the non-regression setting were also investigated by \cite{yung2025moment}. We highlight that the above theorem is significantly stronger than previous results for the regression setting, e.g. \cite{Fong2023}, which only show a.s. weak convergence of the joint predictive distribution on $(Y,X)$. In particular, we have convergence of the \textit{conditional} moment functions, and the total variation result is much stronger than a.s. weak convergence as it means the limiting measure has a density $p_\infty(y,x)$, following \citet{berti2013exchangeable}.  The multi-output version of the above Theorem \ref{th:pred_asymp} is very similar and outlined in Appendix \ref{app:mv_pred_asymp}.

Under a smoothness condition on $s(x,x')$, we can show a stronger result that the martingale posterior has support over smooth mean and variance functions. For a Lipschitz domain $D$ and integer $r$, let $H^{r}(D) = W^{\,r,2}(D)$ denote the Sobolev space, where $H^r(D)$ is a Hilbert space with norm $\|\cdot\|_{H^r}$. 
\begin{assumption}[Sobolev regularity]\label{as:sob}
Let $\mathcal{X}=\overline D$, where $D\subset\mathbb{R}^p$
is a bounded Lipschitz domain. For some integer $r \geq 1$,
suppose that all partial derivatives of $s(x,x')$ with respect
to its first input $x$, up to and including total order $r$,
exist and are jointly continuous in
$(x,x')\in\mathcal X\times\mathcal X$. 
\end{assumption}
\begin{theorem}\label{th:pred_sobolev}
Under Algorithm \ref{alg:gauss} and Assumptions \ref{as:bounded_sxx}-\ref{as:sob}, there exist random functions $\mu_\infty \in H^r(D)$ and $\sigma_\infty^2 \in H^r(D)$, with $\sigma_\infty^2(x) > 0$ for Lebesgue-almost all $x$, such that
\begin{align*}
\|\mu_\infty - \mu_i \|_{H^r} \to 0, \quad \|\sigma_\infty^2 - \sigma_i^2\|_{H^r} \to 0 \quad \textnormal{a.s.}
\end{align*}
\end{theorem}
The control of the Sobolev norm follows from the nice property of the Gaussian having finite moments of all orders. We note that $r$ controls the smoothness of the resulting Sobolev space, which in turn is driven by the smoothness of $s(x,x')$. For $r > p/2$, it is well known that the Sobolev space $H^r$ can be embedded into (H\"{o}lder) continuous spaces \citep{leoni2017first}, e.g. $\mu_\infty$ and $\sigma_\infty^2$ are equal almost-everywhere to continuous representatives. Furthermore, if $r > k + p/2$ for some positive integer $k$, then functions in $H^r(D)$ have representatives with $k$ continuous derivatives. 

Unsurprisingly, the Gaussian choice of $s(x,x')$ in Example \ref{ex:Gaussian_cop} results in very smooth posterior samples, as we outline in the below corollary. 
\begin{corollary}\label{cor:inf}
   Under the assumptions of Theorem \ref{th:pred_sobolev}, for the Gaussian choice of $s(x,x')$ as in Example \ref{ex:Gaussian_cop}, posterior samples of $\mu_\infty$ and $\sigma_\infty^2$ have infinitely differentiable representatives a.s.
\end{corollary}
We also have the following extension to the heavy-tailed setting, which holds as the SU distribution has well-defined moments despite being heavy-tailed. 
\begin{corollary}\label{cor:SU}
Theorems \ref{th:pred_asymp} and \ref{th:pred_sobolev} both hold under the same assumptions if $p_i(y \mid x) = p^{\textnormal{SU}}(y; \mu_i(x), \sigma_i^2(x),\delta)$ for fixed $\delta > 0$ and the moment updates (\ref{eq:mean_SU}) and (\ref{eq:var_SU}) are used instead, and the total variation limit is $p^{\textnormal{SU}}(y; \mu_\infty(x), \sigma_\infty^2(x),\delta)$. 
\end{corollary}

\subsection{Frequentist asymptotics}\label{sec:freq}

In this section, we investigate the frequentist properties of our proposed recursive updates under i.i.d. samples $(Y_i,X_i) \iid P^*$. We let $\mu_*$, $\sigma_*^2$ and $\Sigma_*$ denote the true mean, variance and covariance matrix functions in the respective single and multi-output settings. The main difficulty in establishing consistency is that the mean and variance updates in (\ref{eq:mean_gauss}) and (\ref{eq:var_gauss}) are {coupled} in a nonlinear fashion. Other nonparametric recursive estimates for the mean function  \citep{ying2008online, lei2018convergence} rely on a linear update without a data-dependent variance component, and are thus much easier to study compared to our case.
Like other proofs for recursive predictive methods \citep{Tokdar2009, Hahn2018}, our results rely on the construction of an almost-supermartingale \citep{Robbins1971} to show convergence, followed by a consistency argument. However, due to the semiparametric nature of the update, novel techniques are needed. 

Our main finding is that our recursive moment updates are consistent under very weak assumptions on $P^*$, even if $p^*(y \mid x)$ is not Gaussian, and this extends to the multivariate case under covariate-dependent MAR.  The tradeoff of this generality is that consistency is measured using a norm induced by the similarity function $s(x,x')$, rather than in a stronger $L^1$ or $L^2$ sense. For comparison, in the simpler setting of single-output online mean function estimation, \citet{lei2018convergence} obtain a.s. $L^2$ consistency but under much stronger assumptions than our setting, e.g. requiring a bounded outcome space and the mean function living in a reproducing kernel Hilbert space.

\subsubsection{Single-output}
 We begin with the single-output case before extending to the general multi-output setting. To that end, we state some additional required assumptions. 
\begin{assumption}[DGP]\label{as:dgp}
Let $Y \in \mathbb{R}$, and suppose  $P_X^*$ has compact support $\mathcal{X}$ and Lebesgue density $p_X^*$,  and $E_{P^*}[Y^4] < \infty$ and $E_{P^*}\left[\left|\log \sigma_*^2(X)\right|\right]  < \infty$.  
\end{assumption}
\begin{assumption}[Bandwidth/learning rate]\label{as:freq_lr}
Let $\rho \in (0,1)$ and $\alpha_i = \alpha_0(i+1)^{-1}$, where {$0 < \alpha_0 < \min\{1,(2\rho^2S)^{-1}\}$} for $S$ defined in Assumption \ref{as:bounded_sxx}.
\end{assumption}
\begin{assumption}[Regularity of kernel]\label{as:regularity}
    The kernel $k(x;\psi)$ is jointly continuous in $(x,\psi)$, and $\inf_{x\in\mathcal X} E_\Psi[k(x;\Psi)]>0$. 
\end{assumption}
\begin{assumption}[Completeness of kernel] \label{as:complete}
The family $\{k(x;\psi):\psi \in \textnormal{supp}(\Pi)\}$ is complete in the sense of Definition \ref{def:complete} and its support contains $\mathcal{X}$.  
\end{assumption}
A sufficient condition for the integrability of $\log \sigma_*^2(x)$ in Assumption \ref{as:dgp} is that $\sigma_*^2(x)$ is continuous and $\sigma_*^2(x) > 0$ for all $x \in \mathcal{X}$ where $\mathcal{X}$ is compact. We believe that Assumption \ref{as:freq_lr} is stricter than necessary and is a technical artifact in the proof when controlling the almost-supermartingale, and we suspect consistency holds under a more general $\alpha_i = \alpha_0(i^{-m})$ for $0.5 < m \leq 1$. Finally, we see that completeness of the kernel will be responsible for ensuring consistency.

For a function $f \in L^1(P_X^*)$, let $\|\cdot  \|_s : L^1(P_X^*)\to [0,\infty)$ be defined as 
\begin{align*}
    \|f\|_s^2 = \int f(x) \, f(x') \, s(x,x') \, dP_X^*(x)\, dP_X^*(x').
\end{align*}
Next, let us consider the error functions
\begin{align*}
   u_n = \frac{\mu_* - \mu_n}{\sigma_n}, \quad w_n = \frac{\sigma_*^2 + (\mu_* - \mu_n)^2}{\sigma_n^2} - 1\cdot 
\end{align*}
\begin{theorem}\label{th:consist}
Under Assumptions \ref{as:bounded_sxx}, \ref{as:init}, \ref{as:dgp},  \ref{as:freq_lr}, the updates (\ref{eq:mean_gauss}) and (\ref{eq:var_gauss}) satisfy
\begin{align*}
    \|u_n\|_s \to 0, \quad \|w_n\|_s \to 0 \quad P^*\textnormal{-a.s.}
\end{align*}
Furthermore, under Assumptions \ref{as:regularity} and \ref{as:complete}, $\|\cdot \|_s$ is a norm on $L^1(P_X^*)$ and $\|u_n\|_s = \|w_n\|_s= 0$ if and only if  $\mu_n(x) = \mu_*(x)$ and $\sigma_n^2(x) = \sigma_*^2(x)$ for $P_X^*$-almost all $x$.
\end{theorem}
We first highlight that without the completeness condition in Assumption \ref{as:complete}, $\|f\|_s = 0$ may not necessarily give $f = 0$ for $P_X^*$-almost all $x$. Furthermore, one can show that the norm $\|\cdot\|_s$ is weaker than the more familiar $\|\cdot \|_{L^1(P_X^*)}$, due to the smoothing induced by $s(x,x')$. In Appendix \ref{app:norm_str}, we give further discussion on how the norm $\|\cdot\|_s$ is related to weak convergence in the function space. The main technical obstruction to strengthening the convergence is to control high-frequency oscillations in $u_n$ and $w_n$, which is nontrivial due to the nonlinear mean and variance update. Nonetheless, we see that this does not occur in practice, and we leave this for future work. {Finally, we note that these weak norm issues vanish if $\mathcal{X}$ is finite.}

The next corollary extends our consistency result to the generalized moment updates corresponding to the SU distribution given in Section \ref{sec:heavy}.  Note that here we require $\rho_i \to 1$ so the fixed points of the moment updates are the true moment functions, although in practice we recommend using a fixed prequentially tuned $\rho$. 
\begin{corollary}\label{cor:consist_gen}
Fix $\delta>0$ and let
$0<\rho_i<1$ satisfy $1-\rho_i=O(\alpha_i)$.
Under the assumptions of Theorem \ref{th:consist},
with bandwidth sequence $\rho_i$, and the strengthened condition
$0<\alpha_0<\min\{1,(2S)^{-1}\}$,
the conclusions of Theorem \ref{th:consist} hold for the moment updates
(\ref{eq:mean_SU}) and (\ref{eq:var_SU}).
\end{corollary}

\subsubsection{Multi-output}

Importantly, Theorem \ref{th:consist} can be generalized to the multivariate case with missing data. Here, we consider the $q = 2$ case with order-compatible missingness as it is the clearest to present; the general case with any $q$ and arbitrary missingness pattern also holds true, which we leave for Appendix \ref{app:mv_gen_thm}. Let $(Y_i,X_i,O_i) \iid P^*$, and let $S_i$ be the random selection matrix.  We have the multi-output version of Assumption \ref{as:dgp}. 
\begin{assumption}[Multi-output]\label{as:dgp_mv}
    Let $Y \in \mathbb{R}^2$, $O \in \mathcal{O} =\{ \{1\}, \{1,2\}\}$, and suppose $P_X^*$ has compact support $\mathcal{X}$ and Lebesgue density $p_X^*$. Assume $E_{P^*}[\|Y\|^4] < \infty$ and  $E_{P^*}\left[\left|\log \det \Sigma_*(X)\right|\right]  < \infty$. Furthermore, assume covariate-dependent MAR, that is $O \perp Y \mid X$, with $P^*(O = \{1,2\} \mid x) > 0$ for $P_X^*$-almost all $x$. 
\end{assumption}
 For a vector-valued function
$f\in L^1(P_X^*;\mathbb R^q)$ and a matrix-valued function
$F\in L^1(P_X^*;\mathbb R^{q\times q})$, define
$\|f\|_{s}^2 = \sum_{j = 1}^q \|f^{(j)}\|_s^2$ and $\|F\|_{s}^2 = \sum_{j,k = 1}^q \|F^{(j,k)}\|_s^2$. Let us define the vector/matrix-valued error functions
\begin{align*}
    u_n = L_n^{-1}(\mu_* - \mu_n), \quad W_n = L_n^{-1}\left[\Sigma_* + (\mu_* - \mu_n)(\mu_* - \mu_n)^\top\right] L_n^{-\top} - I_q,
\end{align*}
and let the projection matrix be $Q^o = \left[\begin{smallmatrix}
    1 & 0 \\
    0 & 0 
\end{smallmatrix}\right]$ if $o = \{1\}$, and $Q^o = I_2$ if $o = \{1,2\}$. Next, we define the projection-averaged versions of $u_n$ and $W_n$:
\begin{align*}
    \bar{u}_n(x) = \sum_{o \in \mathcal{O}} p^*(o \mid x) \, \left[Q^o \,u_n(x)\right], \quad \overline{W}_n(x) = \sum_{o \in \mathcal{O}} p^*(o \mid x)\, \left[Q^o\, W_n(x) \,Q^o \right],
\end{align*}
where $p^*(o \mid x) = P^*(O =o \mid X= x)$. The above terms are thus averaging the error functions over the possible missingness patterns. We then have the main result. 
\begin{theorem}\label{th:consist_mv}
Under Assumptions \ref{as:bounded_sxx},  \ref{as:freq_lr}, \ref{as:dgp_mv}, \ref{as:mv_init},  the updates (\ref{eq:mean_gauss_missing}) and (\ref{eq:cov_gauss_missing}) satisfy
\begin{align*}
    \|\bar{u}_n\|_s \to 0, \quad \|\overline{W}_n\|_s \to 0 \quad P^*\textnormal{-a.s.}
\end{align*}
Furthermore, under Assumptions \ref{as:regularity} and \ref{as:complete}, $\|\cdot \|_s$ is a norm on the respective spaces $L^1(P_X^*; \mathbb{R}^q)$ and $L^1(P_X^*; \mathbb{R}^{q\times q})$,   and $\|\bar{u}_n\|_s = \|\overline{W}_n\|_s= 0$ if and only if $\mu_n(x) = \mu_*(x)$ and $\Sigma_n(x) = \Sigma_*(x)$ for $P_X^*$-almost all $x$. 
\end{theorem}
In the fully observed case with $P^*(O = \{1,2\} \mid x) = 1$ for $P_X^*$-almost all $x$, it is clear that $\bar{u}_n = u_n$ and $\overline{W}_n = W_n$ as expected, which is the natural multivariate version of Theorem \ref{th:consist}. Otherwise, we note that the measure used for consistency depends on the missingness pattern, where a positive probability of observing $O = \{1,2\}$ for each $x$ ensures that the covariance matrix is identifiable.  The extension to larger $q$ and general ordering follows naturally, which we outline in Appendix \ref{app:mv_gen_thm}.

\section{Illustrations}\label{sec:illustrations}

In this section, we illustrate the performance of our proposed recursive moment (RM) updates in a range of single and multi-output examples.
All recursive updates are implemented in \texttt{JAX} in Python, and
all timings are carried out on an Apple M2 Pro CPU. Additional speed-up for predictive resampling may be obtained with parallelization if desired, but it is sufficiently fast already. Throughout, we choose $P_X$ to be uniform on $\mathcal{X}$ as the covariate space is compact in our examples.

\subsection{Simulations}\label{sec:sim}

All simulations are repeated for 20 seeds, and run-times and the following evaluation metrics are reported. We evaluate test data prediction and the mean squared error for point estimates of $\mu_*$ and $\sigma_*^2$. To evaluate the uncertainty quantification, we follow \cite{wade2025bayesian} and report the empirical coverage  defined as 
$\text{Cov}_{\mu} = \int {1}(\mu_*(x) \in C_\mu(x))\, dP_X^*,$
where $C_\mu(x)$ is the point-wise 95\% credible interval for $\mu(x)$, and $\mu_*(x)$ is the true mean function, where we note that we do not necessarily expect asymptotic calibration. Average interval lengths are defined as $\text{Len}_{\mu} = \int | C_\mu(x)|\, dP_X^*$, and coverage/length can be evaluated on a grid or by averaging over the held-out test set. We report both quantities for $\sigma^2(x)$ as well. 

\textbf{Heavy-tailed\hspace{5mm}} First, we cover the Huber-contaminated setting of Example \ref{ex:kurt}. For comparison, we fit the normalized-weights dependent DP (DDP) mixture of \citet{Antoniano2014} with their suggested hyperparameters. The MCMC sampler is based on the Matlab implementation of \cite{wade2025bayesian}, with a burn-in of 3000 iterations followed by 1000 retained draws with thinning every second iteration. We also compare to the standard GP with the RBF kernel with length-scale optimized through the marginal likelihood, which is implemented in \texttt{sklearn} in Python. Note that coverage for $\sigma^2(\cdot)$ is not reported for the GP since the noise variance is a fixed constant.  Coverage/interval length are estimated on an $x$-grid of size $50$, and test log-likelihood is evaluated on a held-out set of size $2000$.

Table \ref{tab:sim-kurt} shows that RM (SU) has the smallest MSE for both $\mu$ and $\sigma^2$ in both settings, with the best or very competitive test prediction. Only the DDP mixture outperforms in prediction, which is unsurprising as the DGP is a Gaussian mixture. The run-times for the RM are also orders of magnitude smaller than the DDP mixture and copula approaches, and even faster than the GP for $n = 1000$. RM attains good coverage with reasonable interval lengths comparable to the DDP mixture for $\mu$.
For $\sigma^2$, RM (SU) produces intervals roughly half as wide as the DDP mixture, with comparable coverage at $n=200$ but worse coverage at $n=1000$. Finally, we notice that the gap between RM (Gauss) and RM (SU) shrinks as $n$ increases, and RM (Gauss) is still consistent under non-Gaussian noise.

\begin{table*}[t]
\centering
\scriptsize
\setlength{\tabcolsep}{4.5pt}
\renewcommand{\arraystretch}{1.12}
\caption{Heavy-tailed contamination simulation. Boldface marks the smallest MSE and run-time and the largest held-out log score for each sample size; standard errors in parentheses.}
\label{tab:sim-kurt}
\resizebox{\textwidth}{!}{%
\begin{tabular}{lrrrrrrrr}
\toprule
& \multicolumn{2}{c}{Estimation} & \multicolumn{2}{c}{Mean UQ} & \multicolumn{2}{c}{Variance UQ} & \multicolumn{1}{c}{Prediction} & \multicolumn{1}{c}{Run-time} \\
\cmidrule(lr){2-3}\cmidrule(lr){4-5}\cmidrule(lr){6-7}\cmidrule(lr){8-8}\cmidrule(lr){9-9}
Method & \multicolumn{1}{c}{$\mathrm{MSE}_\mu\times 10^{2}$} & \multicolumn{1}{c}{$\mathrm{MSE}_{\sigma^2}$} & \multicolumn{1}{c}{$\mathrm{Cov}_{\mu}$} & \multicolumn{1}{c}{$\mathrm{Len}_{\mu}$} & \multicolumn{1}{c}{$\mathrm{Cov}_{\sigma^2}$} & \multicolumn{1}{c}{$\mathrm{Len}_{\sigma^2}$} & \multicolumn{1}{c}{Avg $\log p(y_{\mathrm{test}})$} & \multicolumn{1}{c}{Total (s)} \\
\midrule
\multicolumn{9}{l}{\textit{$n=200$}} \\
GP & 10.90 (3.09) & 16.60 (1.13) & 0.90 (0.03)\hphantom{0} & 0.95 (0.06)\hphantom{0} & / & / & -2.00 (0.03)\hphantom{0} & \textbf{0.36} \\
DDP & 17.56 (8.04) & 17.89 (3.45) & 0.95 (0.009) & 1.18 (0.06)\hphantom{0} & 0.85 (0.03)\hphantom{0} & 11.35 (2)\hphantom{.000} & -1.05 (0.009) & 362 \\
Cop & 4.15 (0.62) & 10.22 (0.96) & 0.99 (0.006) & 1.24 (0.05)\hphantom{0} & 0.75 (0.04)\hphantom{0} & 2.80 (0.3)\hphantom{00} & -1.10 (0.02)\hphantom{0} & 188 \\
RM (Gauss) & 11.98 (3.91) & 9.56 (1.11) & 0.96 (0.01)\hphantom{0} & 1.50 (0.1)\hphantom{00} & 0.33 (0.04)\hphantom{0} & 1.97 (0.3)\hphantom{00} & -1.74 (0.03)\hphantom{0} & 0.49 \\
RM (SU) & \makebox[0pt][l]{\textbf{4.12 (0.61)}}\hphantom{4.12 (0.61)} & \makebox[0pt][l]{\textbf{7.86 (1.45)}}\hphantom{7.86 (1.45)} & 0.97 (0.009) & 1.18 (0.05)\hphantom{0} & 0.90 (0.03)\hphantom{0} & 6.10 (0.8)\hphantom{00} & \makebox[0pt][l]{\textbf{-1.02 (0.007)}}\hphantom{-1.02 (0.007)} & 0.79 \\
\addlinespace[4pt]
\multicolumn{9}{l}{\textit{$n=1000$}} \\
GP & 2.10 (0.41) & 14.25 (0.30) & 0.92 (0.02)\hphantom{0} & 0.47 (0.02)\hphantom{0} & / & / & -1.93 (0.04)\hphantom{0} & 2.29 \\
DDP & 2.06 (0.31) & 7.25 (0.99) & 0.94 (0.009) & 0.49 (0.02)\hphantom{0} & 0.92 (0.02)\hphantom{0} & 5.54 (0.7)\hphantom{00} & \makebox[0pt][l]{\textbf{-0.96 (0.007)}}\hphantom{-0.96 (0.007)} & 1625 \\
Cop & 1.56 (0.28) & 9.72 (0.57) & 0.94 (0.02)\hphantom{0} & 0.56 (0.02)\hphantom{0} & 0.54 (0.04)\hphantom{0} & 1.37 (0.1)\hphantom{00} & -1.06 (0.01)\hphantom{0} & 189 \\
RM (Gauss) & 2.66 (0.46) & 5.98 (0.91) & 0.89 (0.03)\hphantom{0} & 0.70 (0.04)\hphantom{0} & 0.24 (0.03)\hphantom{0} & 1.00 (0.1)\hphantom{00} & -1.64 (0.03)\hphantom{0} & \textbf{1.04} \\
RM (SU) & \makebox[0pt][l]{\textbf{1.07 (0.21)}}\hphantom{1.07 (0.21)} & \makebox[0pt][l]{\textbf{4.94 (0.61)}}\hphantom{4.94 (0.61)} & 0.94 (0.02)\hphantom{0} & 0.56 (0.01)\hphantom{0} & 0.63 (0.06)\hphantom{0} & 2.36 (0.2)\hphantom{00} & -1.00 (0.008) & 1.94 \\
\bottomrule
\end{tabular}%
}
\end{table*}

\textbf{Multivariate $\mathbf{x}$\hspace{5mm}} Next, we carry out a multivariate covariate example which extends Example 3 of
\citet{wade2025bayesian}. We generate $n=600$ observations with
$X^{(j)}\iid\operatorname{Unif}(-2,2)$, $j=1,\ldots,5$, and $p^*(y \mid x)
     = \mathcal N\!\left(y; m_{\mathrm{mv}}(x),s_{\mathrm{mv}}^2(x)\right)$ where
\[
 t(x)=\sin\!\left(\frac{\pi x^{(1)}\,x^{(2)}}{2}\right),\quad
 m_{\mathrm{mv}}(x)=
 \begin{cases}1,&t(x)\leq0, \\-1,&t(x)>0,\end{cases}
 \quad
 s_{\mathrm{mv}}(x)=
 \begin{cases}0.1,&t(x)\leq0,  \\1,&t(x)>0.\end{cases}
\]
Note that $X^{(3)},X^{(4)},X^{(5)}$ are inactive covariates. Here we use similar settings as in Example 3 of \citet{wade2025bayesian} for the DDP mixture, with the same normalized weights hyperparameters for each of the $5$ covariate dimensions. Coverage and interval length are now estimated on the held-out test set of size $2000$ instead of a grid. We do not report the copula results as the memory/run-time requirements are too high due to the larger test set required for predictive resampling. 

As RM and GP involve automatic hyperparameter tuning, they can identify that only the first two covariates are relevant. This is much more challenging for the DDP mixture, as the marginal likelihood is intractable and each MCMC run takes a significant amount of time; the spurious covariates are likely responsible for the poor performance of the DDP mixture. Again, we see that the RM approaches are very competitive for estimation and prediction, with RM (Gauss) attaining the best MSE for $\sigma^2$ as the noise is Gaussian but RM (SU) still scoring the best for prediction. The gap in run-time to the DDP case is even more noticeable in this setting due to the multivariate covariates, with RM run-times beating even the GP.

\begin{table*}[t]
\centering
\scriptsize
\setlength{\tabcolsep}{4.5pt}
\renewcommand{\arraystretch}{1.12}
\caption{Multivariate simulation. Boldface marks the smallest MSE and run-time and the largest held-out log score for each sample size; standard errors in parentheses.}
\label{tab:sim-mv}
\resizebox{\textwidth}{!}{%
\begin{tabular}{lrrrrrrrr}
\toprule
& \multicolumn{2}{c}{Estimation} & \multicolumn{2}{c}{Mean UQ} & \multicolumn{2}{c}{Variance UQ} & \multicolumn{1}{c}{Prediction} & \multicolumn{1}{c}{Run-time} \\
\cmidrule(lr){2-3}\cmidrule(lr){4-5}\cmidrule(lr){6-7}\cmidrule(lr){8-8}\cmidrule(lr){9-9}
Method & \multicolumn{1}{c}{$\mathrm{MSE}_\mu\times 10^{1}$} & \multicolumn{1}{c}{$\mathrm{MSE}_{\sigma^2}\times 10^{1}$} & \multicolumn{1}{c}{$\mathrm{Cov}_{\mu}$} & \multicolumn{1}{c}{$\mathrm{Len}_{\mu}$} & \multicolumn{1}{c}{$\mathrm{Cov}_{\sigma^2}$} & \multicolumn{1}{c}{$\mathrm{Len}_{\sigma^2}$} & \multicolumn{1}{c}{Avg $\log p(y_{\mathrm{test}})$} & \multicolumn{1}{c}{Total (s)} \\
\midrule
\multicolumn{9}{l}{\textit{$n=600$}} \\
GP & \makebox[0pt][l]{\textbf{2.22 (0.05)}}\hphantom{2.22 (0.05)} & 2.62 (0.03) & 0.81 (0.01)\hphantom{0} & 1.25 (0.04)\hphantom{0} & / & / & -1.26 (0.006) & 2.51 \\
DDP & 11.49 (0.18) & 5.04 (0.14) & 0.83 (0.004) & 2.95 (0.04)\hphantom{0} & 0.53 (0.007) & 2.65 (0.05)\hphantom{0} & -1.63 (0.01)\hphantom{0} & 10988 \\
RM (Gauss) & 2.35 (0.05) & \makebox[0pt][l]{\textbf{1.68 (0.07)}}\hphantom{1.68 (0.07)} & 0.85 (0.01)\hphantom{0} & 1.52 (0.04)\hphantom{0} & 0.39 (0.009) & 0.69 (0.02)\hphantom{0} & -1.08 (0.01)\hphantom{0} & \textbf{1.57} \\
RM (SU) & 2.31 (0.04) & 2.69 (0.24) & 0.90 (0.006) & 1.74 (0.03)\hphantom{0} & 0.63 (0.01)\hphantom{0} & 2.15 (0.1)\hphantom{00} & \makebox[0pt][l]{\textbf{-0.92 (0.01)\hphantom{0}}}\hphantom{-0.92 (0.01)\hphantom{0}} & 2.09 \\
\bottomrule
\end{tabular}%
}
\end{table*}

\subsection{Real data analysis}
We now analyze a large-scale spatiotemporal weather dataset within the multi-output regression setting based on the curated WEATHER-5K dataset of \citet{han2026benchmarking}. We study the annual average temperature (TMP), dew point (DEW) and sea level pressure (SLP) readings from $1033$ weather stations based in the contiguous United States from 2014-2023, resulting in $n = 10330$ observations. The multi-output regression task is thus to regress the $q = 3$ weather outcomes given the weather station's longitude and latitude, and the year of measurement ($p = 3$). Our model is particularly suitable here, as the correlations between TMP, DEW and SLP vary noticeably over the US geography, e.g. \citet{gentilli1955estimating}. Further details on data preprocessing can be found in Appendix \ref{app:weather}.

To evaluate our method, we hold out 20\%  of the dataset stratified by year as test data. Among the remaining 80\% training observations ($n = 8264$), we randomly omit 20\% of DEW measurements for weather stations with  $\text{longitude}>-100^\circ$ to simulate covariate-dependent missing outcomes. We tune the hyperparameters of our approach using $1$ data permutation, fit to 10 data permutations, and use the output ordering $\{\text{TMP, SLP, DEW}\}$ so the missingness is order-compatible. For predictive resampling, we again use $N = n + 4000$ and obtain $B = 1000$ samples. 

As baselines, we compare to a full-rank intrinsic coregionalization model (ICM) implemented in \texttt{GPflow} in Python, and the more flexible GP autoregressive regression model (GPAR,  \citet{requeima2019gaussian}) implemented in \texttt{PyTorch} in Python. For ICM, exact inference is infeasible due to the $O((qn)^3)$ complexity, so we only consider sparse variational inference with $m$ inducing points. For GPAR, exact inference is possible but costs an expensive $O(qn^3)$, and we also consider approximate inference with inducing points.  The authors' implementation does not allow for variational optimization of the inducing points, so we choose a uniform grid of $m' = 10 \times l \times l$ points for time, longitude and latitude within the support.

\begin{figure}[!h]
    \centering
    \includegraphics[width=\linewidth]{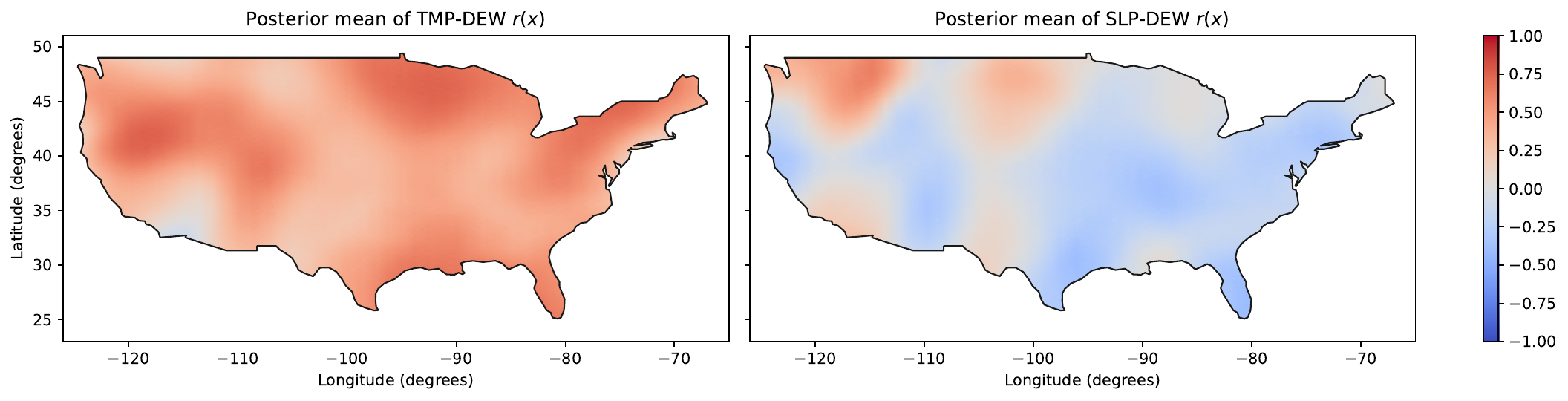}
    
    \caption{Posterior mean of the correlation function $r(x)$ between  TMP-DEW (left) and SLP-DEW (right) in the year 2021 under the recursive moment updates}
    \label{fig:weather_corr}
\end{figure}

\begin{figure}[!h]
    \centering
    \includegraphics[width=\linewidth]{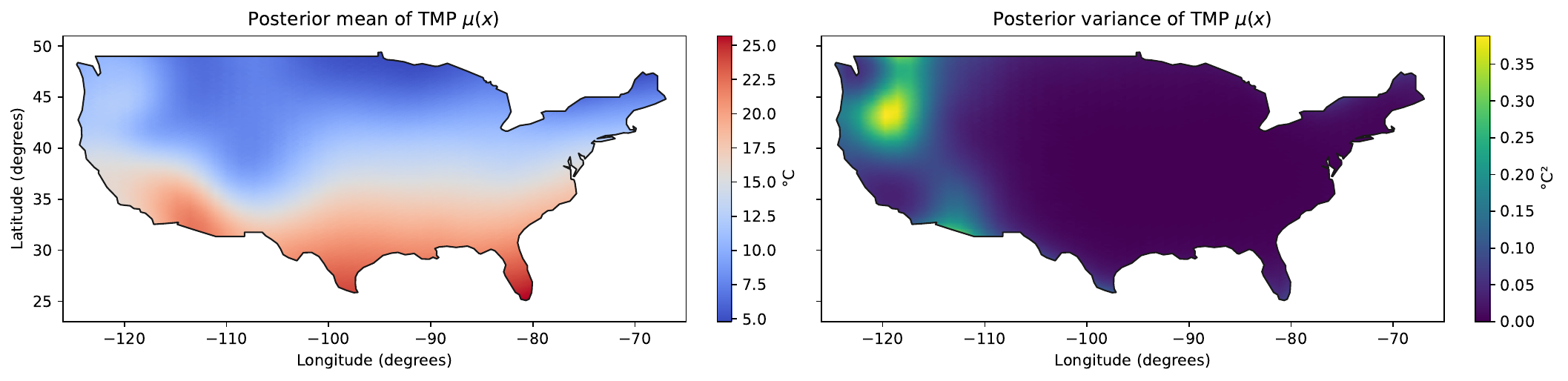}
    
    \caption{Posterior mean (left) and variance (right) of the mean function $\mu(x)$ for TMP in the year 2021 under the recursive moment updates}
    \label{fig:weather_tmp}
    
\end{figure}

Table \ref{tab:weather_results} reports the average log predictive density separately on the 2066 held-out full data points and the 1139 held-out DEW measurements (conditioned on TMP and SLP). We also report run-times, where the fit column corresponds to hyperparameter tuning and fitting, and the total time includes additional method-specific prediction time which we outline shortly. Figure \ref{fig:weather_corr} plots the posterior mean of the outcome correlation function $r(x)$ between TMP-DEW and SLP-DEW across varying longitude/latitude in the year 2021. Figure \ref{fig:weather_tmp} similarly plots the posterior mean and variance of the mean function $\mu(x)$ of TMP, where the uncertain regions are driven by a larger heteroscedastic variance. 
\begin{table}[!h]
\scriptsize
\setlength{\tabcolsep}{7pt}
\renewcommand{\arraystretch}{1.12}
\caption{Weather data results. ICM is run with 5000 optimizer steps. RM (Independent) plugs in the same hyperparameters as RM (Multi-output), so we do not report run-times.
Total run-time includes method-specific prediction time on the $50 \times 50$ longitude-latitude grid for the year 2021.}
\centering
\label{tab:weather_results}
\begin{tabular}{lrrrr}
\toprule
& \multicolumn{2}{c}{Avg test log pdf}
& \multicolumn{2}{c}{Run-time (s)} \\
\cmidrule(lr){2-3}\cmidrule(lr){4-5}
Method
& \multicolumn{1}{c}{Full}
& \multicolumn{1}{c}{DEW}
& \multicolumn{1}{c}{Fit}
& \multicolumn{1}{c}{Total} \\
\midrule
RM (Multi-output) & \textbf{-4.136} & -1.076 & \textbf{50} & \textbf{111} \\
RM (Independent) \vspace{1mm} & -4.321 & -1.239 & / & / \\
ICM ($m=64$) & -5.661 & -1.476 & 177 & 178 \\
ICM ($m=128$) & -5.496 & -1.410 & 431 & 431 \\
ICM ($m=256$) \vspace{1mm} & -5.377 & -1.360 & 1137 & 1137 \\
GPAR ($l=10$) & -5.093 & -1.101 & 52 & 163 \\
GPAR ($l=20$) & -4.758 & -1.048 & 1063 & 1671 \\
GPAR (exact) & -4.389 & \textbf{-1.010} & 1362 & 2479 \\
\bottomrule
\end{tabular}
\end{table}
We highlight that the RM multi-output approach has the best test log likelihood for the held-out dataset, and is very competitive with the best approach (exact GPAR) for the missing DEW measurements. As a sanity check, the multi-output RM noticeably improves performance compared to 3 independent RM fits using the same hyperparameters. RM's strong performance is likely due to the ability to capture heterogeneous covariance structure between the outcomes, e.g. Figure \ref{fig:weather_corr}. The ICM is unable to capture this structure, and while GPAR is more flexible, it does not allow for a full covariate-dependent outcome covariance structure as the order-dependence is an assumption about the DGP, e.g. the first outcome in the ordering is assumed to have homoscedastic noise. This highlights a stark advantage of our approach: RM is fully flexible and consistent under weak assumptions, whereas GP approaches require complex constructions to attain flexibility.  We further note that GP performance is very sensitive to approximation, as we see large deterioration in performance for GPAR as $m'$ decreases, and a similar but less drastic trend for ICM.

The main strength however of the RM approach is the rapid computation time. RM's exact fit time is faster than GP approximations with even a small number of inducing points, and is orders of magnitude faster than the exact GP. For RM's total run-time, we include the time for predictive resampling on a $50 \times 50$ longitude-latitude grid for the year 2021 as illustrated in Figures \ref{fig:weather_corr} and  \ref{fig:weather_tmp}. Estimating $\mu(x)$ for GPAR requires Monte Carlo sampling for each $x$ (using 200 samples) on the same longitude-latitude grid in 2021, so this is taken into account for GPAR's total time. We further note that estimating the posterior variance of $\mu(x)$ for GPAR would require additional computation due to the intractable autoregressive structure. On the other hand, ICM prediction is quick after fitting due to the tractable structure, but the tradeoff is a lack of flexibility.

\section{Discussion}\label{sec:discussion}

In this paper, we have introduced a class of recursive moment updates for Bayesian nonparametric regression, which covers a wide array of settings including heteroscedastic, heavy-tailed and multi-output settings. In addition to being flexible with theoretical guarantees, the approach is expedient and does not require MCMC for uncertainty quantification. An interesting direction for future methodological research is exploration of more complex choices of the similarity function $s(x,x')$, perhaps incorporating machine learning or dimension-reduction techniques. Another ambitious extension is to allow covariate-dependent skewness by matching the third moment. Another important direction for future research is a strengthening of the theoretical results, including stronger consistency guarantees and convergence rates, although new technical tools will be needed. 

\section*{Acknowledgments}
Edwin Fong receives funding from the Research Grants Council of Hong Kong through the Early Career Scheme (Grant 27304424) and the General Research Fund (Grant 17306925). The authors thank Joshua Rooijakkers for helpful discussions. ChatGPT Astra was used to assist with coding, proofreading, and reviewing the presentation of technical proofs during the preparation of this manuscript. The authors take full responsibility for the content.

\section*{Data availability}
All datasets used in this paper are publicly available. Code for reproducing the  results can be found at \url{https://github.com/edfong/recursive_moments}.

\begingroup
\setstretch{1.0}
\bibliographystyle{abbrvnat}
\bibliography{paper-ref}
\endgroup

\newpage
\begin{appendices}
\counterwithin{assumption}{section}
\counterwithin{lemma}{section}
\counterwithin{theorem}{section}
\counterwithin{proposition}{section}

\section{Notation}

We begin with some notation choices for the Appendix to aid in proof readability. A repeatedly used term is $\alpha_i(x,X_i)$, which we will write as the shorthand
\begin{align*}
    \beta_i(x) = \alpha_i(x,X_i) = \frac{\alpha_i s(x,X_i)}{1-\alpha_i + \alpha_i s(x,X_i)}\cdot
\end{align*}
As we will be working with functions, we will sometimes omit the dependence on the test point $x$, e.g. we can write the single-output updates as
\begin{align*}
\mu_i &= \mu_{i-1} + \beta_i \, \sigma_{i-1} \,\rho Z_i, \qquad \sigma_i^2 = \sigma_{i-1}^2\left(1 + \beta_i\,\rho^2 \left[(1-\beta_i)Z_i^2 - 1 \right]\right). 
\end{align*}
In some cases, we will evaluate a function $f$ at both $x$ and $x'$, in which case we will use the shorthand $f(x') = f'$. Finally, whenever $\alpha_i$ is observed with no arguments, it will be assumed to be the $O(i^{-m})$ scalar sequence and not the function sequence.

We use the notation $x\lesssim y$ to denote the existence of a large constant $C$ such that $x \leq C y$. In the cases where an explicit constant is helpful, we will use $C$ which may take on different values in different parts of the proofs. Similarly, we use $R_n$ to denote remainders that may take on different values. 

\section{Proposition proofs}

\subsection{Proof of Proposition \ref{prop:SU_update}}\label{app:gen_th}

It is clearer to first present the form of the mean and variance update under a general transformation $g_{\lambda}$, before substituting the SU case. 
\begin{proposition}\label{prop:gen_update}
Let $P(\,\cdot\,; \mu,\sigma,\lambda)$ for $\mu \in \mathbb{R}$  and $\sigma > 0$ denote the family induced by $Y = \mu + \sigma g_{\lambda}(\varepsilon)$ with $\varepsilon \sim \mathcal{N}(0,1)$,
where $g_{\lambda}:\mathbb{R} \to \mathbb{R}$ is a strictly increasing bijection satisfying $E[g_{\lambda}(\varepsilon)] = 0$ and $E[g_{\lambda}(\varepsilon)^2] = 1$ for all $\lambda \in \Lambda$. 
 Let 
 $$P_{i-1}(y \mid x) = P\left(y; \mu_{i-1}(x), \sigma_{i-1}(x),\lambda\right)$$
 for fixed $\lambda \in \Lambda$. Then the mean and variance function updates corresponding to the nonparametric copula update take the form
    \begin{align*} 
    \mu_i &= \mu_{i-1} + \beta_i \, \sigma_{i-1}\, R^{(1)}_{\lambda}(\varepsilon_i), \\
\sigma_i^2
&=
\sigma_{i-1}^2
\left[
1
+
\beta_i \left\{R^{(2)}_{\lambda}(\varepsilon_i) -1
-
\beta_i\, R^{(1)}_{\lambda}(\varepsilon_i)^2
\right\}\right],
    \end{align*}
    where the update terms take the form
 \begin{align*}
 R^{(j)}_{\lambda}(\varepsilon_i) = E_{W\sim \mathcal{N}(0,1)}\left[g_{\lambda}\left(\sqrt{1-\rho^2}W + \rho \varepsilon_i\right)^j\right]
\end{align*}
    and $\varepsilon_i= g_{\lambda}^{-1}(Z_i)$ with 
    $$Z_i = \frac{Y_i - \mu_{i-1}(X_i)}{\sigma_{i-1}(X_i)}\cdot $$
    Furthermore, $\varepsilon_i \sim \mathcal{N}(0,1)$ under predictive resampling.
\end{proposition}

\begin{proof}
The copula update for $u_i = P_i(y \mid x)$ is
\begin{equation*}
    u_i
    =
    \left(
        1-\beta_i\right) u_{i-1}
        +
        \beta_i \,
        H_\rho\left(u_{i-1}, v_{i-1}\right)
\end{equation*}
where $v_{i-1} = P_{i-1}(Y_i\mid X_i)$ and
\begin{align*}
    H_\rho(u,v) = \Phi\left\{\frac{\Phi^{-1}(u) - \rho \Phi^{-1}(v)}{\sqrt{1-\rho^2}}\right\}\cdot
\end{align*}
First, we note that
\begin{align*}
    u_{i-1} = P\left(\mu_{i-1} + \sigma_{i-1}\, g_{\lambda}(\varepsilon) \leq y\right) =\Phi\left\{g_{\lambda}^{-1}\left(\frac{y-\mu_{i-1}(x)}{\sigma_{i-1}(x)}\right)\right\}\cdot 
\end{align*}
A similar argument gives $v_{i-1} = \Phi(\varepsilon_i)$. Therefore, we have
\begin{align*}
    H_\rho(u_{i-1},v_{i-1}) = \Phi\left\{\frac{g_{\lambda}^{-1}\left(\frac{y- \mu_{i-1}}{\sigma_{i-1}}\right) -\rho \varepsilon_i}{\sqrt{1-\rho^2}} \right\}.
\end{align*}
 Now let $W \sim \mathcal{N}(0,1)$ and define
\begin{align*}
    \widetilde{Y}_i = \mu_{i-1} + \sigma_{i-1} \,g_{\lambda}\left(\sqrt{1-\rho^2}\,W + \rho \varepsilon_i\right).
\end{align*}
We thus have $\text{Pr}(\widetilde{Y}_i\leq y ) =H_\rho(u_{i-1},v_{i-1})$. The mean of $P_i(\, \cdot \mid x)$ is then equal to
\begin{align*}
    \mu_i = \left(
        1-\beta_i\right) \mu_{i-1}(x) + \beta_i \, E\left[\widetilde{Y}_i\right]= \mu_{i-1} + \beta_i\, \sigma_{i-1}\, R_{\lambda}^{(1)}(\varepsilon_i).
\end{align*}
For the conditional second moment function $m_i^{(2)}$, we similarly have
\begin{align*}
    m_i^{(2)}(x) &= \left(
        1-\beta_i\right) m_{i-1}^{(2)} + \beta_i\,E\left[\widetilde{Y}_i^2\right]\\
        &= m_{i-1}^{(2)} + \beta_i\left[\sigma_{i-1}^2 \left\{R_{\lambda}^{(2)}(\varepsilon_i) - 1\right\}+  2 \mu_{i-1} \, \sigma_{i-1}\,R_{\lambda}^{(1)}(\varepsilon_i)\right].
\end{align*}
The identity $\sigma_i^2 = m_i^{(2)} - \mu_i^2$ then gives us the update for the variance. The predictive resampling result follows directly from the invertibility of $g_{\lambda}$.

We end with a technical point, and show that finite $E[g_{\lambda}(\varepsilon)^2]$ is sufficient to give $R_{\lambda}^{(j)}(\varepsilon_i) < \infty$ for $j \in \{1,2\}$. One can see that for $V_i \sim \mathcal{N}(\rho \varepsilon_i, 1-\rho^2)$, we have
\begin{align*}
    R_{\lambda}^{(2)}(\varepsilon_i)  &= E\left[g_{\lambda}(V_i)^2 \right]= \int g_{\lambda}(v)^2 \, \mathcal{N}(v;0,1)\, \frac{\mathcal{N}(v; \rho \varepsilon_i, 1-\rho^2)}{\mathcal{N}(v;0,1)}\, dv.
\end{align*}
Since $0 < \rho < 1$, the density ratio is bounded (as a function of $v$) for each $\varepsilon_i$,  so $E[g_{\lambda}(\varepsilon)^2] < \infty$ implies $R_{\lambda}^{(2)}(\varepsilon_i) < \infty$. Cauchy-Schwarz then gives  $R_{\lambda}^{(1)}(\varepsilon_i) < \infty$.
\end{proof}
Proposition \ref{prop:SU_update} in the main paper then follows from directly calculating $R_{\delta}^{(j)}(\varepsilon_i)$, which we demonstrate for $j = 1$. Here, for $\delta > 0$ and $\omega = \exp(1/\delta^2)$, we have
\begin{align}\label{eq:g_delt}
    g_{\delta}(\varepsilon) = \frac{\sinh\left(\frac{\varepsilon}{\delta}\right)}{\sqrt{\frac{1}{2}(\omega^2 - 1)}}
\end{align}
Next, we note that
\begin{align*}
E\left[\sinh\left(\frac{V_i}{\delta}\right)\right] = \frac{1}{2}\left\{E\left[\exp\left(\frac{V_i}{\delta}\right)\right] - E\left[\exp\left(-\frac{V_i}{\delta}\right)\right]\right\}
\end{align*}
where $V_i \sim \mathcal{N}(\rho\varepsilon_i,1-\rho^2)$ as before. The two expectations correspond to Gaussian moment generating functions evaluated at $t = 1/\delta$ and $t = -1/\delta$ respectively, so 
\begin{align*}
    E\left[\sinh\left(\frac{V_i}{\delta}\right)\right] &= \frac{1}{2}\left\{ \exp\left(\frac{\rho\varepsilon_i}{\delta} + \frac{1-\rho^2}{2\delta^2}\right) - \exp\left(-\frac{\rho\varepsilon_i}{\delta} + \frac{1-\rho^2}{2\delta^2}\right)\right\}\\
    &= \omega^{(1-\rho^2)/2} \sinh\left(\frac{\rho \varepsilon_i}{\delta}\right),
\end{align*}
which gives the desired form for $R_{\delta}^{(1)}(\varepsilon_i)$. A similar argument using the identity $\sinh^2(x) = \{\cosh(2x)- 1\}/2$ gives the stated form for $R_{\delta}^{(2)}(\varepsilon_i)$. 

For the $\delta \to \infty$ limit, let us write $D_{\delta} = \sqrt{\frac{1}{2}(\omega^2 - 1)}$. First, we note that
\begin{align*}
    \delta D_{\delta} = \sqrt{\frac{\exp\left({2}/{\delta^2}\right) -1}{2/\delta^2}} \to 1.
\end{align*}
Next, for $Z_i \neq 0$, we have
\begin{align*}
    \varepsilon_i = (\delta  D_{\delta}) \,Z_i\,  \frac{\sinh^{-1}\left(D_{\delta} Z_i\right)}{D_{\delta}Z_i} \to Z_i
\end{align*}
following the earlier limit and $\lim_{t\to 0}\, \sinh^{-1}(t)/t = 1$ since $D_{\delta} \to 0$. The $Z_i = 0$ case is immediate. Finally, we write
\begin{align*}
    R_\delta^{(1)}(\varepsilon_i) =  \rho \varepsilon_i \,  \omega^{(1-\rho^2)/2} \,\frac{1}{\delta D_{\delta}} \,\frac{\sinh(\rho\varepsilon_i/\delta)}{\rho\varepsilon_i/\delta} \to \rho Z_i
\end{align*}
where we have used the previous limits and $\lim_{t\to 0}\, \sinh(t)/t = 1$ as $\rho \varepsilon_i/\delta \to 0$. 

For $R_{\delta}^{(2)}(\varepsilon_i)$, we can write
\begin{align*}
    R_{\delta}^{(2)}(\varepsilon_i) = \frac{\omega^{2(1-\rho^2)}-1 }{\omega^2 - 1} + \omega^{2(1-\rho^2)}\,\frac{\sinh^2\left(\rho\varepsilon_i/\delta\right)}{D_{\delta}^2}
\end{align*}
using the identity $\cosh(2x) - 1 = 2\sinh^2(x)$. The first term converges to $1-\rho^2$ as $\lim_{t \to 1} \,(t^a -1)/(t-1) = a$. For the second term, we apply a similar technique:
\begin{align*}
\frac{\sinh^2\left(\rho\varepsilon_i/\delta\right)}{D_{\delta}^2} = \rho^2 \varepsilon_i^2 \, \frac{1}{(\delta D_{\delta})^2} \,\left\{\frac{\sinh(\rho\varepsilon_i/\delta)}{\rho\varepsilon_i/\delta}\right\}^2 \to \rho^2 Z_i^2.
\end{align*}
Combining this gives $R_{\delta}^{(2)}(\varepsilon_i)  \to 1-\rho^2 + \rho^2 Z_i^2$.

\subsection{Proof of Proposition \ref{prop:SU_robust}}
In this section, we provide a more precise characterization of the behaviour of $R^{(1)}_{\delta}(\varepsilon_i)$ and $R^{(2)}_{\delta}(\varepsilon_i)$ as $|Z_i| \to \infty$, from which Proposition \ref{prop:SU_robust} follows.
\begin{proposition}
    For fixed $\rho \in (0,1)$ and $\delta > 0$ and taking $|Z_i|\to \infty$, (\ref{eq:R_delt}) satisfies 
    \begin{align*}
        \lim_{|Z_i| \to \infty}\,  \frac{|R_\delta^{(1)}(\varepsilon_i)|}{ C_{\delta,\rho}|Z_i|^\rho} &= 1,\quad 
         \lim_{|Z_i| \to \infty}\,  \frac{R_\delta^{(2)}(\varepsilon_i)}{\omega^{1-\rho^2}C_{\delta,\rho}^2 |Z_i|^{2\rho}} = 1,
    \end{align*}
    where $C_{\delta,\rho} = 2^{\rho-1} \omega^{(1-\rho^2)/2}D_{\delta}^{\rho - 1}$.
\end{proposition}
\begin{proof}
We again begin with $R_\delta^{(1)}(\varepsilon_i)$ first, where the only non-constant terms are
\begin{align*}
    \sinh\left\{\rho \sinh^{-1}\left(D_{\delta} Z_i\right)\right\}.
\end{align*}
Since the above term is odd as a function of $Z_i$, we have
\begin{align*}
    |R_{\delta}^{(1)}(\varepsilon_i)| = \frac{\omega^{(1-\rho^2)/2}}{D_{\delta}}   \sinh\left\{\rho \sinh^{-1}\left(D_{\delta} |Z_i|\right)\right\}.
\end{align*}
Now, we have
\begin{align*}
    \sinh^{-1}\left(D_{\delta}|Z_i|\right) &= \log\left(2D_{\delta} |Z_i|\right) + o(1),\\
    \sinh\{\rho \sinh^{-1}(x)\} &= \frac{1}{2}\exp\left(\rho \sinh^{-1}(x)\right)\left[1 - \exp\left(-2\rho \sinh^{-1}(x)\right)\right].
\end{align*}
Combining the above gives
\begin{align*}
     \sinh\left\{\rho \sinh^{-1}\left(D_{\delta} |Z_i|\right)\right\} = 2^{\rho - 1} D_{\delta}^{\rho} \,|Z_i|^{\rho}\left\{1 + o(1) \right\}.
\end{align*}
This then gives the desired form of $C_{\delta,\rho}$. For $R_{\delta}^{(2)}(\varepsilon_i)$, a similar argument gives
$$\cosh\left\{2\rho \sinh^{-1}\left(D_{\delta} |Z_i|\right)\right\} = 2^{2\rho -1 } D_{\delta}^{2\rho} |Z_i|^{2\rho}\{1 + o(1)\},$$
from which the result for $R_{\delta}^{(2)}(\varepsilon_i)$ follows.
\end{proof}
For the $\rho \to 1$ limit with fixed $Z_i$ and $\delta > 0$, continuity gives
\begin{align*}
    \lim_{\rho \to 1} \sinh\left\{\rho \sinh^{-1}\left(D_{\delta}Z_i\right) \right\} = D_{\delta} Z_i
\end{align*}
since $D_{\delta}$ and $Z_i$ are constants. Since $\lim_{\rho \to 1}\omega^{(1-\rho^2)/2} = 1$, the $D_{\delta}$ constants cancel and we obtain $\lim_{\rho \to 1} R_{\delta}^{(1)}(\varepsilon_i) = Z_i$. For $R_\delta^{(2)}(\varepsilon_i)$, continuity gives 
\begin{align*}
    \lim_{\rho \to 1} R_{\delta}^{(2)}(\varepsilon_i) = \frac{\cosh\left\{2 \sinh^{-1}\left(D_{\delta}Z_i\right) \right\} - 1}{\omega^2 - 1}
\end{align*}
and the identity $\cosh(2x) - 1 = 2\sinh^2(x)$ then gives $ \lim_{\rho \to 1} R_{\delta}^{(2)}(\varepsilon_i) = Z_i^2$.

\subsection{Proof of Proposition \ref{prop:Zi}}

This follows from standard properties of the multivariate Gaussian. The random variable (r.v.) $Y \sim \mathcal{N}(\mu,\Sigma)$ for arbitrary $(\mu, \Sigma)$ may be written as
\[
Y=\mu+L Z,\qquad Z\sim\mathcal N(0,I_q),
\]
where $L$ is the Cholesky factor of $\Sigma$, and $Z = L^{-1}(Y - \mu)$. Hence, for each $j$,
\begin{align}\label{eq:Yj}
Y^{(j)}
=
\mu^{(j)}
+
\sum_{k=1}^{j-1}L^{(j,k)}Z^{(k)}
+
L^{(j,j)}Z^{(j)}.
\end{align}
Moreover, since $L^{(1:j-1,1:j-1)}$ is lower triangular with positive diagonal,
 $Z^{(1:j-1)}$ and $Y^{(1:j-1)}$ are deterministic one-to-one functions of each other.
Since the components of $Z$ are i.i.d., it follows that
\[
(Y^{(j)}\mid Y^{(1:j-1)}) \sim  
\mathcal N\left(
\mu^{(j)}+\sum_{k=1}^{j-1}L^{(j,k)}Z^{(k)},
\,\left(L^{(j,j)}\right)^2
\right).
\]
Therefore, the conditional density $p_{i-1}(y^{(j)} \mid y^{(1:j-1)}, x)$ has mean function
\[
\widetilde{\mu}_{i-1}^{(j)}
\left(x;y^{(1:j-1)}\right)
=
\mu_{i-1}^{(j)}(x)
+
\sum_{k=1}^{j-1}
L_{i-1}^{(j,k)}(x) \, z^{(k)},
\]
and standard deviation function
\[
\widetilde{\sigma}_{i-1}^{(j)}
\left(x;y^{(1:j-1)}\right)
=
L_{i-1}^{(j,j)}(x).
\]
Evaluating at $x=X_i$ and $y=Y_i$ and rearranging (\ref{eq:Yj}) then gives
\[
Z_i^{(j)}
=
\frac{
Y_i^{(j)}
-
\widetilde{\mu}_{i-1}^{(j)}
\left(X_i;Y_i^{(1:j-1)}\right)
}{
\widetilde{\sigma}_{i-1}^{(j)}
\left(X_i;Y_i^{(1:j-1)}\right)
}\cdot
\]
\subsection{Proof of Proposition \ref{prop:mean_var_missing}}

For completeness, we begin by deriving the moment update under the fully observed case.  The conditional distribution of the multivariate Gaussian from Proposition \ref{prop:Zi} gives
\begin{align*}
    P_{i-1}(y^{(j)} \mid y^{(1:j-1)},x) = \Phi(z^{(j)})
\end{align*}
where $z^{(j)}$ is the $j$-th component of the vector $z = L^{-1}_{i-1}(y - \mu_{i-1})$. Similarly, we have $v_{i-1}^{(j)}=\Phi(Z_i^{(j)})$.
Since
$p_{i-1}(y\mid x)
=
|L_{i-1}|^{-1}\prod_{j=1}^q \phi(z^{(j)})$,
the copula-weighted component in (\ref{eq:condit_copula_mv}) is a PDF of the form
\begin{align*}
\widetilde{p}_i(y \mid x) &= p_{i-1}(y\mid x)
\prod_{j=1}^q
c_\rho\left(
\Phi(z^{(j)}),\Phi(Z_i^{(j)})
\right)\\
&=
|L_{i-1}|^{-1}
\prod_{j=1}^q
\phi\left(
z^{(j)};\rho Z_i^{(j)},1-\rho^2
\right),
\end{align*}
where we have used the standard conditional Gaussian identity
\[
\phi(z)\, c_\rho\{\Phi(z),\Phi(Z_i)\}
=
\phi(z;\rho Z_i,1-\rho^2).
\]
If we define
\[
\widetilde Y_i
=
\mu_{i-1}
+
L_{i-1}
\left\{
\rho Z_i+\sqrt{1-\rho^2}\,W
\right\},
\qquad
W\sim\mathcal N(0,I_q),
\]
then it is clear that $\widetilde{Y}_i \sim \widetilde{p}_i(\, \cdot \mid x)$ through a change of variables argument between $y$ and $z$. Following the same moment argument used in Proposition \ref{prop:SU_update}, we have
\begin{align*}
    E[\widetilde Y_i ] = \mu_{i-1} + d_i, \quad E[\widetilde Y_i \, \widetilde Y_i^\top ] &= (1-\rho^2) \Sigma_{i-1} + [\mu_{i-1} + d_i][\mu_{i-1} + d_i]^\top,
\end{align*}
where we write $d_i(x) = \rho L_{i-1}(x) Z_i$. Computing the moments gives
\begin{align}\label{eq:supp_mv_mean}
    \mu_i &= (1-\beta_i)\mu_{i-1} + \beta_i  E[\widetilde Y_i ] = \mu_{i-1} + \beta_i \,d_i
\end{align}
and for the second moment matrix $M_i^{(2)}$ we have
\begin{align}
    M_i^{(2)} &= (1-\beta_i)M_{i-1}^{(2)}  + \beta_i E[\widetilde Y_i \, \widetilde Y_i^\top ] \nonumber \\
    &= M_{i-1}^{(2)} + \beta_i\left[\mu_{i-1} d_i^\top + d_i \mu_{i-1}^\top + d_i d_i^\top - \rho^2 \Sigma_{i-1}\right]. \label{eq:supp_m2_mean}
\end{align}
Now, we take the expectation over $(Y_i^{(M_i)} \mid Y_i^{(O_i)},X_i)$. For the realized $O_i$, we treat $S_i$ as a fixed selection matrix and apply Gaussian conditioning under the predictive law
$p_{i-1}(\,\cdot\mid X_i)$. By MAR, the resulting conditional law
given the observed outcomes also applies when conditioning on
the missingness pattern, that is 
\begin{align*}
p_{i-1}\left(y^{(M_i)} \mid Y_i^{(O_i)},O_i,X_i\right) =p_{i-1}\left(y^{(M_i)} \mid Y_i^{(O_i)},X_i\right).
\end{align*}
The key is that under the Gaussian assumption for $p_{i-1}(\,\cdot \mid x)$, the conditional density $p_{i-1}(Y_i^{(M_i)} \mid Y_i^{(O_i)},X_i)$ is Gaussian. 

Instead of working with $y_i^{(M_i)}$ however, we directly consider the conditional moments of $Z_i$. Under the predictive model, we have $Z_i \mid X_i \sim \mathcal{N}(0, I_q)$.  Next, given $X_i$, we note that observing $Y_i^{(O_i)}$ is equivalent to observing 
$$
A_i Z_i = S_i[Y_i - \mu_{i-1}(X_i)].$$
A convenient approach is to notice that 
\begin{align*}
    \begin{bmatrix}
        Z_i\\
        A_i Z_i
    \end{bmatrix}\mid X_i \sim \mathcal{N}\left(\begin{bmatrix}
        0_q\\
        0_{|O_i|}
    \end{bmatrix}, \begin{bmatrix}
        I_q & A_i^\top\\
        A_i & A_i A_i^\top 
    \end{bmatrix}\right),
\end{align*}
then Gaussian conditioning gives
\begin{align*}
    Z_i \mid A_i Z_i, X_i \sim \mathcal{N}\left(\overline{Z}_i, I_q - Q_i \right).
\end{align*}
We now take the conditional expectation of (\ref{eq:supp_mv_mean}) and (\ref{eq:supp_m2_mean}) over $Z_i \mid Y_i^{(O_i)}, X_i$. Let
\begin{align*}
    \overline{d}_i = E[d_i \mid Y_i^{(O_i)}, X_i] = \rho L_{i-1}\overline{Z}_i. 
\end{align*}
We then have the mean update as
\begin{align*}
    {\mu}_i =  \mu_{i-1} + \beta_i\overline{d}_i
\end{align*}
and the second moment as
\begin{align*}
    {M}_i^{(2)} &= M_{i-1}^{(2)} + \beta_i\left[\mu_{i-1} \overline{d}_i^\top + \overline{d}_i \mu_{i-1}^\top + \rho^2 L_{i-1}(I - Q_i)L_{i-1}^\top + \overline{d}_i \overline{d}_i^\top - \rho^2 \Sigma_{i-1}\right]\\
    &= M_{i-1}^{(2)} + \beta_i\left[\mu_{i-1} \overline{d}_i^\top + \overline{d}_i \mu_{i-1}^\top - \rho^2 L_{i-1}Q_iL_{i-1}^\top + \overline{d}_i \overline{d}_i^\top\right].
\end{align*}
Computing $\Sigma_i = M_i^{(2)} - \mu_i \mu_i^\top$ then gives the form in the main paper.

\subsection{Proof of Proposition \ref{prop:missing_invar}}
First, we note that under $Y \sim p_{i-1}(\,\cdot \mid  x= X_i)$, we can write
\begin{align*}
    Y = \mu_{i-1}(X_i) + L_{i-1}(X_i) \, U_{i-1}, \quad  U_{i-1} \sim \mathcal{N}(0,I_q).
\end{align*}
Next, we note that the observed component of $Y$ can be written as
\begin{align*}
    Y^{(O_i)} = S_i\mu_{i-1}(X_i) + A_i U_{i-1}, 
\end{align*}
so conditioning on $Y^{(O_i)} = y^{(O_i)}$ is equivalent to conditioning on  $A_i U_{i-1}  =  a$ where
$a:= y^{(O_i)} - S_i \mu_{i-1}(X_i).$
Gaussian conditioning with $(U_{i-1}, A_i U_{i-1})$ gives 
\begin{align*}
U_{i-1} \mid A_i U_{i-1} = a \sim \mathcal{N}\left(A_i^\top (A_i A_i^\top)^{-1}a, I - Q_i\right).
\end{align*}
Since $Y^{(M_i)}$ is an affine function of $U_{i-1}$, the above distribution characterizes $p_{i-1}(y^{(M_i)} \mid y^{(O_i)}, x = X_i)$. Next,  $Y \sim p_i(\,\cdot \mid x = X_i)$ can be expressed as a function of the previous time step's mean/covariance:
\begin{align*}
    Y = \mu_{i-1}(X_i) + L_{i-1}(X_i)\, U_i, \quad  U_i \sim \mathcal{N}\left(\beta_i(X_i)\rho \overline{Z}_i, C_i
\right),
\end{align*}
where $C_i = I+
\beta_i(X_i)\rho^2
[
\{1-\beta_i(X_i)\}\overline{Z}_i\overline{Z}_i^\top
-
Q_i
]$.
We will now use key properties of the projection matrix, that is $Q_i^\top = Q_i$,   $Q_i Q_i = Q_i$, and $A_i  Q_i = A_i$. Next, we can decompose $U_i$ into two components:
\begin{align*}
    U_i = Q_i U_i + \varepsilon_i, \quad \varepsilon_i = (I - Q_i) U_i
\end{align*}
where we will now show $\varepsilon_i \sim \mathcal{N}(0, I - Q_i)$ and $\varepsilon_i \perp Q_i U_i$. To verify these properties, it is clear that $(Q_i U_i, \varepsilon_i)$ are jointly Gaussian, and
\begin{align*}
    E[\varepsilon_i] &= (I - Q_i)\beta_i(X_i) \rho \overline{Z}_i = 0,\\
        E[\varepsilon_i \varepsilon_i^\top] &= (I - Q_i) \,C_i \,(I - Q_i) = (I - Q_i) (I - Q_i) = I - Q_i,
\end{align*}
where we have used $C_i(I - Q_i) = I - Q_i$, 
following $Q_i \overline{Z}_i = \overline{Z}_i$ and $Q_i Q_i = Q_i$. The independence property follows from the joint Gaussianity and
\begin{align*}
\text{Cov}(Q_i U_i, \varepsilon_i)= Q_i C_i (I - Q_i) = 0
\end{align*}
again following the above identity for $C_i (I - Q_i)$. 

Since $A_i U_i = A_i Q_i U_i$, we immediately have that $A_i U_i \perp \varepsilon_i$. Furthermore, $Q_i U_i = A_i^\top (A_i A_i^\top)^{-1}A_i U_i$, so we have
\begin{align*}
    U_i \mid A_i U_i = a \, \sim \mathcal{N}\left(A_i^\top (A_i A_i^\top)^{-1} a, I - Q_i\right).
\end{align*}
Since $(U_{i-1} \mid A_i U_{i-1} = a) \overset{d}{=} (U_{i} \mid A_i U_{i} = a)$, and $Y^{(M_i)}$ is the same affine function of $U_i$, we have 
$p_i(y^{(M_i)} \mid y^{(O_i)}, x = X_i) = p_{i-1}(y^{(M_i)} \mid y^{(O_i)}, x = X_i)$ for all $y \in \mathbb{R}^q$. 

We now strengthen to the case for a generic $x\in \mathcal{X}$ under order-compatible missingness where $O_i = \{1,\ldots,q_i\}$ for some $q_i \leq q$. Let $A_i(x) = S_i L_{i-1}(x)$ for a generic $x \in \mathcal{X}$. In this setting, the Cholesky form of $L_{i-1}(x)$ gives
\begin{align*}
    Q_i(x) := A_i(x)^\top (A_i(x) A_i(x)^\top)^{-1} A_i(x) =  \begin{bmatrix}
        I_{q_i} & 0\\
        0 & 0
    \end{bmatrix}
\end{align*}
which is now independent of $x$. Importantly, this holds for $Q_i = Q_i(X_i)$.

The first part of the proof regarding $p_{i-1}(\, \cdot \mid x)$ remains unchanged, where we now have $a := y^{(O_i)} - S_i \mu_{i-1}(x)$ and we are conditioning on $A_i(x)\,  U_{i-1} = a$ where $A_i(x) = S_i L_{i-1}(x)$. Then, we have
\begin{align*}
    U_{i-1} \mid A_i(x) \, U_{i-1} = a \sim \mathcal{N}\left(A_i(x)^\top (A_i(x) A_i(x)^\top)^{-1} a, I - Q_i\right)
\end{align*}
where we have used $Q_i(x) = Q_i$. For the second part of the proof with the updated $p_i(\,\cdot \mid x)$, we have
\begin{align*}
    Y = \mu_{i-1}(x) + L_{i-1}(x)\, U_i(x), \quad  U_i(x) \sim \mathcal{N}\left(\beta_i(x)\rho \overline{Z}_i, C_i(x)
\right),
\end{align*}
where $C_i(x) = I+
\beta_i(x)\rho^2
[
\{1-\beta_i(x)\}\overline{Z}_i\overline{Z}_i^\top
-
Q_i
]$.
The rest of the proof remains identical, as order-compatibility implies
\begin{align*}
    A_i(x) \, Q_i = A_i(x), \quad Q_i(x) = Q_i.
\end{align*}
 We can then use the same decomposition $U_i(x) = Q_i  U_i(x) + \varepsilon_i(x)$ with $\varepsilon_i(x) = (I - Q_i) U_i(x)$, and thus show that $(U_i(x) \mid A_i(x) \, U_i(x)) \overset{d}{=} (U_{i-1} \mid A_i(x)\, U_{i-1})$ which gives us the result for any $x \in \mathcal{X}$ under order-compatibility.

\section{Predictive asymptotics proofs}

\subsection{Preliminary lemmas}

We begin with some intermediate lemmas that will be useful in other proofs.
\begin{lemma}\label{lem:sxx_bound}
    Under Assumption \ref{as:bounded_sxx}, we have $\sup_{x,x' \in \mathcal{X}} s(x,x') \leq S$.
\end{lemma}
\begin{proof}
    Cauchy-Schwarz applied to the definition of $s(x,x')$ gives $$s(x,x') \leq \sqrt{s(x,x)\, s(x',x')} \leq S,$$
    where $S$ is finite by Assumption \ref{as:bounded_sxx}.
\end{proof}

\begin{lemma}\label{lem:alphaxx_bound}
Under Assumptions \ref{as:bounded_sxx} and \ref{as:lr}, we have
\begin{align*}
\sup_{x,x' \in \mathcal{X}}\alpha_i(x,x') \lesssim i^{-m}.
\end{align*}
\end{lemma}
\begin{proof}
It is clear from Assumption \ref{as:bounded_sxx} and Lemma \ref{lem:sxx_bound} that
\begin{align*}
    \sup_{x,x'}\alpha_i(x,x') \leq \frac{\alpha_i S}{1- \alpha_i}\cdot
\end{align*}
Since $\alpha_i=O(i^{-m})$ with $m>0$, we have $\alpha_i\to0$, and hence $(1-\alpha_i)^{-1}=O(1)$. Since $\alpha_i < 1$ for all $i$, we can find a sufficiently large constant for all $i \geq n+1$.
\end{proof}
We include another lemma that is useful when managing the variance term.

\begin{lemma}\label{lem:var_bound}
    Let $Z \sim \mathcal{N}(0,1)$, $0 \leq a < 1$ and $0 < \rho < 1$. Then we have
    \begin{align*}
        E\left[\left\{1 + a\rho^2 \left[(1-a)Z^2 - 1 \right]\right\}^2\right] \leq 1.
    \end{align*}
    Furthermore, for any $z \in \mathbb{R}$, we also have
    \begin{align*}
        1 + a\rho^2 \left[(1-a)z^2 - 1 \right] \geq 1- \rho^2 > 0. 
    \end{align*}
\end{lemma}
\begin{proof}
    Expanding the square gives
    \begin{align*}
   E\left[\left\{1 + a\rho^2 \left[(1-a)Z^2 - 1 \right]\right\}^2\right] &= 1 - 2a^2 \rho^2 + a^2 \rho^4\left(2 - 4a + 3a^2\right)\\
   & \leq 1 + a^2 \rho^4(3a^2 - 4a)
    \end{align*}
    where the second line uses $\rho < 1$. Finally, we have $(3a^2 - 4a) \leq  0$ for $0 \leq a<1$. The second part of the lemma follows by taking $z = 0$ and again using $0 \leq a < 1$. 
\end{proof}

\subsection{Proof of Theorem \ref{th:pred_asymp}}
We follow the setting of \citet[Proposition 1]{fong2025bayesian},  and in a sense extend the setting of \cite{garelli2024asymptotics} to the functional case. Let $B = L^2(\mathcal{X})$ denote the Hilbert space  where $\mathcal{X}$ is equipped with the Lebesgue measure. Let $Z_i \iid \mathcal{N}(0,1)$ be i.i.d. Gaussian r.v.s for $i \geq n+1$ (independent of $\{X_i\}_{i \geq n+1}$), and let $\{\mathcal{F}_i\}_{i \geq n}$ be our filtration with $\mathcal{F}_i = \sigma( X_{n+1},Z_{n+1}, \ldots, X_i, Z_i)$ with $\mathcal{F}_n = \{\emptyset, \Omega\}$. Let $L^2(B)$ be the space of Bochner measurable functions with $E[\|f\|_B^2] < \infty$ and norm $\|f\|^2_{L^2(B)} = E\left[\|f\|_B^2\right]$. Unless otherwise stated, (a.s.) convergence will be in the $B$ norm.

We begin by proving convergence of the mean and variance functions. For any fixed sequence $Z_{1:n}$, we can see that the estimates $\mu_n$ and $\sigma_n^2$ lie in $B$ given the initialization from Assumption \ref{as:init}. 
As $L^2(\mathcal{X})$ is separable, following a similar argument to \citet[Lemma A8]{fong2025bayesian}, the random functions $\mu_i$ and $\sigma_i^2$ are Bochner $\mathcal{F}_i$-measurable and $\{\mu_i\}_{i \geq n}$ is a $B$-valued martingale (given integrability). As in \cite{fong2025bayesian}, it suffices to check pointwise expectations through  `scalarization' \citep{Pisier2016}, e.g. by checking $E[\mu_{i+1}(x) \mid \mathcal{F}_i] = \mu_i(x)$ for all $x \in \mathcal{X}$ followed by a Riesz representation argument to verify the martingale condition. 

It thus suffices to verify that $\{\mu_i\}_{i \geq n}$ is bounded in $L^2(B)$ so we can use  \citet[Theorem 2.9]{Pisier2016} for convergence. Since $\mu_i^2(x) \leq m_i^{(2)}(x)$, we immediately have
\begin{align*}
    \sup_i E\left[\|\mu_i \|_B^2\right] \leq \sup_i E\left[\int_{\mathcal{X}} m_i^{(2)}(x) \, dx\right] = \int_{\mathcal{X}} m_n^{(2)}(x) \,dx < \infty,
\end{align*}
where $m_n^{(2)} \in L^1(\mathcal{X})$ follows from $\mu_n \in L^2(\mathcal{X})$ and $\sigma_n^2 \in L^1(\mathcal{X})$, where the latter follows from $\sigma_n^2 \in L^2(\mathcal{X})$ and $\mathcal{X}$ having finite Lebesgue measure. As a result, $\{\mu_i\}_{i \geq n}$ is bounded in $L^2(B)$, and thus converges a.s. to $\mu_\infty \in B$.

The variance function is not a martingale, but can be decomposed into a martingale and a summable term. Specifically, we have
\begin{align}
    \sigma_i^2 &= \sigma_{i-1}^2 + \underbrace{\beta_i \, \sigma_{i-1}^2 \, \rho^2 \left(Z_i^2 - 1\right)}_{\Delta M_i} - \underbrace{\beta_i^2 \sigma_{i-1}^2 \rho^2  Z_i^2}_{\Delta R_i}=\sigma_n^2 + \sum_{j = n+1}^i \Delta M_j - \sum_{j = n+1}^i \Delta R_j.\label{eq:app_var_decomp}
\end{align}
Note that $M_i = \sigma_n^2 + \sum_{j = n+1}^i \Delta M_j$ is now a $B$-valued martingale, and $R_i = \sum_{j = n+1}^i \Delta R_j$ is an error term. First, we have
\begin{align*}
    E\left[\|\Delta M_i\|^2_B \mid \mathcal{F}_{i-1}\right] \lesssim  i^{-2m}\|\sigma_{i-1}^2 \|_B^2 
\end{align*}
where we have used $E[(Z_i^2 - 1)^2] = 2$. Lemma \ref{lem:var_bound} then gives 
$$\sup_{i\geq n} E\left[\|\sigma_{i}^2 \|_B^2\right] \leq  \|\sigma_{n}^2 \|_B^2 < \infty$$
which gives $E\left[\|\Delta M_i\|^2_B \right] = O(i^{-2m})$.
This is summable as $m > 0.5$, so martingale orthogonality gives that $\{M_i\}_{i \geq n}$ is bounded in $L^2(B)$ and converges a.s. to $M_\infty \in B$.

For the remainder term, we note that $\|\Delta R_i\|_B \lesssim Z_i^2 \, \|\sigma_{i-1}^2\|_B \, i^{-2m}$,
and a similar argument gives $E\left[\|\Delta R_i\|_B\right] =O(i^{-2m})$,
so $\sum_{i = n+1}^\infty E\left[\|\Delta R_i\|_B\right] < \infty$ which implies $\sum_{i = n+1}^\infty \|\Delta R_i\|_B < \infty$ a.s.
Since $B$ is complete and $R_i$ is a Cauchy sequence a.s., it converges a.s. to $R_\infty \in B$. Together, this implies $\sigma_i^2$ converges a.s. to $\sigma_\infty^2 \in B$.

Finally, we verify the positivity of $\sigma_\infty^2$. Following a similar argument to \cite{garelli2024asymptotics}, we take the log of the variance update,
\begin{align*}
    \log \sigma_i^2 = \log \sigma_{i-1}^2 + \log\left(1 + U_i\right),
\end{align*}
where we write $U_i = \beta_i\,\rho^2 \left[(1-\beta_i)Z_i^2 - 1 \right].$ The above can be handled in a similar fashion through a decomposition into a martingale and error term. We write
\begin{align*}
    \log \sigma_i^2 = \log \sigma_{i-1}^2 + \Delta M_i' + \Delta R'_i
\end{align*}
where $\Delta M_i' = \beta_i \, \rho^2(Z_i^2 - 1)$  and
\begin{align*}
    \Delta R'_i &= \log\left(1 + U_i\right) - \Delta M_i' = \left\{\log\left(1 + U_i\right) - U_i \right\} - \rho^2 \beta_i^2 Z_i^2.
\end{align*}
We can use the same earlier argument to show the martingale term converges a.s., and  $\log \sigma_n^2 \in B$ as $1-\rho^2 \leq  1+U_i \leq 1 +\rho^2 Z_i^2$ as $\log \sigma_0 \in B$ from Assumption \ref{as:init}. It thus suffices to show $R_i'  = \sum_{j = n+1}^i \Delta R_j'$ is a Cauchy sequence. The term $\rho^2 \beta_i^2 Z_i^2$  can be handled similarly to $\Delta R_i$ earlier. Next, for sufficiently large $i$, we have $U_i \geq - \rho^2 K i^{-m}\geq  -\frac{1}{2}$ for all $x \in \mathcal{X}$. Combined with $|\log(1+x) - x| \leq x^2$ for $x > -1/2$, we have
\begin{align*}
|\log(1+U_i) - U_i| \leq U_i^2 \lesssim i^{-2m} \left(1+ Z_i^2\right)^2.
\end{align*}
Since the bound does not depend on $x$ and $\int_{\mathcal{X}} dx <  \infty$, we have
\begin{align*}
    \|\log(1+U_i) - U_i\|_B \lesssim i^{-2m} \left(1+ Z_i^2\right)^2.
\end{align*}
Finally, as $E[(1+Z_i^2)^2] = 6$, we have $\sum_{i = n+1}^\infty E\left[\|\Delta R_i'\|_B\right] < \infty$ giving $\sum_{i = n+1}^\infty \|\Delta R_i'\|_B < \infty$ a.s. So $R_i'$ is a Cauchy sequence and converges to $R'_\infty \in B$ a.s. Together, this implies $\log \sigma_i^2$ converges to $\ell_\infty \in B$ a.s.

We now show $\sigma_\infty^2 = \exp\left(\ell_\infty\right) > 0$ for Lebesgue-almost all $x \in \mathcal{X}$ a.s. We will work on the probability-one set on which $\sigma_i^2$ and $\log \sigma_i^2$ converge in $L^2(\mathcal{X})$. Since $L^2(\mathcal{X})$ convergence implies convergence in (Lebesgue) measure, we have that both $\sigma_i^2 \to \sigma_\infty^2$ and $\log \sigma_i^2 \to \ell_\infty$ in measure.
Continuity of $\exp(\cdot)$ gives $\sigma_i^2 \to \exp\left(\ell_\infty\right)$
in measure. Finally, uniqueness of limits in measure gives $\sigma_\infty^2 = \exp\left(\ell_\infty\right)$ Lebesgue-almost everywhere. Since $\ell_\infty \in L^2(\mathcal{X})$, it must be finite Lebesgue-almost everywhere, so $\sigma^2_\infty(x) > 0$ for Lebesgue-almost all $x\in \mathcal{X}$. This gives our desired positivity a.s.

Finally, we tackle the total variation convergence. Again we work on the appropriate probability-one set on which $L^2(\mathcal{X})$ convergence occurs. Let
\begin{align*}
    D_i(x) = \frac{1}{2}\int \left|\mathcal{N}\{y; \mu_i(x), \sigma_i^2(x)\} - \mathcal{N}\{y; \mu_\infty(x), \sigma_\infty^2(x)\} \right| \, dy.
\end{align*}
For each $x$, continuity of the Gaussian density and Scheff\'{e}'s lemma give $D_i(x) \to 0$ if $(\mu_i(x), \sigma_i^2(x)) \to(\mu_\infty(x), \sigma_\infty^2(x))$ and $\sigma_\infty^2(x) > 0$. $L^2(\mathcal{X})$ convergence of $(\mu_i, \sigma_i^2)$  gives $(\mu_i,\sigma_i^2) \to (\mu_\infty,\sigma_\infty^2)$ in $P_X$-measure, following Assumption \ref{as:design}. Together with the positivity of $\sigma_\infty^2$ for $P_X$-almost all $x$, we have $D_i \to 0$ in $P_X$-measure from the continuity of total variation distance as a function of the Gaussian parameters. Moreover, \(0\leq D_i\leq 1\). Hence, for every \(\varepsilon>0\),
\begin{align*}
\int D_i(x)\,dP_X(x)
&\leq
\varepsilon+
P_X(D_i>\varepsilon)
\end{align*}
where $P_X(D_i>\varepsilon) \to 0$. Letting \(\varepsilon\to 0\) gives $\int D_i(x)\,dP_X(x)\to 0$ as desired. 

\subsubsection{Proof of Corollary \ref{cor:SU} (Theorem \ref{th:pred_asymp} part)}
Let $g_{\delta}(\varepsilon)$ be defined as in (\ref{eq:g_delt}). The proof is largely similar to that of Theorem \ref{th:pred_asymp}, so we focus on the differences. We replace $\rho Z_i$ with $R_{\delta}^{(1)}(\varepsilon_i)$ and $\rho^2 (Z_i^2 - 1)$ with $R_{\delta}^{(2)}(\varepsilon_i)-1$. Under predictive resampling, $\varepsilon_i \iid \mathcal{N}(0,1)$, which gives 
\begin{align*}
    E\left[R_{\delta}^{(j)}(\varepsilon_i) \mid \mathcal{F}_{i-1}, X_i\right] = E_{W' \sim \mathcal{N}(0,1)}\left[g_{\delta}(W')^j\right] = \begin{dcases}
        0 &\quad \text{if } j = 1, \\
        1 &\quad \text{if } j = 2,
    \end{dcases}
\end{align*}
since $W' = \sqrt{1-\rho^2} W + \rho \varepsilon_i \sim \mathcal{N}(0,1)$. We rely on the following property of $g_{\delta}$.
\begin{lemma}\label{lem:g_delta_4thmoment}
    For fixed $\delta > 0$, the transformation $g_{\delta}$ satisfies
    \begin{align*}
      \kappa_{\delta} :=  E_{W' \sim \mathcal{N}(0,1)}\left[g_{\delta}(W')^4\right] < \infty.
    \end{align*}
    Furthermore, we have
    \begin{align*}
        E\left[\left\{R_{\delta}^{(1)}(\varepsilon_i)\right\}^4 \mid \mathcal{F}_{i-1}, X_i\right]\leq \kappa_{\delta}, \quad E\left[\left\{R_{\delta}^{(2)}(\varepsilon_i)\right\}^2 \mid \mathcal{F}_{i-1}, X_i\right]\leq \kappa_{\delta}.
    \end{align*}
\end{lemma}

\begin{proof}
The Gaussian moment generating function gives $\kappa_{\delta} = \frac{1}{2}{(\omega^4 + 2\omega^2 + 3)} < \infty$.
The remaining bounds follow from Jensen's inequality, since $R_{\delta}^{(j)}(\varepsilon_i) = E[g_{\delta}(W')^j \mid \varepsilon_i]$ for $W' = \sqrt{1-\rho^2}W + \rho\varepsilon_i$ with $W\sim \mathcal{N}(0,1)$ independently.
\end{proof}
\begin{lemma}\label{lem:SU_varbound}
Let $\varepsilon \sim \mathcal{N}(0,1)$, $0 \leq a < 1$, $0 < \rho < 1$ and $\delta > 0$. Then there exists a finite $C$ such that
\begin{align*}
    E\left[\left(1 + a\left\{R_{\delta}^{(2)}(\varepsilon) -1 - a R_{\delta}^{(1)}(\varepsilon)^2\right\}\right)^2\right] \leq 1 + C a^2.
\end{align*}
\end{lemma}

\begin{proof}
Jensen's inequality gives $R_{\delta}^{(1)}(\varepsilon)^2 \leq R_{\delta}^{(2)}(\varepsilon)$, so
\begin{align*}
    0 < 1-a
    &\leq 1 + a\left\{R_{\delta}^{(2)}(\varepsilon)-1-aR_{\delta}^{(1)}(\varepsilon)^2\right\}
    \leq 1 + a\left\{R_{\delta}^{(2)}(\varepsilon)-1\right\}.
\end{align*}
Since the above term is positive, we can square and take expectations, where the square of the upper bound has expectation
\begin{align*}
    1 + a^2 E\left[\left\{R_{\delta}^{(2)}(\varepsilon)-1\right\}^2\right]
    \leq 1 + (\kappa_{\delta}-1)a^2,
\end{align*}
where we have used $E[R_{\delta}^{(2)}(\varepsilon)] = 1$ and Lemma \ref{lem:g_delta_4thmoment}.
\end{proof}
Lemma \ref{lem:g_delta_4thmoment} replaces the Gaussian fourth moment bound, whereas Lemma \ref{lem:SU_varbound} replaces Lemma \ref{lem:var_bound}. The argument for the convergence $\mu_i \to \mu_\infty$ is unchanged.

For the variance, Lemmas \ref{lem:alphaxx_bound} and \ref{lem:SU_varbound} give
\begin{align*}
    E[\|\sigma_i^2\|_B^2 \mid \mathcal{F}_{i-1}]
    &\leq \,(1+Ci^{-2m})\|\sigma_{i-1}^2\|_B^2
\end{align*}
which when iterated gives $\sup_i E[\|\sigma_i^2\|_B^2]<\infty$. The variance decomposition now has
\begin{align*}
    \Delta M_i &= \beta_i\sigma_{i-1}^2\left\{R_{\delta}^{(2)}(\varepsilon_i)-1\right\}, \quad \Delta R_i = \beta_i^2\sigma_{i-1}^2 R_{\delta}^{(1)}(\varepsilon_i)^2,
\end{align*}
where Lemma \ref{lem:g_delta_4thmoment} can be used to control the norms to show $\sigma_i^2 \to \sigma_\infty^2$ a.s.

To verify positivity, write
\begin{align*}
    U_i=\beta_i\left\{R_{\delta}^{(2)}(\varepsilon_i)-1-\beta_i R_{\delta}^{(1)}(\varepsilon_i)^2\right\}.
\end{align*}
As in the Gaussian case, we decompose the log-variance update as
\begin{align*}
    \log\sigma_i^2=\log\sigma_{i-1}^2+\Delta M_i'+\Delta R_i',
\end{align*}
where $\Delta M_i'=\beta_i\{R_{\delta}^{(2)}(\varepsilon_i)-1\}$ and
\begin{align*}
    \Delta R_i'=\{\log(1+U_i)-U_i\}-\beta_i^2R_{\delta}^{(1)}(\varepsilon_i)^2.
\end{align*}
Again,  $U_i\geq-\beta_i\geq-1/2$ for sufficiently large $i$ uniformly in $x$, and the rest of the convergence proof is identical.  Finally, for fixed $\delta>0$, the SU density is continuous in $(\mu,\sigma^2)$ whenever $\sigma^2>0$, so the total variation argument follows. 

Finally, we highlight that the entire proof extends to general transformations $g_{\lambda}$ if analogous versions of Lemmas \ref{lem:g_delta_4thmoment} and \ref{lem:SU_varbound} and continuity of the density hold.

\subsubsection{Extension to the multi-output case}\label{app:mv_pred_asymp}
For the multi-output setting, we use the Euclidean norm $\|\cdot \|$ for vectors and the Frobenius norm $\| \, \cdot \, \|_F$ for matrices. We begin with a statement of the multi-output version of Assumption \ref{as:init}, followed by the multi-output extension of Theorem \ref{th:pred_asymp}. 
\begin{assumption}[Initialization]\label{as:mv_init}
The initial estimates for the recursion are $\mu_0(x)=0$ and $\Sigma_0(x) = I_q$.
\end{assumption}
\begin{theorem}\label{th:pred_asymp_mv}
Consider the multivariate version of Algorithm \ref{alg:gauss} with updates (\ref{eq:mean_gauss_mv}) and (\ref{eq:var_gauss_mv}). Under Assumptions \ref{as:bounded_sxx}, \ref{as:lr}, \ref{as:mv_init} and \ref{as:design}, there exist random  $\mu_\infty \in L^2(\mathcal{X}; \mathbb{R}^q)$ and $\Sigma_\infty \in L^2(\mathcal{X}; \mathbb{R}^{q \times q})$, with $\Sigma_\infty(x) \succ 0$ for Lebesgue-almost all $x \in \mathcal{X}$, such that
\begin{align*}
\int_{\mathcal{X}}\left\|\mu_\infty(x) - \mu_i(x)\right\|^2 dx \to 0, \quad \int_{\mathcal{X}}\left\|\Sigma_\infty(x) - \Sigma_i(x)\right\|_F^2 dx \to 0 \quad \textnormal{a.s.}
\end{align*}
Furthermore, we have the total variation convergence
\begin{align*}
    \frac{1}{2}\int_{\mathcal{X}} \left[\int \left|p_i(y \mid x) - \mathcal{N}\{y; \mu_\infty(x),\Sigma_\infty(x)\} \right| dy\,\right] p_X(x) \, dx \to 0 \quad \textnormal{a.s.}
\end{align*}
\end{theorem}

\begin{proof}
    Again, the proof is practically identical to that of Theorem \ref{th:pred_asymp}, so we focus on the differences.
    We use $B$ interchangeably to denote $L^2(\mathcal{X}; \mathbb{R}^q)$ and $L^2(\mathcal{X}; \mathbb{R}^{q \times q})$, with respective norms $\| \mu\|_B^2 = \int_{\mathcal{X}} \|\mu(x)\|^2\, dx$ for a vector function $\mu$, and $\| \Sigma\|_B^2 = \int_{\mathcal{X}} \|\Sigma(x)\|_F^2\, dx$ for a matrix function $\Sigma$. 

    We will rely on a few lemmas. The first is for handling the variance increments.
\begin{lemma}\label{lem:frob_gauss_bound}
    Let $Z \sim \mathcal{N}(0,I_q)$ and $\Sigma$ be a positive definite matrix with Cholesky factor $L$. Then we have
    \begin{align*}
        E\left[\left\|L (Z Z^\top - I) L^\top\right\|_F^2\right] \leq (q+1) \|\Sigma\|_F^2.
    \end{align*}
\end{lemma}
\begin{proof}
The matrix is symmetric so its squared Frobenius norm is the trace of the matrix squared, giving
\begin{align*}
    \left\|L (Z Z^\top - I) L^\top\right\|_F^2 = \operatorname{tr}\left\{(L ZZ^\top L^\top)^2\right\} - 2\operatorname{tr}(\Sigma LZZ^\top L^\top) + \operatorname{tr}(\Sigma^2).
\end{align*}
The second term has expectation
\begin{align*}
  E\left[\operatorname{tr}(\Sigma LZZ^\top L^\top)\right]   =\operatorname{tr}(\Sigma^2)
\end{align*}
from linearity of the trace. Note that $LZ$ is a vector, so we have $\operatorname{tr}\left\{(L ZZ^\top L^\top)^2\right\} = (Z^\top L^\top L Z)^2$. Furthermore, $LZ \sim \mathcal{N}(0,\Sigma)$, so standard Gaussian quadratic form identities \citep[Section 8.2.2]{petersen2012matrix} then give
\begin{align}\label{eq:gauss_quad}
E[Z^\top L^\top LZ]=\operatorname{tr}\Sigma, \quad &\operatorname{Var}\left[Z^\top L^\top LZ\right] =2\|\Sigma\|_F^2, \end{align}
which implies $E\left[(Z^\top L^\top L Z)^2\right] = 2\operatorname{tr}(\Sigma^2) + (\operatorname{tr}\Sigma)^2$.
Together, this gives
\begin{align*}
    E\left[\left\|L(ZZ^\top-I)L^\top\right\|_F^2\right]
    &=  (\operatorname{tr}\Sigma)^2 +  \operatorname{tr}(\Sigma^2) \leq(q+1)\|\Sigma\|_F^2,
\end{align*}
where we used $(\operatorname{tr}\Sigma)^2 = \langle \Sigma, I \rangle_F^2 \leq q\|\Sigma\|_F^2$ following Cauchy-Schwarz.
\end{proof}
    The second is a multivariate version of Lemma \ref{lem:var_bound}.
    
\begin{lemma}\label{lem:var_bound_mv}
    Let $Z \sim \mathcal{N}(0,I_q)$, $0 \leq a < 1$, $0 < \rho < 1$ and let $\Sigma$ be a positive definite matrix with Cholesky factor 
    $L$. Let us define
    \begin{align*}
        \Sigma' = L\left[I + a\rho^2\left\{(1-a) Z Z^\top - I \right\}\right]L^\top.
    \end{align*}
    Then we have 
    $E[\left(\operatorname{tr}\Sigma' \right)^2] \leq \left(\operatorname{tr}\Sigma \right)^2$. Furthermore, for any $z \in \mathbb{R}^q$, we have
    \begin{align*}
            I+a\rho^2\left\{(1-a)zz^\top-I\right\}\succeq(1-\rho^2)I.
    \end{align*}
\end{lemma}

\begin{proof}
By linearity and cyclicity of the trace,
\begin{align*}
    \operatorname{tr}\Sigma'=(1-a\rho^2)\operatorname{tr}\Sigma
    +a\rho^2(1-a)Z^\top L^\top LZ.
\end{align*}
Consequently, (\ref{eq:gauss_quad}) gives
\begin{align*}
    E[(\operatorname{tr}\Sigma')^2] = \operatorname{Var}[\operatorname{tr}\Sigma'] +  E[\operatorname{tr}\Sigma']^2
    &=
      2a^2\rho^4(1-a)^2\|\Sigma\|_F^2 + (1-a^2\rho^2)^2(\operatorname{tr}\Sigma)^2\\
    &\leq
      \left\{1-2a^2\rho^2
      +a^2\rho^4(2-4a+3a^2)\right\}
      (\operatorname{tr}\Sigma)^2\\
    &\leq(\operatorname{tr}\Sigma)^2.
\end{align*}
The second line uses $\|\Sigma\|_F^2\leq(\operatorname{tr}\Sigma)^2$
for positive semidefinite matrices, and the final inequality follows
from the same scalar calculation as in Lemma \ref{lem:var_bound}. The second part of the lemma again follows from setting $z = 0$ and noting that $0 \leq a < 1$.
\end{proof}
    To begin, we note that $\Sigma_i$ is positive definite at every finite step, since
\begin{align*}
    I+\beta_i\rho^2\left\{(1-\beta_i)Z_iZ_i^\top-I\right\}\succeq(1-\rho^2)I
\end{align*}
from the second part of Lemma \ref{lem:var_bound_mv}. The mean and second moment matrix $\Sigma_i(x)+\mu_i(x)\mu_i(x)^\top$ are pointwise martingales, and taking the trace of the latter gives
\begin{align*}
    \sup_{i\geq n}E[\|\mu_i\|^2_{B}]
    \leq\int_{\mathcal{X}}\left\{\|\mu_n(x)\|^2+\operatorname{tr}\Sigma_n(x)\right\}\,dx<\infty,
\end{align*}
giving $\mu_i\to\mu_\infty$ a.s. Next, we again decompose into $\Sigma_i=\Sigma_{i-1}+\Delta M_i-\Delta R_i$, where
\begin{align*}
    \Delta M_i &= \beta_i\rho^2L_{i-1}(Z_iZ_i^\top-I)L_{i-1}^\top,
    \quad \Delta R_i = \beta_i^2\rho^2L_{i-1}Z_iZ_i^\top L_{i-1}^\top.
\end{align*}
Lemmas \ref{lem:alphaxx_bound} and
 \ref{lem:frob_gauss_bound} together give $E[\|\Delta M_i\|_B^2\mid\mathcal{F}_{i-1}]
    \lesssim i^{-2m}\|\Sigma_{i-1}\|_B^2$. For the remainder, (\ref{eq:gauss_quad}) gives
\begin{align*}
    E[\|LZZ^\top L^\top\|_F^2]
    &=2\|\Sigma\|_F^2+(\operatorname{tr}\Sigma)^2 \leq(q+2)\|\Sigma\|_F^2,
\end{align*}
which gives $E[\|\Delta R_i\|_B^2\mid\mathcal{F}_{i-1}]
    \lesssim i^{-4m}\|\Sigma_{i-1}\|_B^2$.
Next, we bound $\sup_{i\geq n}E[\|\Sigma_i\|_B^2]$.
Applying Lemma \ref{lem:var_bound_mv} conditionally on
$\mathcal{F}_{i-1}$ and $X_i$ gives
\begin{align*}
    E[(\operatorname{tr}\Sigma_i(x))^2
      \mid \mathcal{F}_{i-1},X_i]
    \leq (\operatorname{tr}\Sigma_{i-1}(x))^2.
\end{align*}
Since $\|\Sigma_i(x)\|_F^2\leq
(\operatorname{tr}\Sigma_i(x))^2$, integrating and iterating gives
\begin{align*}
    \sup_{i\geq n}E[\|\Sigma_i\|_B^2]
    &\leq \int_{\mathcal{X}}
        (\operatorname{tr}\Sigma_n(x))^2\,dx \leq q\|\Sigma_n\|_B^2<\infty,
\end{align*}
so $\Sigma_i\to\Sigma_\infty$ a.s. Finally, we verify $\Sigma_\infty \succ 0$, using $\log\det\Sigma_i$ in place of $\log\sigma_i^2$. Let
\begin{align*}
    U_i&=\beta_i\rho^2
        \left\{(1-\beta_i)Z_iZ_i^\top-I\right\}, \quad \log\det\Sigma_i
    =\log\det\Sigma_{i-1}+\log\det(I+U_i).
\end{align*}
 We decompose the update as $\log\det\Sigma_i=\log\det\Sigma_{i-1}+\Delta M_i'+\Delta R_i'$, where
\begin{align*}
    \Delta M_i'
    &=\beta_i\rho^2(\|Z_i\|^2-q), \quad \Delta R_i'
    =\left\{\log\det(I+U_i)-\operatorname{tr}U_i\right\}
      -\beta_i^2\rho^2\|Z_i\|^2.
\end{align*}
Since $E[\|Z_i\|^2]=q$ and
$\operatorname{Var}(\|Z_i\|^2)=2q$, the terms $\Delta M_i'$ and $\beta_i^2\rho^2\|Z_i\|^2$ are handled easily.
For the remaining term, all eigenvalues of $U_i$ are at least
$-\beta_i\rho^2\geq-1/2$ for sufficiently large $i$,
uniformly in $x$. Applying $|\log(1+u)-u|\leq u^2$
to each eigenvalue,
\begin{align*}
    |\log\det(I+U_i)-\operatorname{tr}U_i|
    &\leq \sum_{j=1}^q\lambda_j^2
     =\|U_i\|_F^2
     \lesssim i^{-2m}(1+\|Z_i\|^2)^2.
\end{align*}
Hence $\log\det\Sigma_i\to\ell_\infty$
a.s. The positivity argument is then identical. Since $\Sigma_i\to\Sigma_\infty$ in Lebesgue measure,
$\Sigma_\infty$ is symmetric positive semidefinite
Lebesgue-almost everywhere. Continuity of the determinant and exponential give
$\det\Sigma_\infty=\exp(\ell_\infty)>0$ Lebesgue-almost everywhere. Thus $\Sigma_\infty \succ 0$ Lebesgue-almost everywhere a.s. The total variation convergence proof is then identical.
\end{proof}

\subsection{Proof of Theorem \ref{th:pred_sobolev}}
The result is primarily driven by the smoothness of $\alpha_i(\cdot,x')$ and properties of Sobolev spaces. The proof is similar to that of Theorem \ref{th:pred_asymp} with more care in managing norms due to the higher-order derivatives. To that end, we begin with some useful lemmas. 
\begin{lemma}\label{lem:alpha_sob}
    Under Assumptions \ref{as:bounded_sxx}, \ref{as:lr} and \ref{as:sob}, we have
    \begin{align*}
        \sup_{x' \in \mathcal{X}} \|\alpha_i(\,\cdot\,,x')\|_{W^{r,\infty}} \lesssim  i^{-m}.
    \end{align*}
\end{lemma} 
\begin{proof}
Since continuous functions on compact spaces are bounded, Assumption \ref{as:sob} gives the existence of a finite $S_r$ such that $\sup_{x' \in \mathcal{X}} \|s(\,\cdot\,,x')\|_{W^{r,\infty}} \leq S_r$.
    Next, for $0<a<1$ and $s \geq 0$, consider the function
    \begin{align*}
        f_a(s) = \frac{s}{1 - a + as}
    \end{align*}
    which satisfies, for $ 1 \leq k \leq r$,
    \begin{align*}
       |f^{(k)}_a(s)|= \frac{k!(1-a)a^{k-1}}{(1- a + as)^{k+1}}\leq \frac{k!}{(1-a)^{k+1}}\cdot
    \end{align*}
    Since $\alpha_i\to0$ and $\alpha_i<1$ for every $i$, we have $\inf_{i\geq n+1}(1-\alpha_i)>0$. Thus these derivatives are uniformly bounded, as is the $k = 0$ term, since $f_{\alpha_i}(s)\leq S_r/(1-\alpha_i)$. For a multi-index $\nu$ with $1\leq|\nu|\leq r$, Fa\`{a} di Bruno's formula expresses the derivative $\partial^\nu[f_{\alpha_i}\circ s(\cdot,x')]$ as a finite linear combination of terms
    \begin{align*}
        f_{\alpha_i}^{(k)}\{s(x,x')\}
        \prod_{j=1}^k\partial^{\nu_j}s(x,x')\qquad \text{where }
         \sum_{j=1}^k\nu_j=\nu,\quad |\nu_j|\geq1.
    \end{align*}
    Every factor is uniformly bounded in $i$, $x$ and $x'$, so
    \begin{align*}
        \sup_{i \geq n+1}\,\sup_{x' \in \mathcal{X}}\|f_{\alpha_i} \circ s(\cdot,x') \|_{W^{r,\infty}} < \infty.
    \end{align*}
    Multiplying by $\alpha_i=O(i^{-m})$ gives the result, since $\alpha_i(\cdot,x')=\alpha_i \, f_{\alpha_i}\circ s(\cdot,x')$.
\end{proof}

Another useful lemma is the following.
\begin{lemma}\label{lem:multip}
    Let $f \in W^{r,\infty}(D)$ and $g \in W^{r,2}(D)$. Then we have
    \begin{align*}
        \|fg\|_{W^{r,2}} \lesssim  \|f\|_{W^{r,\infty}} \,\|g\|_{W^{r,2}}, \quad 
        \|f^2\|_{W^{r,\infty}} \lesssim \|f\|_{W^{r,\infty}}^2.
    \end{align*}
\end{lemma}
\begin{proof}
  We follow standard multi-index notation. For each multi-index $\nu$ with $|\nu|\leq r$, the Leibniz product rule and H\"{o}lder's inequality give
    \begin{align*}
        \|\partial^\nu(fg)\|_{L^2}
        &\leq\sum_{\gamma\leq\nu}\binom{\nu}{\gamma}
        \|\partial^\gamma f\|_{L^\infty}
        \|\partial^{\nu-\gamma}g\|_{L^2}\lesssim\|f\|_{W^{r,\infty}}\|g\|_{W^{r,2}},
    \end{align*}
    Summing the squares over $|\nu|\leq r$ gives the first bound. The same argument with $L^\infty$ in place of $L^2$ and $g=f$ gives the second.
\end{proof}

For any fixed observed sample, the initialization, smoothness of $\beta_i$ and lower bound on the variance for each $i$ give $\mu_n,\sigma_n,\sigma_n^2\in H^r(D)$.  First, we show that $\sigma_i^2$ is bounded in $L^2(H^r(D))$ and converges a.s. Following (\ref{eq:app_var_decomp}), we decompose
\begin{align*}
    \sigma_i^2 = \sigma_{i-1}^2 + \Delta M_i - \Delta R_i.
\end{align*}
Using $E[(Z_i^2-1)^2]=2$ and $E[Z_i^4]=3$, Lemmas \ref{lem:alpha_sob} and \ref{lem:multip} give
\begin{align}
    E[\|\Delta M_i\|_{H^r}^2\mid\mathcal{F}_{i-1},X_i]
    &\lesssim i^{-2m}\|\sigma_{i-1}^2\|_{H^r}^2,\notag\\
    E[\|\Delta R_i\|_{H^r}^2\mid\mathcal{F}_{i-1},X_i]
    &\lesssim i^{-4m}\|\sigma_{i-1}^2\|_{H^r}^2.\label{eq:app_sob_increments}
\end{align}
Expanding the squared norm of the update gives
\begin{align*}
    \|\sigma_i^2\|_{H^r}^2
    &= \|\sigma_{i-1}^2\|_{H^r}^2
       + \|\Delta M_i\|_{H^r}^2
       + \|\Delta R_i\|_{H^r}^2\\
    &\quad + 2\langle \sigma_{i-1}^2,\Delta M_i\rangle_{H^r}
       - 2\langle \sigma_{i-1}^2,\Delta R_i\rangle_{H^r} - 2\langle \Delta M_i,\Delta R_i\rangle_{H^r}.
\end{align*}
Since $\Delta M_i$ has conditional mean 0, the term $\langle \sigma_{i-1}^2,\Delta M_i\rangle_{H^r}$ vanishes under the expectation. Applying Cauchy-Schwarz to the remaining cross terms gives
\begin{align}
     E\left[\| \sigma_i^2\|_{H^r}^2 \mid \mathcal{F}_{i-1},X_i\right] \leq (1+ C i^{-2m}) \| \sigma_{i-1}^2\|_{H^r}^2.\label{eq:app_sob_var_recursion}
\end{align}
Taking expectations and iterating thus gives that $\sigma_i^2$ is bounded in $L^2(H^r(D))$. Convergence of $\sigma_i^2$ then follows easily from a similar argument to Theorem \ref{th:pred_asymp} based on the convergence of the martingale and Cauchy sequence. Taking expectations in (\ref{eq:app_sob_increments}) and using the fact that $\sigma_i^2$ is bounded in $L^2(H^r(D))$ gives  $\sigma_i^2 \to \sigma_\infty^2 $ a.s.

The mean function is already a martingale, so we just need to bound
\begin{align*}
\| \mu_i - \mu_{i-1}\|_{H^r}^2 &= Z_i^2 \rho^2\| \beta_i \, \sigma_{i-1} \|_{H^r}^2 \lesssim Z_i^2 \, i^{-2m} \|\sigma_{i-1}\|^2_{H^r}.
\end{align*}
The task is now to show $\sigma_{i-1}$ is bounded in $L^2(H^r(D))$. We once again use a Taylor expansion argument and a martingale and residual decomposition. To that end, we require the following intermediate lemma. 
\begin{lemma}\label{lem:sqrt_remainder}
    For $z \in \mathbb{R}$, $0 \leq  a< 1$ and $0 < \rho < 1$, consider the function
\begin{align*}
    g(a,z) = \sqrt{1 + \rho^2 a\left[(1-a)z^2 - 1\right]}.
\end{align*}
A Taylor expansion around $a = 0$ gives
\begin{align*}
    g(a,z) = 1 + \frac{1}{2}\rho^2 a(z^2 - 1) + a^2 q(a,z), \qquad 
   \max_{0 \leq k \leq r}\, \sup_{a \in [0,1)} \left|\frac{\partial^k}{\partial a^k}\, q(a,z) \right| \leq P_r(z)
\end{align*}
for the polynomial $P_r(z)=C_r(1+z^2)^{r+2}$, with $C_r$ depending only on $r$ and $\rho$.
\end{lemma}
\begin{proof}
Let $h(a,z) = 1+\rho^2 a[(1-a)z^2 - 1]$. Then $h(a,z) \geq 1-\rho^2>0$ from Lemma \ref{lem:var_bound}, and its first two derivatives in $a$ are bounded in absolute value by $2\rho^2(1+z^2)$, with all higher derivatives being 0. Fa\`{a} di Bruno's formula therefore gives
\begin{align*}
    \sup_{a\in[0,1]}|\partial_a^j g(a,z)|
    \leq C_j(1+z^2)^j,\qquad 1\leq j\leq r+2,
\end{align*}
since the derivatives of the square-root function are bounded on $[1-\rho^2,\infty)$. For $a > 0$, Taylor's formula with integral remainder gives 
\begin{align*}
    q(a,z) &= a^{-2}\int_0^a(a-t)(\partial_a^2g)(t,z)\,dt=\int_0^1(1-t)\,(\partial_a^2g)(ta,z)\,dt.
\end{align*}
The latter form is also valid for $a = 0$.
Differentiating under the integral for $0\leq k\leq r$, which is permitted since the integrand and its derivatives are jointly continuous for each $z$, gives
\begin{align*}
    |\partial_a^k q(a,z)|
    &=\left|\int_0^1(1-t)t^k\,
      (\partial_a^{k+2}g)(ta,z)\,dt\right|\leq C_r(1+z^2)^{r+2},
\end{align*}
which proves the bound.
\end{proof}
 Now, we can write
\begin{align*}
    \sigma_i = \sigma_{i-1}+ \underbrace{\frac{1}{2}\rho^2 \beta_i\sigma_{i-1}(Z_i^2 - 1)}_{A_i} + \underbrace{\beta_i^2\, \sigma_{i-1} \, q(\beta_i,Z_i)}_{B_i}.
\end{align*}
Expanding the squared norm gives
\begin{align*}
    \|\sigma_i\|_{H^r}^2 &=  \|\sigma_{i-1}\|_{H^r}^2 + 2 \langle \sigma_{i-1}, A_i\rangle_{H^r} + 2\langle\sigma_{i-1},B_i \rangle_{H^r}\\
    &+ \|A_i\|_{H^r}^2 + 2 \langle A_i, B_i\rangle_{H^r} + \|B_i\|_{H^r}^2.
\end{align*}
Fa\`{a} di Bruno's formula and Lemmas \ref{lem:alpha_sob} and \ref{lem:sqrt_remainder} give 
$\|q(\beta_i,Z_i)\|_{W^{r,\infty}}\lesssim P_r(Z_i).$  Thus Lemmas \ref{lem:alpha_sob} and \ref{lem:multip} give
\begin{align*}
    \|A_i\|_{H^r}
    &\lesssim i^{-m}|Z_i^2-1|\,\|\sigma_{i-1}\|_{H^r},\quad \|B_i\|_{H^r}
    \lesssim i^{-2m}P_r(Z_i)\,\|\sigma_{i-1}\|_{H^r}.
\end{align*}
The term involving $\langle\sigma_{i-1},A_i\rangle_{H^r}$ has conditional expectation zero. Again, Cauchy-Schwarz bounds the other cross terms, and $E[P_r(Z_i)^2]$ and $E[(Z_i^2-1)^2]$ are finite because of the Gaussian having all moments. We thus have
\begin{align*}
     E[\|\sigma_i\|_{H^r}^2 \mid \mathcal{F}_{i-1},X_i] \leq (1 + C i^{-2m}) \,   \|\sigma_{i-1}\|_{H^r}^2.
\end{align*}
Taking expectations and iterating then gives $\sup_{i\geq n}E[\|\sigma_i\|_{H^r}^2]<\infty$. We thus have that $\mu_i$ is bounded in $L^2(H^r(D))$ and thus converges to some $\mu_\infty\in H^r(D)$ a.s. Finally, $H^r(D)$ convergence implies $L^2(D)$ convergence, so uniqueness of the latter limit identifies $\sigma_\infty^2$ with the positive limit in Theorem \ref{th:pred_asymp}.

\subsubsection{Proof of Corollary \ref{cor:inf}}
For the Gaussian similarity in Example \ref{ex:Gaussian_cop}, $s(x,x')$ is infinitely differentiable jointly in $(x,x')$. Compactness of $\mathcal X$ therefore gives Assumption \ref{as:sob} for every integer $r\geq1$. Applying Theorem \ref{th:pred_sobolev} for each $r$ and intersecting the resulting countably many probability-one events, uniqueness of the $L^2(D)$ limits gives
\begin{align*}
    \mu_\infty,\sigma_\infty^2\in\bigcap_{r=1}^{\infty}H^r(D)
    \quad\textnormal{a.s.}
\end{align*}
For each integer $k\geq0$, choose an integer $r>k+p/2$. Sobolev embedding then gives $C^k(D)$ representatives of both limits. For each limit, these representatives agree everywhere on $D$, since they are continuous and agree almost everywhere. Hence both limits have $C^\infty(D)$ representatives a.s.

\subsubsection{Proof of Corollary \ref{cor:SU} (Theorem \ref{th:pred_sobolev} part)}
We follow the proof of Theorem \ref{th:pred_sobolev}, using the moment identities and variance decomposition from the proof of Corollary \ref{cor:SU} (Theorem \ref{th:pred_asymp} part), with fixed $\delta>0$. Lemmas \ref{lem:g_delta_4thmoment}, \ref{lem:alpha_sob}, and \ref{lem:multip} give the same conditional increment bounds as equation (\ref{eq:app_sob_increments}). Since $\Delta M_i$ still has conditional mean 0, the same cross-term argument gives (\ref{eq:app_sob_var_recursion}), and iteration yields $\sup_{i\geq n}E[\|\sigma_i^2\|_{H^r}^2]<\infty$. The martingale and Cauchy sequence arguments then give $\sigma_i^2\to\sigma_\infty^2$ a.s. in $H^r(D)$.

For the mean, we again need a bound on $\sigma_i$ in $L^2(H^r(D))$. The square-root argument requires higher moments, which follow from the same Jensen argument as Lemma \ref{lem:g_delta_4thmoment}. For every integer $k\geq1$,
\begin{align*}
    E\left[\left\{R_{\delta}^{(2)}(\varepsilon_i)\right\}^k\right]
    \leq E[|g_{\delta}(W')|^{2k}]<\infty,
    \qquad W'\sim\mathcal{N}(0,1),
\end{align*}
where the finiteness comes from the existence of the Gaussian moment generating function. Using the definition of $\beta_i$ and the argument of Lemma \ref{lem:alphaxx_bound}, there exists a positive constant $b<1$ such that $\beta_i(x)< b$ uniformly in $i\geq n+1$ and $x\in\mathcal X$. As in Lemma \ref{lem:SU_varbound}, $R_{\delta}^{(1)}(\varepsilon)^2\leq R_{\delta}^{(2)}(\varepsilon)$, so for $0\leq  a< b$,
\begin{align}\label{eq:SU_var_lowerbound}
    h(a,\varepsilon) := 1+a\left\{R_{\delta}^{(2)}(\varepsilon)-1\right\}
    -a^2R_{\delta}^{(1)}(\varepsilon)^2
    \geq1-b>0.
\end{align}
The two derivatives in $a$ of $h(a,\varepsilon)$ are bounded in absolute value by a constant times $1+R_{\delta}^{(2)}(\varepsilon)$, and all higher derivatives vanish. For $g(a,\varepsilon) = \sqrt{h(a,\varepsilon)}$, the same Taylor and Fa\`{a} di Bruno argument as in Lemma \ref{lem:sqrt_remainder} therefore gives
\begin{align*}
   g(a,\varepsilon) &=1+\frac{a}{2}\left\{R_{\delta}^{(2)}(\varepsilon)-1\right\}
      +a^2q(a,\varepsilon),\\
    \max_{0\leq k\leq r}&\sup_{a\in [0,b)}
    |\partial_a^k q(a,\varepsilon)|
    \leq C\left[1+R_{\delta}^{(2)}(\varepsilon)\right]^{r+2}.
\end{align*}
The bound has finite second moment by the preceding moment calculation. Since\\
$E\left[R_{\delta}^{(2)}(\varepsilon_i)-1\right]=0$, the same squared-norm expansion for $\sigma_i$, now with $R_{\delta}^{(2)}(\varepsilon_i)-1$ in place of $\rho^2(Z_i^2-1)$, gives 
\begin{align*}
    E[\|\sigma_i\|_{H^r}^2 \mid \mathcal{F}_{i-1},X_i]
    \leq(1+Ci^{-2m})\|\sigma_{i-1}\|_{H^r}^2.
\end{align*}
The rest of the proof is identical.

\section{Frequentist asymptotics proofs}

\subsection{Notation and preliminary lemmas}

In this section, we will assume throughout that $(Y_n, X_n) \iid P^*$ for $n \geq 1$, and let $\mathcal{F}_n = \sigma\left(Y_1,X_1,\ldots, Y_n, X_n\right)$.  We note that by Assumption \ref{as:dgp}, we have
\begin{align*}
|\mu_*(x)|< \infty, \quad  0 < \sigma_*^2(x) < \infty  \quad \text{for } P_X^*\text{-almost all } x,
\end{align*}
and $\mu_* \in L^2(P_X^*)$, $\sigma_*^2 \in L^1(P_X^*)$. Finiteness and integrability follow from 
$$\mu_*(x)^2 \leq E[Y^2 \mid X = x], \quad \sigma_*^2(x) \leq E[Y^2 \mid X = x],$$ 
and $E[Y^2] < \infty$ follows from $E[Y^4] < \infty$. Strict positivity of $\sigma_*^2(x)$ $P_X^*$-almost everywhere follows from $E[|\log \sigma_*^2(X) |]< \infty$.

\subsubsection{Error functions}
For the single-output case, a key term is the pointwise (random) error function
\begin{align*}
    r_n(y,x) = \frac{y - \mu_n(x)}{\sigma_n(x)}
\end{align*}
for $y \in \mathbb{R}$ and $x \in \mathcal{X}$. We then have  $Z_n= r_{n-1}(Y_n, X_n)$, where $Z_n$ is not necessarily distributed as $\mathcal{N}(0,1)$, and the error functions are clearly equal to
\begin{align}\label{eq:Ezn}
u_{n-1}(x) &= E[Z_n \mid  \mathcal{F}_{n-1}, X_n = x], \quad  w_{n-1}(x) = E[Z_n^2 - 1 \mid \mathcal{F}_{n-1}, X_n = x].
\end{align}
For convenience, we will also define the intermediate term
\begin{align*}
    v_n(x) = \frac{\sigma_*^2(x)}{\sigma_n^2(x)}\cdot 
\end{align*}
Since $0\leq\beta_i(x)<1$, the initialization and updates imply that \(\mu_n\) and \(\sigma_n^2\) are bounded on \(\mathcal X\) for each $n$, with $\inf_{x\in\mathcal X}\sigma_n^2(x)
\geq (1-\rho^2)^n>0$.
Together with the integrability of $\mu_*$ and $\sigma_*^2$ established earlier, this gives \(u_n\in L^2(P_X^*)\) and \(w_n\in L^1(P_X^*)\) for every finite \(n\) $P^*$-a.s. All subsequent identities involving these error functions are understood to hold \(P_X^*\)-almost everywhere.
\subsubsection{Learning rates}
Next, we have the following uniform upper bounds to handle  $\beta_n(x)$.
\begin{lemma}\label{lem:betan_bounds}
    Under Assumptions \ref{as:bounded_sxx} and \ref{as:freq_lr}, we have
    \begin{align*}
        \beta_n^2(x)  \lesssim  \alpha_n^2, \quad \left|\beta_n(x) - \alpha_n s(x,X_n) \right| \lesssim \alpha_n^2
    \end{align*}
     where the constants are uniform over $n$, $x$ and $X_n$.
\end{lemma}
\begin{proof}
    The first part follows from the same argument as Lemma \ref{lem:alphaxx_bound}. For the second,
    \begin{align*}
        \beta_n(x) - \alpha_n s(x,X_n)  = \alpha_n^2\, \frac{ s(x,X_n) \{1 - s(x,X_n)\}}{1- \alpha_n + \alpha_n s(x,X_n)}\cdot
    \end{align*}
    The denominator is bounded away from $0$, and Assumption \ref{as:bounded_sxx} gives the result.
\end{proof}

\subsubsection{Similarity norm}
Let $\mathcal{S}$ denote the integral operator,
\begin{align*}
    (\mathcal{S}f)(x) = \int \, f(x') \, s(x,x')\, dP_X^*(x'),
\end{align*}
and let $\langle f, g \rangle_s$ denote the bilinear form
\begin{align*}
    \langle f, g \rangle_s = \int f(x) \, g(x')\, s(x,x') \, dP_X^*(x) \, dP_X^*(x').
\end{align*}
For shorthand, we will write $L^1_* = L^1(P_X^*)$, with the respective norm \\ 
$\| f \|_{L^1_*} = \int |f(x)|\, dP_X^*(x).$  We now demonstrate that the function $\|\cdot \|_s$ is a norm.
\begin{lemma}\label{lem:norm_s}
Under Assumption \ref{as:bounded_sxx}, $\| \cdot \|_s: L^1_* \to \mathbb{R}$ is a seminorm, and 
\begin{align*}
        \|\mathcal{S} f\|_\infty \lesssim \|f\|_s, \quad \left|\langle f,g \rangle_s\right| \lesssim \| f\|_{L^1_*}\, \|g\|_{L^1_*}.
\end{align*}
If additionally Assumptions \ref{as:dgp}, \ref{as:regularity} and \ref{as:complete} hold, then $\|\cdot \|_s$ is a norm. 
\end{lemma}

\begin{proof}
Let $f$ and $g$ be functions in $L^1_*$. We begin with the bilinear form result:
\begin{align*}
    \left| \langle f, g\rangle_s \right| &\leq \int\int |f(x)|\, |g(x')| |s(x,x')| dP_X^*(x)\, dP_X^*(x')\leq S \|f\|_{L_*^1} \, \|g\|_{L_*^1}
\end{align*}
by Lemma \ref{lem:sxx_bound}.  Next, the kernel form of $s(x,x')$ immediately gives us non-negativity, as Fubini's theorem gives
\begin{align}\label{eq:norm_s_mixture}
    \| f\|_s^2 = \int \, (\eta f)(\psi)^2\, d\Pi(\psi) \geq 0, \quad 
 (\eta f)(\psi) =\int \frac{f(x)\, k(x;\psi)}{E_{\Psi}[k(x;\Psi)]} \, dP_X^*(x).
\end{align}
The scaling property of a norm holds immediately. For the triangle property,
\begin{align*}
    \left|\langle f, g\rangle_s \right|   = \left|\int (\eta f)(\psi) \, (\eta g)(\psi) \,d\Pi(\psi) \right|\leq \|f\|_s \, \|g \|_s,
\end{align*}
where we have used Cauchy-Schwarz. Next, we note that
\begin{align*}
    \| f + g\|_s^2 &= \|f\|_s^2 + \|g\|_s^2 + 2 \langle f,g\rangle_s \\
    &\leq \|f\|_s^2 + \|g\|_s^2 + 2 \|f\|_s \|g\|_s = \left(\| f\|_s + \|g \|_s\right)^2.
\end{align*}
Taking the square root then gives the triangle inequality, so $\|\cdot \|_s$ is a seminorm. For the first inequality, we note that
\begin{align*}
    (\mathcal{S}f)(x) = \int \frac{k(x;\psi)}{E_\Psi[k(x;\Psi)]}\,(\eta f)(\psi) \, d\Pi(\psi),  
\end{align*}
so Cauchy-Schwarz applied to the above gives
\begin{align*}
    (\mathcal{S}f)(x)^2 \leq \int \left[\frac{k(x;\psi)}{E_\Psi[k(x;\Psi)]}\right]^2\, d\Pi(\psi)\, \int (\eta f)(\psi')^2 \, d\Pi(\psi') = s(x,x)\, \|f\|_s^2.
\end{align*}
Taking the supremum of both sides gives $\| (\mathcal{S}f)^2\|_\infty \leq S \|f\|_s^2$. 

For the norm result, let $\|f\|_s = 0$. By (\ref{eq:norm_s_mixture}), we have $(\eta f)(\psi)^2 = 0$ for $\Pi$-almost all $\psi$, which can be strengthened to hold for all $\psi \in \text{supp}(\Pi)$ if $\eta f$ is continuous. Let $\psi_j \to \psi$ be some convergent sequence, and note that
\begin{align*}
\left|
\frac{f(x)\, k(x;\psi_j)}{E_\Psi[k(x;\Psi)]}
\right|
\le
\frac{\sup_{x \in \mathcal{X},j \geq 1}|k(x;\psi_j)|}
{\inf_{x\in\mathcal X}E_\Psi[k(x;\Psi)]}
\,|f(x)| \lesssim |f(x)|
\end{align*}
where the denominator is bounded away from 0 by Assumption \ref{as:regularity}, and joint continuity of $k$, compactness of $\mathcal{X}$ and convergence of $\psi_j$ ensure the numerator is finite.
 Since $f \in L_*^1$ and $k(x;\psi)$ is jointly continuous, we can apply dominated convergence to show $(\eta f)(\psi_j) \to (\eta f)(\psi)$, so $\eta f$ is continuous and thus $(\eta f)(\psi) = 0$ for all $\psi \in \text{supp}(\Pi)$.  
 By completeness and Tonelli's theorem, we then have
 \begin{align*}
0 &=\int\!\left[\int_{\mathbb{R}^p} \left\{\mathbf{1}(x \in \mathcal{X})\,
\frac{|f(x)|p_X^*(x)}{E_\Psi[k(x;\Psi)]}\right\}
\,k(x;\psi)\,dx\right]d\Pi(\psi)  =\|f\|_{L^1_*},
 \end{align*}
 where completeness ensures the inner integral is $0$ for each $\psi$. We thus have $f(x) = 0$ for $P_X^*$-almost all $x$. The converse direction is immediate. 
\end{proof}
\subsection{Proof of Theorem \ref{th:consist}}
We will now study the convergence of the energy $K_n = \int \,K^X_n(x)\,  dP_X^*(x)$  where
\begin{align*}
  K^X_n = 2\text{KL}\left[\mathcal{N}\left(\mu_*, \sigma_*^2\right), \,\mathcal{N}\left(\mu_n, \sigma_n^2\right) \right]= w_n - \log v_n,
\end{align*}
where $\text{KL}(f,g)$ is the Kullback-Leibler (KL) divergence between the densities $f$ and $g$. Note that we are not measuring the KL divergence from the true $p^*(y \mid x)$. The above choice of $K_n^X$ is especially compatible with our moment updates, allowing us to utilize the almost-supermartingale theory of \cite{Robbins1971}. We note that the factor $2$ is for convenience, and that Assumptions \ref{as:init} and \ref{as:dgp} give
\begin{align*}
    K_0 = E[Y^2]-1  - E[\log\sigma_*^2(X)] < \infty.
\end{align*}
While we will show the convergence of $K_n$ to a finite (random) limit $P^*$-a.s., it is not necessarily 0. The key term that we will show goes to 0 $P^*$-a.s. is
\begin{align*}
T_n  &= 2\rho \|u_n\|_s^2 + \rho^2 \|w_n\|_s^2,
\end{align*}
where clearly $T_n \geq 0$ from Lemma \ref{lem:norm_s}.

\subsubsection{Technical lemmas}
In this section, we outline some useful technical lemmas. First, we show a lemma on the growth of $r_n(y, x)$, which we use to control the residual terms in the almost-supermartingale. The assumptions are slightly stronger than needed but convenient. 
\begin{lemma}\label{lem:rn}
Under Assumptions \ref{as:bounded_sxx}, \ref{as:init} and \ref{as:freq_lr}, the following holds for all $(y,x) \in \mathbb{R} \times \mathcal{X}$: 
\begin{align*}
    r_n^2(y,x) \lesssim \left[1+ r_0^2(y,x)\right](n+1)^{\alpha_0\rho^2 S}.
\end{align*}
\end{lemma}

\begin{proof}
 Let $r \in \mathbb{R}$, $z \in \mathbb{R}$, $0 < \rho <1$ and $0 \leq a <1$. Consider the term
\begin{align*}
   h(z) =  \frac{(r - \rho a z)^2}{1+\rho^2 a[(1-a)z^2 - 1]}\cdot
\end{align*}
Differentiating with respect to $z$ gives
\begin{align*}
    h'(z) = \frac{-2\rho a(r-\rho a z)\left\{1 - \rho^2 a + \rho(1-a) r z \right\}}{\left\{1+\rho^2 a[(1-a)z^2 - 1] \right\}^2}\cdot
\end{align*}
For $r \neq 0$, we can solve for $h'(z) = 0$ and verify that $h(z)$ is maximized at 
\begin{align*}
    z_* = -\frac{1-\rho^2 a}{\rho(1-a)r}
\end{align*}
which when substituted gives the upper bound
\begin{align*}
    h(z) \leq h(z_*) = \frac{r^2}{1- \rho^2 a} + \frac{a}{1-a}\cdot
\end{align*}
When $r = 0$, we have
\begin{align*}
    h(z) = \frac{a}{1-a} \, \frac{\rho^2 a(1-a) z^2}{1+\rho^2 a[(1-a)z^2 - 1]}  \leq \frac{a}{1-a}
\end{align*}
so the upper bound holds for all $r \in \mathbb{R}$.

Substituting $r = r_{n-1}(y,x)$, $z = Z_n$ and $a = \beta_n(x)$ then gives the recursion
\begin{align*}
    r_n^2(y,x) \leq \frac{r_{n-1}^2(y,x)}{1-\rho^2 \beta_n(x)} + \frac{\beta_n(x)}{1 - \beta_n(x)}\cdot 
\end{align*}
Iterating the above gives
\begin{align*}
    r_n^2(y,x) \leq r_0^2(y,x)\prod_{i = 1}^n \frac{1}{1-\rho^2 \beta_i(x)} + \sum_{i = 1}^n \frac{\beta_i(x)}{1 - \beta_i(x)} \prod_{j = i+1}^n \frac{1}{1-\rho^2 \beta_j(x)}\cdot  
    \end{align*}
Next, the definition of $\beta_j(x)$ gives
\begin{align*}
   \frac{\beta_j(x)}{1 - \rho^2\beta_j(x)} \leq  \frac{\beta_j(x)}{1 - \beta_j(x)}\leq \frac{\alpha_0 S}{j+ 1- \alpha_0}\cdot
\end{align*}
The product term then satisfies the following for $0 \leq i \leq n$:
\begin{align*}
    \prod_{j = i+1}^n \frac{1}{1-\rho^2 \beta_j(x)} \leq \exp\left\{\sum_{j = i+1}^n \frac{\rho^2 \beta_j(x)}{1- \rho^2 \beta_j(x)}\right\}  \leq \left(\frac{n+1}{i + 1-\alpha_0}\right)^{ \alpha_0\rho^2 S}.
\end{align*}
Together, this gives
\begin{align*}
    r_n^2(y,x) &\lesssim r_0^2(y,x)\, (n+1)^{\alpha_0 \rho^2 S} + \alpha_0 S (n+1)^{\alpha_0 \rho^2 S} \,\sum_{i = 1}^n i^{-(1+\alpha_0 \rho^2 S)}\\
    &\lesssim \left[1+ r_0^2(y,x)\right](n+1)^{\alpha_0\rho^2 S}
\end{align*}
since $\sum_{i = 1}^\infty i^{-(1+\alpha_0 \rho^2 S)} < \infty$, and Assumption \ref{as:init} ensures $r_0(y,x)$ is well-defined.
\end{proof}

We first collect bounds on the standardized mean/variance increments. Let
\begin{align*}
    \Delta_{\mu} = \rho \beta_n Z_n, \quad \Delta_{\sigma^2} = \beta_n \rho^2 \left\{(1-\beta_n)Z_n^2 - 1\right\},
\end{align*}
so that $\mu_n = \mu_{n-1} + \sigma_{n-1}\Delta_{\mu}$ and $\sigma_n^2 = \sigma_{n-1}^2(1 + \Delta_{\sigma^2})$. We then have the following.
\begin{lemma}\label{lem:Delta_mu_sigma_bounds}
Under Assumptions \ref{as:bounded_sxx}, \ref{as:init}, \ref{as:dgp} and \ref{as:freq_lr}, we have
\begin{align}\label{eq:Delta_mu_sigma_bounds}
    |\Delta_{\mu}| &\lesssim \alpha_n |Z_n|, \quad |\Delta_{\sigma^2}| \lesssim \alpha_n(1+Z_n^2).
\end{align}
Furthermore, we have
\begin{align}
    \left|E[\Delta_{\mu} \mid \mathcal{F}_{n-1}] - \alpha_n \rho\,\mathcal{S}u_{n-1}\right|
    &\lesssim \alpha_n^2 E[(1+Z_n^2)^2 \mid \mathcal{F}_{n-1}], \label{eq:EDelta_mu_bounds2}\\
    \left|E[\Delta_{\sigma^2} \mid \mathcal{F}_{n-1}] - \alpha_n \rho^2\,\mathcal{S}w_{n-1}\right|
    &\lesssim \alpha_n^2 E[(1+Z_n^2)^2 \mid \mathcal{F}_{n-1}]. \label{eq:EDelta_sigma_bounds2}
\end{align}
\end{lemma}

\begin{proof}
Since $0\leq\beta_n<1$, Lemma \ref{lem:betan_bounds} immediately gives (\ref{eq:Delta_mu_sigma_bounds}) and
\begin{align*}
    \left|\Delta_{\mu} - \alpha_n s(\cdot,X_n)\rho Z_n\right|
    &\lesssim \alpha_n^2(1+Z_n^2)^2, \\
    \left|\Delta_{\sigma^2} - \alpha_n s(\cdot,X_n)\rho^2(Z_n^2-1)\right|
    &\lesssim \alpha_n^2(1+Z_n^2)^2, 
\end{align*}
where we used $\Delta_{\sigma^2}=\rho^2\beta_n(Z_n^2-1)-\rho^2\beta_n^2Z_n^2$ for the second inequality. We use the common upper bound $(1+Z_n^2)^2$ for convenience. Next, (\ref{eq:Ezn}) gives
\begin{align*}
    E[s(\cdot,X_n)Z_n \mid \mathcal{F}_{n-1}] &= \mathcal{S}u_{n-1},
\quad E[s(\cdot,X_n)(Z_n^2-1) \mid \mathcal{F}_{n-1}] = \mathcal{S}w_{n-1},
\end{align*}
and Jensen's inequality gives (\ref{eq:EDelta_mu_bounds2}) and (\ref{eq:EDelta_sigma_bounds2}).
\end{proof}

Next, we show a key technical lemma that controls the change in the error functions $w_n$ and $u_n$.
\begin{lemma}\label{lem:delta_w_u_ineq}
Under Assumptions \ref{as:bounded_sxx}, \ref{as:init}, \ref{as:dgp} and \ref{as:freq_lr}, we have
\begin{align*}
    u_n - u_{n-1} &= -\Delta_{\mu} - \frac{1}{2}\Delta_{\sigma^2}u_{n-1} + R_n^u,\\
    w_n - w_{n-1} &= -2\Delta_{\mu}u_{n-1} - \Delta_{\sigma^2}(1+w_{n-1}) + R_n^w,
\end{align*}
where $|R_n^w| + |R_n^u| \lesssim \alpha_n^2(1+Z_n^2)^2(1+ |u_{n-1}|+|w_{n-1}|).$ Furthermore,
\begin{align*}
    |u_n - u_{n-1}| + |w_n - w_{n-1}|  \lesssim \alpha_n(1+Z_n^2)(1+ |u_{n-1}|+|w_{n-1}|).
\end{align*}
\end{lemma}

\begin{proof}
We start with $w_n$. The updates give
\begin{align}
    w_n-w_{n-1}
    &= \frac{-2\Delta_{\mu}u_{n-1}+\Delta_{\mu}^2-\Delta_{\sigma^2}(1+w_{n-1})}{1+\Delta_{\sigma^2}} \label{eq:delta_w_exact}\\
    &= -2\Delta_{\mu}u_{n-1}-\Delta_{\sigma^2}(1+w_{n-1})+R_n^w, \nonumber
\end{align}
where
\begin{align*}
    R_n^w = \frac{\Delta_{\mu}^2+\Delta_{\sigma^2}^2(1+w_{n-1})+2\Delta_{\mu}\Delta_{\sigma^2}u_{n-1}}{1+\Delta_{\sigma^2}}\cdot
\end{align*}
Using $1+\Delta_{\sigma^2}\geq1-\rho^2$ from Lemma
\ref{lem:var_bound} and the triangle inequality gives
\begin{align*}
    |R_n^w|
    &\leq \frac{\Delta_{\mu}^2
    +\Delta_{\sigma^2}^2(1+|w_{n-1}|)
    +2|\Delta_{\mu}\Delta_{\sigma^2}||u_{n-1}|}{1-\rho^2}\\
    &\lesssim \alpha_n^2(1+Z_n^2)^2
    (1+|u_{n-1}|+|w_{n-1}|),
\end{align*}
where the last line follows from (\ref{eq:Delta_mu_sigma_bounds}). For the first-difference bound, (\ref{eq:delta_w_exact}) gives
\begin{align*}
    |w_n-w_{n-1}|
    &\leq \frac{2|\Delta_{\mu}||u_{n-1}|+\Delta_{\mu}^2
    +|\Delta_{\sigma^2}|(1+|w_{n-1}|)}{1-\rho^2}\\
    &\lesssim \alpha_n(1+Z_n^2)
    (1+|u_{n-1}|+|w_{n-1}|),
\end{align*}
using (\ref{eq:Delta_mu_sigma_bounds}) and
$\Delta_{\mu}^2\lesssim\alpha_n^2Z_n^2\leq\alpha_n Z_n^2$. Next, for $u_n$, we have
\begin{align*}
    u_n-u_{n-1}
    &=\frac{u_{n-1}-\Delta_{\mu}}{\sqrt{1+\Delta_{\sigma^2}}}-u_{n-1}=-\Delta_{\mu}-\frac{1}{2}\Delta_{\sigma^2}u_{n-1}+R_n^u,
\end{align*}
where
\begin{align*}
    R_n^u
    &=u_{n-1}\left(\frac{1}{\sqrt{1+\Delta_{\sigma^2}}}-1+\frac{1}{2}\Delta_{\sigma^2}\right) -\Delta_{\mu}\left(\frac{1}{\sqrt{1+\Delta_{\sigma^2}}}-1\right).
\end{align*}
Taylor's theorem gives
\begin{align*}
    \left|\frac{1}{\sqrt{1+t}}-1\right|\lesssim|t|, \quad
    \left|\frac{1}{\sqrt{1+t}}-1+\frac{t}{2}\right|\lesssim t^2
\end{align*}
for $t\geq-\rho^2>-1$, with constants depending on $\rho$. Thus (\ref{eq:Delta_mu_sigma_bounds}) gives
\begin{align*}
    |R_n^u|
    &\lesssim |u_{n-1}||\Delta_{\sigma^2}|^2+|\Delta_{\mu}||\Delta_{\sigma^2}| \lesssim \alpha_n^2(1+Z_n^2)^2(1+ |u_{n-1}|+|w_{n-1}|).
\end{align*}
Finally, writing

\begin{align*}
    u_n-u_{n-1}
    =u_{n-1}\left(\frac{1}{\sqrt{1+\Delta_{\sigma^2}}}-1\right)
    -\frac{\Delta_{\mu}}{\sqrt{1+\Delta_{\sigma^2}}}
\end{align*}
and using the same Taylor bound and $1+\Delta_{\sigma^2}\geq1-\rho^2$ gives
\begin{align*}
    |u_n-u_{n-1}|
    &\lesssim |u_{n-1}||\Delta_{\sigma^2}|+|\Delta_{\mu}| \lesssim \alpha_n(1+Z_n^2)(1+|u_{n-1}|). \qedhere
\end{align*}
\end{proof}
Finally, we provide a useful upper bound on the error terms.

\begin{lemma}\label{lem:error_bounds}
    The error terms satisfy
    \begin{align*}
    1 + |u_n| +  |w_n| \lesssim 1 + K_n^X.
    \end{align*}
\end{lemma}

\begin{proof}
Since $\log v\leq v/2$ for $v>0$, we have
\begin{align*}
    1+K_n^X=u_n^2+v_n-\log v_n
    \geq u_n^2+\frac{v_n}{2}\cdot 
\end{align*}
Using $|u_n|\leq1+u_n^2$ and
$|w_n|=|u_n^2+v_n-1|\leq1+u_n^2+v_n$ gives
\begin{align*}
    1+|u_n|+|w_n|
    \lesssim 1+u_n^2+v_n
    \lesssim 1+K_n^X
\end{align*}
where the last inequality uses $K_n^X \geq 0$.
\end{proof}

\subsubsection{Main proof}
We are now ready to show the main convergence result.
\begin{theorem}\label{th:amart_convergence}
Under Assumptions \ref{as:bounded_sxx}, \ref{as:init}, \ref{as:dgp},  \ref{as:freq_lr}, the updates (\ref{eq:mean_gauss}) and (\ref{eq:var_gauss}) give
\begin{align*}
K_n \to K_\infty < \infty \quad \textnormal{and} \quad  \sum_{n = 1}^\infty \alpha_n T_{n-1} < \infty \quad P^*\textnormal{-a.s.}
\end{align*}
\end{theorem}

\begin{proof}
Using $\log(1+t)\leq t$ for $t>-1$ and Lemma \ref{lem:delta_w_u_ineq}, we have
\begin{align*}
    K_n^X-K_{n-1}^X
    &=\log(1+\Delta_{\sigma^2})+(w_n-w_{n-1})\\
    &\leq\Delta_{\sigma^2}+(w_n-w_{n-1})=-2\Delta_{\mu}u_{n-1}-\Delta_{\sigma^2}w_{n-1}+R_n^w.
\end{align*}
We now take the conditional expectation and integrate over $P_X^*$. Applying (\ref{eq:EDelta_mu_bounds2}) and (\ref{eq:EDelta_sigma_bounds2}), the leading terms integrate to $-\alpha_nT_{n-1}$. The combined error is bounded by
$$C\alpha_n^2E[(1+Z_n^2)^2\mid\mathcal F_{n-1}]
\left\|1+|u_{n-1}|+|w_{n-1}|\right\|_{L_*^1},$$
following Lemmas \ref{lem:delta_w_u_ineq} and \ref{lem:Delta_mu_sigma_bounds}. Next, Lemma \ref{lem:error_bounds} and $\|K_{n-1}^X\|_{L_*^1} = K_{n-1}$ give
\begin{align*}
    E[K_n\mid\mathcal{F}_{n-1}]-K_{n-1}
    \leq-\alpha_nT_{n-1}
    +C\alpha_n^2E[(1+Z_n^2)^2\mid\mathcal{F}_{n-1}](1+K_{n-1}).
\end{align*}
Next, Lemma \ref{lem:rn} and the initialization from Assumption \ref{as:init} give
\begin{align*}
Z_n^2 \lesssim \left(1 + Y_n^2 \right)n^{\alpha_0 \rho^2 S}.
\end{align*}
Assumption \ref{as:dgp} then gives $E[(1+Y_n^2)^2] < \infty$, so
\begin{align*}
   \alpha_n^2 E\left[(1+Z_n^2)^2 \mid \mathcal{F}_{n-1}\right] \lesssim n^{-2(1-\alpha_0 \rho^2 S)},
\end{align*}
and under the upper bound on $\alpha_0 < \min\{1,1/(2\rho^2 S)\}$ from Assumption \ref{as:freq_lr}, we have
\begin{align*}
    \sum_{n = 1}^\infty \alpha_n^2 E\left[(1+Z_n^2)^2 \mid \mathcal{F}_{n-1}\right]< \infty.
\end{align*}
From the seminorm property in Lemma \ref{lem:norm_s}, $T_n \geq 0$, and clearly $K_n \geq 0$ and $K_0 < \infty$. We can thus apply the almost-supermartingale convergence theorem of \cite{Robbins1971}, giving us the final result.
\end{proof}

While the above theorem shows that $\liminf_n T_n = 0$ $P^*$-a.s. since $\sum_{n = 1}^\infty \alpha_n = \infty$, we now show convergence of $T_n$ in order to strengthen this to $T_n \to 0$ $P^*$-a.s. 
\begin{theorem}\label{th:convergence2}
    Under Assumptions \ref{as:bounded_sxx}, \ref{as:init}, \ref{as:dgp},  \ref{as:freq_lr}, the updates (\ref{eq:mean_gauss}) and (\ref{eq:var_gauss}) give
    \begin{align*}
        T_n \to 0 \quad P^*\textnormal{-a.s.}
    \end{align*}
\end{theorem}

\begin{proof}
We will use a second almost-supermartingale. Let $\Delta_u := u_n - u_{n-1}$ and $\Delta_w := w_n - w_{n-1}$. First, we write
\begin{align*}
    T_n - T_{n-1} &=4 \rho \langle u_{n-1}, \Delta_u \rangle_s + 2 \rho^2 \langle w_{n-1},  \Delta_w \rangle_s + 2\rho \| \Delta_u \|_s^2 + \rho^2 \|\Delta_w\|_s^2.
\end{align*}
We first handle the remainder terms, starting with $R_n^u$. Lemmas \ref{lem:norm_s} and \ref{lem:delta_w_u_ineq} give
\begin{align*}
    |\langle u_{n-1},R_n^u\rangle_s|
    &\lesssim\alpha_n^2(1+Z_n^2)^2\left\|1+|u_{n-1}|+|w_{n-1}|\right\|_{L^1_*}\,\,\left\|u_{n-1}\right\|_{L^1_*}\\
    &\lesssim\alpha_n^2(1+Z_n^2)^2(1+K_{n-1})^2,
\end{align*}
where the second inequality follows from Lemma \ref{lem:error_bounds}. Since $\|\Delta_u\|_s^2 = \langle \Delta_u, \Delta_u \rangle_s$,  the second part of Lemma \ref{lem:delta_w_u_ineq} gives a similar bound. Handling  $ |\langle w_{n-1},R_n^w\rangle_s|$ and $\|\Delta_w\|_s^2$ in the same way gives
\begin{align*}
    |\langle u_{n-1},R_n^u\rangle_s|+|\langle w_{n-1},R_n^w\rangle_s|+\|\Delta_u\|_s^2+\|\Delta_w\|_s^2\lesssim\alpha_n^2(1+Z_n^2)^2(1+K_{n-1})^2.
\end{align*}
For the leading terms, Lemma \ref{lem:delta_w_u_ineq} gives
\begin{align*}
    E[\Delta_w\mid\mathcal{F}_{n-1}]
    &=-2u_{n-1}E[\Delta_{\mu}\mid\mathcal{F}_{n-1}]
    -(1+w_{n-1})E[\Delta_{\sigma^2}\mid\mathcal{F}_{n-1}]
    +E[R_n^w\mid\mathcal{F}_{n-1}],\\
    E[\Delta_u\mid\mathcal{F}_{n-1}]
    &=-E[\Delta_{\mu}\mid\mathcal{F}_{n-1}]
    -\frac{1}{2}u_{n-1}E[\Delta_{\sigma^2}\mid\mathcal{F}_{n-1}]
    +E[R_n^u\mid\mathcal{F}_{n-1}].
\end{align*}
Substituting (\ref{eq:EDelta_mu_bounds2}) and (\ref{eq:EDelta_sigma_bounds2}) introduces additional errors bounded pointwise by 
$$C\alpha_n^2E[(1+Z_n^2)^2\mid\mathcal{F}_{n-1}](1+|u_{n-1}|+|w_{n-1}|),$$
and their inner products with $u_{n-1}$ and $w_{n-1}$ can be controlled as for the other remainders. 
Handling $E\left[(1+Z_n^2)^2 \mid \mathcal{F}_{n-1}\right]$ as in Theorem \ref{th:amart_convergence}, we have
\begin{align}
    E[T_n \mid \mathcal{F}_{n-1}] - T_{n-1} &\leq -4 \alpha_n  \rho^2 \int \left(\mathcal{S}u_{n-1}\right)^2 \, dP_X^* \nonumber\\
    &\quad -2\alpha_n\rho^4 \int (1+w_{n-1}) \left(\mathcal{S} w_{n-1}\right)^2\, dP_X^* \nonumber\\
    &\quad -6 \alpha_n\rho^3 \int u_{n-1}\left(\mathcal{S}w_{n-1}\right)\left(\mathcal{S}u_{n-1}\right) \, dP_X^* \label{eq:Et_t}\\
    &\quad + C n^{-2(1-\alpha_0 \rho^2 S)}\, \left(1 + K_{n-1}\right)^2. \nonumber
\end{align}
The first two terms in the above are non-positive, so we only need summability of the remaining two terms. The last term is handled by Assumption \ref{as:freq_lr} as before and by $K_n \to K_\infty < \infty$ $P^*$-a.s. from Theorem \ref{th:amart_convergence}. For the third term (\ref{eq:Et_t}), we have
\begin{align*}
    \left|\int u_{n-1}\left(\mathcal{S}w_{n-1}\right)\left(\mathcal{S}u_{n-1}\right) \, dP_X^*  \right| &\leq \|u_{n-1}\|_{L^1_*} \left\|\mathcal{S}w_{n-1} \right\|_\infty \,\left\|\mathcal{S}u_{n-1}\right\|_\infty\\
    &\lesssim (1+K_{n-1}) \|u_{n-1}\|_s \|w_{n-1}\|_s\\
    &\lesssim (1+K_{n-1})\,T_{n-1},
\end{align*}
where the second line follows from Lemmas \ref{lem:error_bounds} and \ref{lem:norm_s}, and the last line follows from $2|ab| \leq a^2 + b^2$. Since $\sum_{n = 1}^\infty \alpha_n T_{n-1} < \infty$ and $K_n\to K_\infty < \infty$ $P^*$-a.s., the absolute value of (\ref{eq:Et_t}) is also summable. We can then apply the almost-supermartingale result again to show $T_n \to T_\infty$ $P^*$-a.s. Finally, because $\liminf_n T_n = 0$, we must have $T_n \to 0$ $P^*$-a.s.
\end{proof}
It is clear that $T_n \to 0$ implies $\|u_n\|_s \to 0$ and $\|w_n\|_s \to 0$. The final statement then follows from $\|\cdot\|_s$ being a norm from Lemma \ref{lem:norm_s}.

\subsection{Proof of Corollary \ref{cor:consist_gen}}
For the SU case, the proof is a natural extension of the previous section, where we rely on $1-\rho_i = O(\alpha_i)$ to approximate the SU update with the Gaussian update at $\rho = 1$. The main difference is that
\begin{align*}
    \Delta_\mu = \beta_n R_{\delta}^{(1)}(\varepsilon_n), \quad \Delta_{\sigma^2} = \beta_n\left\{R_{\delta}^{(2)}(\varepsilon_n) - \beta_n R_{\delta}^{(1)}(\varepsilon_n)^2  -1 \right\}
\end{align*}
Note that both $R_{\delta}^{(1)}(\varepsilon_n)$ and $R_{\delta}^{(2)}(\varepsilon_n)$ now depend on a changing $\rho_n$, but we use the same notation as the main paper for convenience. The definitions of $r_n(y,x)$, $u_n$, $w_n$, and $K_n$ all remain identical, and we set $\rho = 1$ for $T_n$ so $T_n = 2 \|u_n\|_s^2 + \|w_n\|_s^2$.

We begin by showing that Lemma \ref{lem:rn} extends to this setting. First,
\begin{align*}
    r_n^2= \frac{\left(r_{n-1} -  \Delta_\mu \right)^2}{1 + \Delta_{\sigma^2}}
\end{align*}
where we omit the dependence on $(y,x)$. Since $R_{\delta}^{(1)}(\varepsilon_n)^2 \leq R_{\delta}^{(2)}(\varepsilon_n)$,
\begin{align*}
    r_n^2 \leq  \frac{(r_{n-1} - a z)^2}{1 + a \left[(1- a)z^2 -1 \right]}
\end{align*}
where $z = R_{\delta}^{(1)}(\varepsilon_n)$ and $a = \beta_n$. The right-hand side is exactly $h(z)$ with $\rho = 1$, and the rest of the proof is identical as it only relies on $a < 1$ and not on $\rho < 1$, so 
\begin{align}\label{eq:rn_SU}
    r_n^2(y,x) \lesssim \left[1+r_0^2(y,x)\right](n+1)^{\alpha_0 S}.
\end{align}

Next, we will need to obtain the equivalent of Lemma \ref{lem:Delta_mu_sigma_bounds}.  As a reminder, we have $Z_n = r_{n-1}(Y_n,X_n)$. The following technical lemma is the main new ingredient.
\begin{lemma}\label{eq:Rndelt_bound}
    For $0<\rho_n\leq 1$, we have
    \begin{align*}
        \left| R_{\delta}^{(1)}(\varepsilon_n)\right| \lesssim |Z_n|, \quad   R_{\delta}^{(2)}(\varepsilon_n) \lesssim 1 + Z_n^2. 
    \end{align*}
    Furthermore, we have
    \begin{align*}
\left|R_\delta^{(1)}(\varepsilon_n)-Z_n\right| + \left|R_\delta^{(2)}(\varepsilon_n)-Z_n^2\right|
&\lesssim(1-\rho_n)(1+Z_n^2)^2.
\end{align*}
\end{lemma}

\begin{proof}
    Since $|\sinh(\rho_n t)|\leq|\sinh t|$ for any $t\in\mathbb R$ and $0<\rho_n\leq 1$, we have
$$
\left|R_\delta^{(1)}(\varepsilon_n)\right|
\leq
\frac{\omega^{(1-\rho_n^2)/2}}{D_\delta}|D_\delta Z_n|
\leq\sqrt{\omega}|Z_n|,
$$
where $D_\delta=\sqrt{\frac{1}{2}(\omega^2-1)}$ and $\omega>1$. Similarly, $\cosh(2\rho_n t)\leq\cosh(2t)$ and $\cosh(2t)=1+2\sinh^2(t)$ give
$$
R_\delta^{(2)}(\varepsilon_n)
\leq
\frac{\omega^2\{1+(\omega^2-1)Z_n^2\}-1}{\omega^2-1}
=1+\omega^2Z_n^2.
$$
For the remaining inequalities, hold $Z_n$ fixed, and differentiate with respect to $\rho_n$:
$$
\frac{\partial R_\delta^{(1)}(\varepsilon_n)}{\partial\rho_n}
=
\frac{\omega^{(1-\rho_n^2)/2}}{D_\delta}
\left[
\cosh\left(\frac{\rho_n\varepsilon_n}{\delta}\right)
\sinh^{-1}(D_\delta Z_n)
-\rho_n\log(\omega)
\sinh\left(\frac{\rho_n\varepsilon_n}{\delta}\right)
\right].
$$
To upper bound the above magnitude, we have
$$
\cosh\left(\frac{\rho_n\varepsilon_n}{\delta}\right) \leq \cosh\left(\frac{\varepsilon_n}{\delta}\right)
=\sqrt{1+D_\delta^2Z_n^2}\leq1+D_\delta|Z_n|,
$$
using $\cosh\{\sinh^{-1}(x)\} = \sqrt{1 + x^2}$. Handling $\sinh(\rho_n \varepsilon_n/\delta)$ as before, we have
\begin{align*}
\left|\frac{\partial R_\delta^{(1)}(\varepsilon_n)}{\partial\rho_n}\right|
&\lesssim
(1+|Z_n|)|\sinh^{-1}(D_\delta Z_n)|+|Z_n| \lesssim (1+Z_n^2)
\end{align*}
where we use $|\sinh^{-1}(x)|\leq|x|$ and $2 |Z_n| \leq 1 + Z_n^2$. Next, we have
$$
\frac{\partial R_\delta^{(2)}(\varepsilon_n)}{\partial\rho_n}
=
\frac{\omega^{2(1-\rho_n^2)}}{\omega^2-1}
\left[
2\sinh\left(\frac{2\rho_n\varepsilon_n}{\delta}\right)\sinh^{-1}(D_\delta Z_n)
-4\rho_n\log(\omega)
\cosh\left(\frac{2\rho_n\varepsilon_n}{\delta}\right)
\right],
$$
and using
$
\left|\sinh\left({2\rho_n\varepsilon_n}/{\delta}\right)\right|
\leq
\cosh\left({2\rho_n\varepsilon_n}/{\delta}\right)
\leq1+(\omega^2-1)Z_n^2
$ then gives
$$
\left|\frac{\partial R_\delta^{(2)}(\varepsilon_n)}{\partial\rho_n}\right|
\lesssim
(1+Z_n^2)\{1+|\sinh^{-1}(D_\delta Z_n)|\}
\lesssim(1+Z_n^2)^2.
$$
Both derivative bounds hold uniformly for $0<\rho_n\leq 1$. At $\rho_n = 1$, we have $R_\delta^{(1)}(\varepsilon_n)=Z_n$ and $R_\delta^{(2)}(\varepsilon_n)=Z_n^2$ from Proposition \ref{prop:SU_robust}, so the mean value theorem gives the remaining inequalities.
\end{proof}

Based on the above lemma, we immediately have Lemma \ref{lem:Delta_mu_sigma_bounds} with $\rho = 1$, where we have used $\alpha_n(1-\rho_n) \lesssim \alpha_n^2$ by assumption. To show Lemma \ref{lem:delta_w_u_ineq}, the argument in (\ref{eq:SU_var_lowerbound}) ensures a positive $b < 1$ exists such that $1+\Delta_{\sigma^2}\geq 1- b$ uniformly, so the proof can proceed exactly as before since  (\ref{eq:Delta_mu_sigma_bounds}) holds in this case. Finally, Lemma \ref{lem:error_bounds} is unchanged as the terms are the same. 

The residual bound (\ref{eq:rn_SU}) and Assumption \ref{as:dgp} give
\begin{align*}
\alpha_n^2E[(1+Z_n^2)^2\mid\mathcal F_{n-1}]
\lesssim n^{-2(1-\alpha_0S)},
\end{align*}
which is summable since $\alpha_0S<1/2$ by assumption.
Also, $K_0<\infty$ under the same assumptions.
The proofs of Theorems \ref{th:amart_convergence} and
\ref{th:convergence2} therefore apply with $\rho=1$, giving
\begin{align*}
K_n\to K_\infty<\infty,\qquad
\sum_{n=1}^\infty\alpha_nT_{n-1}<\infty,\qquad
T_n\to0
\quad P^*\textnormal{-a.s.}
\end{align*}
The final statement follows from Lemma \ref{lem:norm_s} as before.

\subsection{Proof of Theorem \ref{th:consist_mv}}\label{app:mv_gen_thm}
We begin by presenting a general version of the theorem for the multi-output case with general missingness patterns and dimensionality $q$. The key difference to the order-compatible result in Theorem \ref{th:consist_mv} is that the projection matrix is data-dependent. We begin with a general version of Assumption \ref{as:dgp_mv}.
\begin{assumption}[Multi-output with general missingness]\label{as:dgp_mv_general}
Let $(Y_n,X_n,O_n)\iid P^*$, where $Y_n\in\mathbb R^q$ and $O_n\in\mathcal O$, where $\mathcal{O}$ is the collection of nonempty subsets of $\{1,\ldots,q\}$. Suppose $P_X^*$ has compact support $\mathcal X$ and Lebesgue density $p_X^*$, $E[\|Y\|^4]<\infty$ and \\
$E[|\log\det\Sigma_*(X)|]<\infty$. Furthermore, assume $O\perp Y\mid X$, with $P^*(O=\{1,\ldots,q\}\mid x)>0$ for $P_X^*$-almost all $x$.
\end{assumption}
As we will need to consider taking expectations over the missing pattern $O$, we will use the superscript $o$ to denote the observed versions of appropriate terms. Let $S^o$ be the selection matrix of size $|o| \times q$, and write
\begin{align*}
    Q^o_n(x) = {A_n^o(x)}^\top \left(A_n^o(x) \,{A_n^o}(x)^\top  \right)^{-1} A_n^o(x),  \qquad A_{n}^o(x) = S^o L_{n}(x).
\end{align*}
 We may drop the dependence on $x$ and write $Q_n^o$ and $A_n^o$ for shorthand. For the observed values as in (\ref{eq:proj_mat}) in the main paper, we would then have 
$$Q_n = Q_{n-1}^{O_n}(X_n), \quad A_n = A_{n-1}^{O_n}(X_n).$$
We will follow the notation of omitting both $o$ and $x$ if referring to the observed values.
We then write the projected error functions as
\begin{align*}
    u_n^o(x) = Q^o_n(x)\, u_n(x), \quad W_n^o(x)=  \, Q^o_n(x)\, W_n(x)\, Q_n^o(x).
\end{align*}
We highlight that the above error functions are still of size $q$ and $q \times q$ respectively, as we are only applying a projection. The averaged error functions are then
\begin{align*}
    \bar{u}_n(x) = \sum_{o \in \mathcal{O}} p^*(o \mid x) \, u_n^o(x), \quad \overline{W}_n(x) = \sum_{o \in \mathcal{O}} p^*(o \mid x)\, W_n^o(x).
\end{align*}
We then have the main theorem for general missingness patterns, with Theorem \ref{th:consist_mv} as the special case when $q=2$ and $O \in \{\{1\},\{1,2\}\}$ a.s. 
\begin{theorem}\label{th:consist_mv_general}
Under Assumptions \ref{as:bounded_sxx}, \ref{as:freq_lr}, \ref{as:mv_init} and \ref{as:dgp_mv_general}, the updates (\ref{eq:mean_gauss_missing})-(\ref{eq:cov_gauss_missing}) satisfy

\begin{align*}
    \|\bar u_n\|_s\to0,\qquad \|\overline W_n\|_s\to0\qquad P^*\textnormal{-a.s.}
\end{align*}
Furthermore, under Assumptions \ref{as:complete} and \ref{as:regularity}, $\|\cdot\|_s$ is a norm on the respective vector- and matrix-valued $L^1(P_X^*)$ spaces, and $\|\bar u_n\|_s=\|\overline W_n\|_s=0$ if and only if $\mu_n(x)=\mu_*(x)$ and $\Sigma_n(x)=\Sigma_*(x)$ for $P_X^*$-almost all $x$.
\end{theorem}

\subsubsection{Proof outline}
The proof follows almost exactly the same technique as the univariate case, with additional technical steps in controlling the remainders due to the projection matrices.  The learning rate bounds, namely Lemma \ref{lem:betan_bounds}, can be applied identically. We let $\|\cdot\|$ and $\|\cdot\|_F$ denote the Euclidean and Frobenius norms for vectors and matrices respectively, and extend $L^1_* = L^1(P_X^*)$, $\langle\,\cdot\,,\,\cdot\,\rangle_s$ and $\mathcal S$ component-wise, so Lemma \ref{lem:norm_s} applies to both types of functions. The residual function is now a vector,
\begin{align*}
    r_n(y,x)&=L_n(x)^{-1}\{y-\mu_n(x)\}, 
\end{align*}
which gives $Z_n = r_{n-1}(Y_n,X_n)$ and $\overline{Z}_n = Q_n Z_n$. The multivariate equivalents of the expected update terms (\ref{eq:Ezn}) are
\begin{align*}
    E[\,\overline Z_n\mid\mathcal F_{n-1},X_n=x]&=\bar u_{n-1}(x),\\
    E[\overline Z_n\overline Z_n^\top-Q_n\mid\mathcal F_{n-1},X_n=x]&=\overline W_{n-1}(x),\nonumber
\end{align*}
where $\mathcal{F}_n = \sigma\left(Y_1,X_1,O_1,\ldots, Y_n, X_n, O_n\right)$. The above follows from conditioning first on $O_n$ and using $O_n\perp Y_n\mid X_n$.

Next, we define the general version of the KL divergence terms.  For mean vector $\mu$ and covariance matrix $\Sigma$, we will use the notation
\begin{align*}
    \mu^o = S^o \mu, \quad \Sigma^o = S^o \Sigma {S^o}^\top,
\end{align*}
which corresponds to choosing the observed coordinates in the mean vector and covariance matrix. Unlike the error function case, the mean vector and covariance matrix have size $|o|$ and $|o| \times |o|$ respectively. We then have
\begin{align*}
    K_n^{X,O}(x,o) &= 2\text{KL}\left[\mathcal{N}\left\{\mu_*^o(x), \Sigma_*^o(x)\right\}, \,\mathcal{N}\left\{\mu_n^o(x), \Sigma_n^o(x)\right\} \right],\\
    K_n^X(x) &= \sum_{o \in \mathcal{O}} p^*(o \mid x) \, K_n^{X,O}(x,o), 
\end{align*}
where $K_n = \int K_n^X(x)\, dP_X^*(x)$ as before. Therefore, $K_n$ is twice the KL divergence averaged over both $O$ and $X$, and will serve as the almost-supermartingale. Naturally, we have 
\begin{align*}
  T_n&=2\rho\|\bar u_n\|_s^2+\rho^2\|\overline W_n\|_s^2.
\end{align*}
A similar argument to the univariate case shows $u_n$ and $W_n$ are suitably integrable, which is preserved by projection/averaging. Likewise, one can verify that $K_0 < \infty$.

\subsubsection{Technical lemmas}
To begin, we will show the multivariate version of Lemma \ref{lem:rn} to bound the residual. 
\begin{lemma}\label{lem:rn_mv}
Under Assumptions \ref{as:bounded_sxx}, \ref{as:freq_lr}, \ref{as:mv_init}, the following holds for all $(y,x) \in \mathbb{R}^q \times \mathcal{X}$: 
\begin{align*}
    \|r_n(y,x)\|^2 \lesssim \left[1+ \|r_0(y,x)\|^2\right](n+1)^{\alpha_0\rho^2 S}
\end{align*}
\end{lemma}
\begin{proof}
    Let $Q$ be an orthogonal projection matrix, so we have $Q \preceq I$. For any $z \in \mathbb{R}^q$, $0 < \rho < 1$ and $0 \leq a < 1$, we thus have
    \begin{align*}
        I + a \rho^2\left\{\left(1 - a\right)zz^\top - Q\right\} \succeq   (1 - a\rho^2)I + a \rho^2\left(1 - a\right)zz^\top. 
    \end{align*}
Next, for some $r \in \mathbb{R}^q$, the above then gives
\begin{align}
     \Delta &=(r - \rho a z)^\top\left[I + a \rho^2\left\{\left(1 - a\right)zz^\top - Q\right\} \right]^{-1}(r - \rho az)\nonumber \\
    &\leq (r - \rho a z)^\top\left[(1 - a\rho^2)I + a \rho^2\left(1 - a\right)zz^\top \right]^{-1}(r - \rho az) \nonumber\\
    &= \frac{1}{{1-a\rho^2} }\left[\|r- \rho a z\|^2- \frac{a \rho^2(1-a)\left(z^\top r - \rho a \|z\|^2\right)^2}{1 - a\rho^2 + a\rho^2(1-a) \|z\|^2}\right]\label{eq:Delta_r_mv}
\end{align}
where we have applied Sherman-Morrison in the last step. Next, for $z \neq 0$, we have
\begin{align*}
  \|r- \rho a z\|^2 &= \|r\|^2  - 2 \rho a z^\top r + \rho^2 a^2 \|z\|^2 = \left[\|r\|^2  - \frac{(z^\top r)^2}{\|z\|^2}\right] + \frac{\left(z^\top r - \rho a \|z\|^2\right)^2}{\|z\|^2} 
\end{align*}
where we have completed the square.  Combining the last term with the fraction in the square bracket of (\ref{eq:Delta_r_mv}) gives
\begin{align*}
    \Delta &\leq  \frac{\|r\|^2 -\frac{(z^\top r)^2}{\|z\|^2}}{1- a\rho^2} + \frac{\left(\frac{z^\top r}{\|z\|} - \rho a \|z\| \right)^2}{ 1 - a\rho^2 + a\rho^2(1-a) \|z\|^2}\cdot
\end{align*}
Note that the second term in the above is exactly $h(\|z\|)$ from the proof of Lemma \ref{lem:rn}, with $(z^\top r)/\|z\|$ as the scalar, so we have the upper bound
\begin{align}\label{eq:Delta_r_mv_ineq}
    \Delta &\leq \frac{\|r\|^2}{1-a \rho^2} + \frac{a}{1-a}\cdot 
\end{align}
For $z = 0$, the above upper bound holds immediately.

We can now show the bound. Dropping the dependence on $(y,x)$, we have
\begin{align*}
    r_n = L_n^{-1}L_{n-1}\left(r_{n-1} - \rho \beta_n \overline{Z}_n\right),
    \end{align*}
    which gives
    \begin{align*}
        r_n^\top r_n = \left(r_{n-1} - \rho \beta_n \overline{Z}_n\right)^\top \left[I + \beta_n \rho^2\left\{\left(1 - \beta_n\right)\overline{Z}_n\overline{Z}_n^\top - Q_n\right\} \right]^{-1}  \left(r_{n-1} - \rho \beta_n \overline{Z}_n\right).
    \end{align*}
    As a result, we can apply (\ref{eq:Delta_r_mv_ineq}) with $r = r_{n-1}$, $a = \beta_n$, $z = \overline{Z}_n$ and $Q = Q_n$, giving
    \begin{align*}
        \|r_n\|^2 \leq \frac{\|r_{n-1}\|^2}{1-\rho^2\beta_n} + \frac{\beta_n}{1 - \beta_n}\cdot
    \end{align*}
    Recursing in the exact same manner as in Lemma \ref{lem:rn} gives us the result.
\end{proof}

Next, we define the standardized multivariate increments:
\begin{align*}
    \Delta_{\mu}=\rho\beta_n\overline Z_n,\quad
    \Delta_{\Sigma}=\beta_n\rho^2
    \left\{(1-\beta_n)\overline Z_n\overline Z_n^\top-Q_n\right\}.
\end{align*}
We then have the following multivariate analogue of Lemma
\ref{lem:Delta_mu_sigma_bounds}.
\begin{lemma}\label{lem:Delta_mu_sigma_mv_bounds}
Under Assumptions \ref{as:bounded_sxx}, \ref{as:freq_lr},
\ref{as:mv_init} and \ref{as:dgp_mv_general}, we have
\begin{align}\label{eq:Delta_mu_sigma_mv_bounds}
    \|\Delta_{\mu}\|
    &\lesssim\alpha_n\|\overline Z_n\|,\quad
    \|\Delta_{\Sigma}\|_F
    \lesssim\alpha_n(1+\|\overline Z_n\|^2).
\end{align}
Furthermore,
\begin{align}
    \left\|E[\Delta_{\mu}\mid\mathcal F_{n-1}]
    -\alpha_n\rho\,\mathcal S\bar u_{n-1}\right\|
    &\lesssim\alpha_n^2
    E[(1+\|\overline Z_n\|^2)^2\mid\mathcal F_{n-1}],
    \label{eq:EDelta_mu_mv_bounds2}\\
    \left\|E[\Delta_{\Sigma}\mid\mathcal F_{n-1}]
    -\alpha_n\rho^2\,\mathcal S\overline W_{n-1}\right\|_F
    &\lesssim\alpha_n^2
    E[(1+\|\overline Z_n\|^2)^2\mid\mathcal F_{n-1}].
    \label{eq:EDelta_sigma_mv_bounds2}
\end{align}
\end{lemma}
\begin{proof}
The proof is the same as the univariate case in Lemma \ref{lem:Delta_mu_sigma_bounds}, using
$\|\overline Z_n\overline Z_n^\top\|_F=\|\overline Z_n\|^2$
and $\|Q_n\|_F\leq\sqrt q$, since $Q_n$ is an orthogonal
projection matrix. 
\end{proof}

Next, we show a technical lemma that is the main novel step for the multivariate extension. Unlike the univariate case, we need to handle the changing projection matrix for $W_n^o$ and $u_n^o$ which changes with $n$. To that end, define the term 
$$D_n := L_{n-1}^{-1}L_n,\quad D_n D_n^\top = I + \Delta_{\Sigma}.$$ 
Furthermore, for a symmetric matrix $A$, let $\mathcal{L}(A)$ denote the lower triangular part of $A$, with the diagonal values halved. The operator $\mathcal{L}$ arises in the study of perturbations of Cholesky factorizations, e.g. \citet{stewart1993perturbation, drmavc1994perturbation}. Heuristically, we have the matrix equivalent of $\sqrt{1+ x} \approx 1 + x/2$ as $\text{Chol}(I + A) \approx I + \mathcal{L}(A)$ when $A$ is small and symmetric. 
\begin{lemma}\label{lem:QnoDn}
Consider the projection term
\begin{align*}
    \Delta_Q = Q_n^o D_n^{-1} - Q_{n-1}^o.
\end{align*}
We have the expansion
 \begin{align}\label{eq:QnoDn_expansion}
   \Delta_Q &= \left[\mathcal{L}(\Delta_{\Sigma})^\top  - Q_{n-1}^o \Delta_{\Sigma} + R_n^{D}\right]Q_{n-1}^o 
    \end{align}
    where  $\|R_n^{D}\|_F \lesssim \|\Delta_{\Sigma}\|_F^2$. In particular, the above form gives $\Delta_Q Q_{n-1}^o = \Delta_Q$. Furthermore, we have the inequalities
    \begin{align*}
    \|Q_n^o D_n^{-1}\|_F \lesssim 1, \qquad\| \Delta_Q\|_F  \lesssim \min\left\{1,\|\Delta_{\Sigma}\|_F\right\}.
\end{align*}
\end{lemma}

\begin{proof}
       Intuitively, $D_n$ is the matrix version of $\sqrt{1+\Delta_{\sigma^2}}$, and we note that $D_n$ is lower triangular and contains positive diagonal entries, so it is the Cholesky factor of $I+ \Delta_{\Sigma}.$    Since
      $A_n^o = A_{n-1}^o D_n,$ we have
    \begin{align*}
        Q_n^o D_n^{-1}&=  {A_n^o}^\top \left(A_n^o \,{A_n^o}^\top  \right)^{-1}  A_{n}^o \, D_n^{-1} =D_n^\top {A_{n-1}^o}^\top\left(\Sigma_n^o\right)^{-1} A_{n-1}^o.
    \end{align*}
    For the inner inverse, we use the identity $A^{-1} =B^{-1} -A^{-1}(A-B)B^{-1}$ \citep[Section 3.2.5]{petersen2012matrix} and $\Sigma_{n}^o = \Sigma_{n-1}^o + A_{n-1}^o \Delta_{\Sigma} (A_{n-1}^o)^\top$ which gives
    \begin{align*}
  (\Sigma_n^o)^{-1} &= (\Sigma_{n-1}^o)^{-1}  -(\Sigma_n^o)^{-1}\left[A_{n-1}^o \Delta_{\Sigma} (A_{n-1}^o)^\top\right] (\Sigma_{n-1}^o)^{-1}.
    \end{align*}
    Substituting this into the above then gives
    \begin{align}
        Q_n^o D_n^{-1}&= D_n^\top Q_{n-1}^o - \left[D_n^\top {(A_{n-1}^o)}^\top (\Sigma_n^o)^{-1}A_{n-1}^o\right] \Delta_{\Sigma} Q_{n-1}^o\nonumber\\
        &= \left[D_n^\top - Q_n^o D_n^{-1}  \Delta_{\Sigma}\right] Q_{n-1}^o. \label{eq:QnoDn}
    \end{align}
    We can then write the above as
    \begin{align*}
    Q_n^o D_n^{-1}  &= \left[I +  \mathcal{L}(\Delta_{\Sigma})^\top  - Q_{n-1}^o \Delta_{\Sigma}  + R_n^{D}\right]Q_{n-1}^o
    \end{align*}
    where $R_n^{D} = \left[D_n - I - \mathcal{L}(\Delta_{\Sigma})\right]^\top  -\Delta_Q\Delta_{\Sigma}$.

    The next step is to bound $\|R_n^{D}\|_F$, which we do separately for the two terms. For the first term, we will use the results of \cite{drmavc1994perturbation} for perturbations of the Cholesky factor. Firstly, we highlight that  $D_n D_n^\top =    I +  \Delta_{\Sigma} \succeq (1-\rho^2) I \succ 0$, so $I +  \Delta_{\Sigma}$ is positive definite. Furthermore, $\Delta_{\Sigma}$ is symmetric, so the setting of \cite{drmavc1994perturbation} applies. Since $D_n$ is the Cholesky factor of $I + \Delta_{\Sigma}$, \citet[Equation (12)]{drmavc1994perturbation} immediately gives
  \begin{align}\label{eq:DnI_frob}
      \|D_n - I \|_F \leq \sqrt{8q + 2\sqrt{q}}\,\,  \|\Delta_{\Sigma}\|_F \lesssim \|\Delta_{\Sigma}\|_F.
  \end{align}
Next, the definition of $D_n D_n^\top$ gives
\begin{align*}
    (D_n - I)(D_n - I)^\top = \Delta_{\Sigma} - (D_n - I) - (D_n - I)^\top,
\end{align*}
and applying $\mathcal{L}$ to both sides gives
\begin{align}\label{eq:chol_first}
    \mathcal{L}\{(D_n - I)(D_n - I)^\top\} = \mathcal{L}(\Delta_{\Sigma}) - (D_n - I),
\end{align}
where we have used the fact that 
$\mathcal{L}\left(A + A^\top\right) = A$ if $A$ is lower triangular.  Note that this is equivalent to a transposed version of \citet[Equation (21)]{drmavc1994perturbation} with $\eta = 1$. Rearranging (\ref{eq:chol_first}), we have
\begin{align*}
    \left\|(D_n - I) - \mathcal{L}(\Delta_{\Sigma}) \right\|_F &= \left\| \mathcal{L}\{(D_n - I)(D_n - I)^\top\} \right\|_F\\
    &\leq \left\| (D_n - I)(D_n - I)^\top\right\|_F\leq \|D_n - I\|_F^2 \lesssim \| \Delta_{\Sigma} \|_F^2
\end{align*}
where the second line follows as $\mathcal{L}$ cannot increase the Frobenius norm. 

For the second term in $R_n^{D}$, substituting (\ref{eq:QnoDn}) gives
\begin{align*}
    \| Q_n^o D_n^{-1} - Q_{n-1}^o\|_F \leq \|D_n - I\|_F + \|D_n^{-1}\|_F \|\Delta_{\Sigma}\|_F,
\end{align*}
where we have used the fact that $\|QA \|_F \leq  \|A\|_F$ if $Q$ is an orthogonal projection matrix.
The term  $\|D_n - I\|_F$ can be handled again by (\ref{eq:DnI_frob}).
For the latter term, since $D_n D_n^\top \succeq (1-\rho^2)I$, we have
\begin{align*}
    \|D_n^{-1}\|_F^2 = \operatorname{tr}\{(D_n D_n^\top)^{-1}\} \leq \frac{q}{1-\rho^2}
\end{align*}
which gives $\|\Delta_Q\|_F  \lesssim \|\Delta_{\Sigma}\|_F$. Together, the above implies $\|R_n^{D}\|_F \lesssim \|\Delta_{\Sigma}\|_F^2$. 

The equality then follows from $Q_{n-1}^o Q_{n-1}^o = Q_{n-1}^o$ since it is an orthogonal projection matrix. For the remaining inequalities, we note that 
 $$\| Q_n^o D_n^{-1}\|_F \leq \|D_n^{-1}\|_F \lesssim 1,$$ since $Q_n^o$ is an orthogonal projection matrix. The final inequality then follows from the just shown $\|\Delta_Q\|_F  \lesssim \|\Delta_{\Sigma}\|_F$ and
 \begin{align*}
         \| Q_n^o D_n^{-1} - Q_{n-1}^o\|_F \leq   \| Q_n^o D_n^{-1}\|_F + \|Q_{n-1}^o\|_F \lesssim 1,
 \end{align*}
where $\|Q_{n-1}^o\|_F = \sqrt{|o|} \leq \sqrt{q}$ as it is an orthogonal projection matrix.
\end{proof}
Next, we will show multivariate extensions of Lemma \ref{lem:delta_w_u_ineq}.
\begin{lemma}\label{lem:delta_w_u_ineq_mv}
Under Assumptions \ref{as:bounded_sxx}, \ref{as:freq_lr},
\ref{as:mv_init} and \ref{as:dgp_mv_general}, we have
\begin{align*}
u_n^o-u_{n-1}^o
    &=\left[\mathcal L(\Delta_\Sigma)^\top
    -Q_{n-1}^o\Delta_\Sigma\right]u_{n-1}^o
    -Q_{n-1}^o\Delta_\mu+R_n^u,\\
    W_n^o-W_{n-1}^o
    &=\left[\mathcal L(\Delta_\Sigma)^\top
    -Q_{n-1}^o\Delta_\Sigma\right]W_{n-1}^o+W_{n-1}^o\left[\mathcal L(\Delta_\Sigma)
    -\Delta_\Sigma Q_{n-1}^o\right]\\
    &\quad -\left[Q_{n-1}^o\Delta_\mu(u_{n-1}^o)^\top +u_{n-1}^o\Delta_\mu^\top Q_{n-1}^o + Q_{n-1}^o\Delta_\Sigma Q_{n-1}^o \right]+R_n^W,
\end{align*}
where the remainders satisfy
\begin{align*}
\|R_n^u\|+\|R_n^W\|_F
\lesssim
\alpha_n^2\left(1+\|\overline Z_n\|^2\right)^2
\left(1+\|u_{n-1}^o\|+\|W_{n-1}^o\|_F\right).
\end{align*}
Furthermore, we have
\begin{align*}
\|u_n^o-u_{n-1}^o\|+\|W_n^o-W_{n-1}^o\|_F
\lesssim
\alpha_n(1+\|\overline Z_n\|^2)
\left(1+\|u_{n-1}^o\|+\|W_{n-1}^o\|_F\right).
\end{align*}
\end{lemma}

\begin{proof}
We handle the $u_n^o$ case first as it is simpler.  We can show that
\begin{align*}
        u_n^o = Q_n^o D_n^{-1}\left(u_{n-1} - \Delta_{\mu}\right).
    \end{align*}
The presence of $Q_n^o D_n^{-1}$ is the main technical challenge, and necessitates Lemma \ref{lem:QnoDn} to bound its change.
 Substituting the expression for $Q_n^o D_n^{-1}$ from Lemma \ref{lem:QnoDn} gives the stated expansion for $u_n^o - u_{n-1}^o$, where $R_n^u = R_n^{D}u_{n-1}^o - \Delta_Q \Delta_\mu$ and
\begin{align*}
    \|R_n^u\| &\leq \|R_n^D\|_F \|u_{n-1}^o\| + \left\|\Delta_Q\right\|_F \| \Delta_{\mu}\|\lesssim \alpha_n^2\left(1 + \|\overline{Z}_n\|^2\right)^2 \left(1 + \|u_{n-1}^o\|\right)
\end{align*}
from Lemma \ref{lem:QnoDn} and (\ref{eq:Delta_mu_sigma_mv_bounds}). For $\|u_n^o - u_{n-1}^o \|$, the update and Lemma \ref{lem:QnoDn} give
\begin{align*}
    \left\|u_n^o - u_{n-1}^o \right\| &\leq \left\|\Delta_Q \right\|_F \|u_{n-1}^o\| + \|Q_n^o D_n^{-1}\|_F \|\Delta_\mu\|\\
    &\lesssim \alpha_n \left(1 + \|\overline{Z}_n\|^2\right)\left(1 + \|u_{n-1}^o\|\right).
\end{align*}
Next, for $W_n^o - W_{n-1}^o$, the definition of $W_n$ gives
    \begin{align*}
        W_n = D_n^{-1} B_n \, D_n^{-\top}, \quad 
        B_n = W_{n-1} - \Delta_{\Sigma} - \Delta_{\mu}u_{n-1}^\top- u_{n-1}\Delta_{\mu}^\top + \Delta_{\mu} \Delta_{\mu}^\top.
    \end{align*}  
    Noting that $\Delta_Q Q_{n-1}^o = \Delta_Q$, we can write
\begin{align*}
    W_n^o=(Q_{n-1}^o+\Delta_Q)Q_{n-1}^o B_n Q_{n-1}^o
    (Q_{n-1}^o+\Delta_Q)^\top.
\end{align*}
Expanding this expression and applying (\ref{eq:QnoDn_expansion}) gives the stated expansion for $W_n^o - W_{n-1}^o$, where $R_n^W$ contains the terms quadratic or of higher order in $(\Delta_{\Sigma}, \Delta_{\mu})$.
Grouping the remainder according to the terms in $B_n$ gives
\begin{align*}
    \|R_n^W\|_F
    &\lesssim
    \left(\|R_n^D\|_F+\|\Delta_Q\|_F^2\right)\|W_{n-1}^o\|_F\\
    &\quad+\left(\|\Delta_Q\|_F+\|\Delta_Q\|_F^2\right)
    \left(\|\Delta_\Sigma\|_F+\|\Delta_\mu\|\|u_{n-1}^o\|\right)
    +\|\Delta_\mu\|^2\\
    &\lesssim
    \alpha_n^2\left(1+\|\overline Z_n\|^2\right)^2
    \left(1+\|u_{n-1}^o\|+\|W_{n-1}^o\|_F\right),
\end{align*}
using Lemma \ref{lem:QnoDn} and
(\ref{eq:Delta_mu_sigma_mv_bounds}). In particular,
$\|\Delta_Q\|_F^2\lesssim\|\Delta_Q\|_F$ since $\|\Delta_Q\|_F \lesssim 1$.

Finally, the exact update and the same bounds give
\begin{align*}
    \|W_n^o-W_{n-1}^o\|_F
    &\lesssim
    \|\Delta_\Sigma\|_F(1+\|W_{n-1}^o\|_F)
    +\|\Delta_\mu\|\|u_{n-1}^o\|+\|\Delta_\mu\|^2\\
    &\lesssim
    \alpha_n(1+\|\overline Z_n\|^2)
    (1+\|u_{n-1}^o\|+\|W_{n-1}^o\|_F),
\end{align*}
where we also used
$\|\Delta_\mu\|^2
=\rho^2\beta_n^2\|\overline Z_n\|^2
\lesssim\alpha_n\|\overline Z_n\|^2$.
\end{proof}
Finally, we have the multivariate extension of Lemma \ref{lem:error_bounds}.
\begin{lemma}\label{lem:error_bounds_mv}
    The error terms satisfy
    
    \begin{align*}
      1 + \|u_n^o \| + \left\| W_n^o\right\|_F \lesssim 1 + K_n^{X,O}.
    \end{align*}
\end{lemma}

\begin{proof}
We begin with the multivariate counterpart to
$1+w_n=u_n^2+v_n$. By definition,
\begin{align*}
W_n^o+Q_n^o
= u_n^o(u_n^o)^\top + Q_n^oL_n^{-1}\Sigma_*L_n^{-\top}Q_n^o.
\end{align*}
Taking traces, using cyclicity, $(Q_n^o)^2=Q_n^o$
and the definition of $Q_n^o$, gives
\begin{align}
\operatorname{tr}(W_n^o+Q_n^o)
&=\|u_n^o\|^2+
\operatorname{tr}\left[
(\Sigma_n^o)^{-1}A_n^oL_n^{-1}
\Sigma_*L_n^{-\top}(A_n^o)^\top
\right]\nonumber\\
&=\|u_n^o\|^2+
\operatorname{tr}\left[(\Sigma_n^o)^{-1}\Sigma_*^o\right],
\label{eq:KL_tr_identity}
\end{align}
where the last equality follows from $A_n^oL_n^{-1}=S^o$.
As in Lemma \ref{lem:error_bounds}, applying
$\log v\leq v/2$ to the eigenvalues of
$(\Sigma_n^o)^{-1}\Sigma_*^o$ gives
\begin{align*}
K_n^{X,O}+|o|
&\geq \|u_n^o\|^2
+\frac12\operatorname{tr}\{(\Sigma_n^o)^{-1}\Sigma_*^o\}=\frac12\left[\|u_n^o\|^2
+\operatorname{tr}(W_n^o+Q_n^o)\right].
\end{align*}
Since $W_n^o+Q_n^o \succeq 0$, its Frobenius norm is upper bounded by its trace, giving
\begin{align*}
\|W_n^o + Q_n^o - Q_n^o\|_F
&\leq\|W_n^o+Q_n^o\|_F+\|Q_n^o\|_F \leq\operatorname{tr}(W_n^o+Q_n^o)+\sqrt{|o|}.
\end{align*}
Together with $\|u_n^o\|\leq1+\|u_n^o\|^2$, this gives
\begin{align*}
1+\|u_n^o\|+\|W_n^o\|_F \lesssim 1+K_n^{X,O},
\end{align*}
where we used $|o|\leq q$ and $K_n^{X,O}\geq0$.
\end{proof}

 \subsubsection{Main proof}
 
The remainder of the proof follows in a largely similar fashion to the univariate case, so we only outline the main differences. From (\ref{eq:KL_tr_identity}), we have
 \begin{align*}
     K_n^{X,O} = \log \frac{|\Sigma_n^o|}{|\Sigma_*^o|} + \operatorname{tr}(W_n^o), \quad     K_n^{X,O} - K_{n-1}^{X,O} = \log \frac{|\Sigma_n^o|}{|\Sigma_{n-1}^o|} + \operatorname{tr}(W_n^o - W_{n-1}^o).
 \end{align*}
Next, we have
\begin{align*}
\log \frac{|\Sigma_n^o|}{|\Sigma_{n-1}^o|} &= \log \left|(\Sigma_{n-1}^o)^{-1} \Sigma_n^o \right| \leq \operatorname{tr}\left[(\Sigma_{n-1}^o)^{-1}\left(\Sigma_n^o- \Sigma_{n-1}^o\right)\right]
\end{align*}
where the inequality follows from $\log(\lambda) \leq \lambda - 1$ applied to the eigenvalues of $(\Sigma_{n-1}^o)^{-1} \Sigma_n^o$. Next, since $\Sigma_{n}^o- \Sigma_{n-1}^o = A_{n-1}^o \Delta_{\Sigma} (A_{n-1}^o)^\top$, we have
\begin{align*}
    \operatorname{tr}\left[(\Sigma_{n-1}^o)^{-1}\left(\Sigma_n^o- \Sigma_{n-1}^o\right)\right] &= \operatorname{tr}\left[(A_{n-1}^o)^\top (\Sigma_{n-1}^o)^{-1}A_{n-1}^o \Delta_{\Sigma} \right] = \operatorname{tr}\left( Q_{n-1}^o \Delta_{\Sigma}\right).
\end{align*}
Next, apply Lemma \ref{lem:delta_w_u_ineq_mv} to $\operatorname{tr}(W_n^o - W_{n-1}^o)$. Using the cyclicity of the trace, and the identities 
$\mathcal{L}(\Delta_\Sigma) + \mathcal{L}(\Delta_\Sigma)^\top = \Delta_\Sigma$ and  $(Q_{n-1}^o)^2 = Q_{n-1}^o$,  we obtain
\begin{align*}
    \operatorname{tr}(W_n^o - W_{n-1}^o) &= -\operatorname{tr}\left(\Delta_{\Sigma} W_{n-1}^o\right) -\operatorname{tr}\left(Q_{n-1}^o \Delta_{\Sigma}\right) - 2\Delta_\mu^\top u_{n-1}^o +\operatorname{tr}(R_n^W), 
\end{align*}
and the inequality on the log determinant ratio gives
\begin{align*}
    K_n^{X,O} - K_{n-1}^{X,O} \leq -\operatorname{tr}\left(\Delta_{\Sigma} W_{n-1}^o\right)  - 2\Delta_\mu^\top u_{n-1}^o +\operatorname{tr}(R_n^W). 
\end{align*}
The other difference is averaging the error bounds over $o$. Lemma \ref{lem:error_bounds_mv} gives
\begin{align}\label{eq:mv_error_average}
    \sum_{o\in\mathcal O}p^*(o\mid x)
    \left\{1+\|u_{n-1}^o(x)\|+\|W_{n-1}^o(x)\|_F\right\}
    \lesssim1+K_{n-1}^X(x).
\end{align}
We now take the conditional expectation of $K_n^{X,O} - K_{n-1}^{X,O}$ and integrate over $o$ and $x$. (\ref{eq:EDelta_mu_mv_bounds2}) and (\ref{eq:EDelta_sigma_mv_bounds2}) give the leading term as $-\alpha_nT_{n-1}$. The remainders are handled exactly as in Theorem \ref{th:amart_convergence}, additionally using $|\operatorname{tr}(R_n^W)|\leq\sqrt q\|R_n^W\|_F$, Lemma \ref{lem:Delta_mu_sigma_mv_bounds} and (\ref{eq:mv_error_average}). Also, $\|\overline Z_n\|\leq\|Z_n\|$ and Lemma \ref{lem:rn_mv} give the same moment bound:
\begin{align}\label{eq:mv_moment_bound}
    \alpha_n^2E[(1+\|\overline Z_n\|^2)^2\mid\mathcal F_{n-1}]
    \lesssim n^{-2(1-\alpha_0\rho^2S)}.
\end{align}
Here $K_n,T_n\geq0$, and we can show $K_0<\infty$ under Assumptions \ref{as:mv_init} and \ref{as:dgp_mv_general}. The same almost-supermartingale argument therefore gives  $K_n\to K_\infty<\infty$ and $\sum_{n=1}^\infty\alpha_n T_{n-1}<\infty$  $P^*$-a.s.
    
The convergence of $T_n$ follows the same strategy as Theorem \ref{th:convergence2}, but with $\bar u_n$ and $\overline W_n$, and applying Jensen's inequality gives 
$$\|\bar{u}_n-\bar{u}_{n-1}\|
\leq\sum_{o\in \mathcal{O}} p^*(o\mid x)\|u_n^o-u_{n-1}^o\|$$
and likewise for $\overline{W}_n$. The squared increments and remainder inner products have the same bounds, by Lemmas \ref{lem:norm_s} and \ref{lem:delta_w_u_ineq_mv}, together with (\ref{eq:mv_error_average}) and (\ref{eq:mv_moment_bound}).  It remains to handle the leading terms. Taking conditional expectations in Lemma \ref{lem:delta_w_u_ineq_mv} replaces $\Delta_\mu$ and $\Delta_\Sigma$ by $\alpha_n\rho\mathcal S\bar u_{n-1}$ and $\alpha_n\rho^2\mathcal S\overline W_{n-1}$ with controllable remainders.
Dropping the non-positive leading terms, we have
\begin{align*}
    E[T_n \mid \mathcal{F}_{n-1}] - T_{n-1} &\leq \int \sum_{o \in \mathcal{O}} p^*(o \mid x) \, h_n(x,o) \, dP_X^*(x) + Cn^{-2(1-\alpha_0\rho^2S)}(1+K_{n-1})^2
\end{align*}
where $h_n(x,o)$ takes the form
\begin{align*}
h_n &= 4\alpha_n\rho^3
(u_{n-1}^o)^\top
\left[
\mathcal L(\mathcal S\overline W_{n-1})
-2(\mathcal S\overline W_{n-1})Q_{n-1}^o
\right]
\mathcal S\bar u_{n-1} 
\\
& \quad +4\alpha_n\rho^4
\operatorname{tr}\Big[
(\mathcal S\overline W_{n-1})
\{\mathcal L(\mathcal S\overline W_{n-1})^\top
-Q_{n-1}^o(\mathcal S\overline W_{n-1})\}
W_{n-1}^o
\Big].
\end{align*}
The first term is the multivariate analogue of the mixed term in (\ref{eq:Et_t}), while the second collects the remaining covariance contributions. Neither term has a definite sign, so we bound both in absolute value.
Again, $\left\|\mathcal{L}(A)\right\|_F \leq \|A\|_F$ and $\|Q_{n-1}^o A\|_F \leq \|A\|_F$, and Cauchy-Schwarz applied to the Frobenius inner product (trace term) gives
\begin{align*}
    \left|h_n\right|&\lesssim \alpha_n \left(\|u_{n-1}^o \| \left\|\mathcal{S}\overline{W}_{n-1} \right\|_F \|\mathcal{S}\bar{u}_{n-1} \|  + \left\|\mathcal{S}\overline{W}_{n-1} \right\|_F^2 \left\|W_{n-1}^o \right\|_F\right)\\
    &\lesssim  \alpha_n \left(\|u_{n-1}^o \|+ \left\|W_{n-1}^o \right\|_F\right)\left(\|\mathcal{S}\bar{u}_{n-1} \|^2 + \left\|\mathcal{S}\overline{W}_{n-1} \right\|_F^2 \right)
\end{align*}
where the second line follows from $2|ab| \leq a^2 + b^2$. Finally, Lemma \ref{lem:norm_s} implies $\|\mathcal{S}\bar{u}_{n-1} \|^2\lesssim \|\bar{u}_{n-1} \|_s^2$ and $\left\|\mathcal{S}\overline{W}_{n-1} \right\|_F^2 \lesssim \|\overline{W}_{n-1}\|_s^2$, which with Lemma \ref{lem:error_bounds_mv} gives
\begin{align*}
   \left|h_n\right| \lesssim \alpha_n(1 + K_{n-1}^{X,O})\left(\|\bar{u}_{n-1} \|_s^2 + \|\overline{W}_{n-1} \|_s^2\right).
\end{align*}
Integrating over $o$ and $x$ gives
\begin{align*}
    E[T_n \mid \mathcal{F}_{n-1}] - T_{n-1} &\leq C\alpha_n (1+K_{n-1})T_{n-1} + Cn^{-2(1-\alpha_0\rho^2S)}(1+K_{n-1})^2.
\end{align*}
 The conclusion $T_n\to0$ $P^*$-a.s. now follows exactly as in Theorem \ref{th:convergence2}.

Lemma \ref{lem:norm_s} applies component-wise under Assumptions  \ref{as:regularity} and \ref{as:complete}. If $T_n = 0$, then $\bar u_n=0$ and $\overline W_n=0$ for $P_X^*$-almost all $x$, giving
\begin{align*}
    0=u_n^\top\bar u_n=\sum_{o\in\mathcal O}p^*(o\mid x)\|u_n^o\|^2,
    \qquad
    0=\operatorname{tr}(W_n\overline W_n)=\sum_{o\in\mathcal O}p^*(o\mid x)\|W_n^o\|_F^2.
\end{align*}
The fully observed pattern with projection $I_q$ has positive probability by Assumption \ref{as:dgp_mv_general}, so $u_n=W_n=0$ gives the final statement. The converse is immediate. 

\subsection{Norm strength}\label{app:norm_str}
We now discuss the strength of the norm $\| \cdot \|_s$, focusing on the single-output case for convenience. Since $\|\mathcal{S}f\|_\infty \lesssim \|f\|_s$ from Lemma \ref{lem:norm_s}, we have thus shown that the smoothed error functions $\mathcal{S} u_n$ and $\mathcal{S} w_n$ uniformly converge to 0 $P^*$-a.s. However, the similarity norm is weaker than the $L^1(P_X^*)$ norm (e.g. $\| f\|_s \lesssim \| f\|_{L^1_*}$), as we may have oscillatory functions which converge to 0 in $\|\,\cdot\,\|_s$ but not in $\|\cdot \|_{L^1_*}$. A concrete example is 
$f_n(x) = \sin(2\pi n x)$ with $P_X^* = \textnormal{Unif}[0,1]$. Since $s(x,x') \leq S$ from Lemma \ref{lem:sxx_bound},  we have $s(x, \cdot) \in L^1_*$, so the Riemann-Lebesgue lemma gives for each $x$
$$(\mathcal{S}f_n)(x) =\int_0^1 s(x,x') \, \sin(2\pi n x') \, dx' \to 0$$
 Thus, $\|f_n\|_s^2 \leq  \int_0^1 \left|(\mathcal{S}f_n)(x) \right|\, dx \to 0$  since $\left|\mathcal{S} f_n\right| \leq S$ is dominated.
However, since $\|f_n\|_{L_*^1} = {2}/{\pi}$, we have convergence in $\|\cdot\|_s$ but not $\|\cdot\|_{L_*^1}$. Nonetheless, we do not see $u_n$ and $w_n$ exhibiting this oscillatory behaviour in practice. 

Convergence in the similarity norm can imply weak convergence in the function
space under additional conditions. Let $L_*^2=L^2(P_X^*)$ and
 $f_n,h\in L_*^2$. We say that $f_n$ converges weakly to $h$ in $L_*^2$, or $f_n\rightharpoonup h$ in $L_*^2$, if
$$
    \int f_n(x)\,g(x)\,dP_X^*(x)
    \longrightarrow
    \int h(x)\,g(x)\,dP_X^*(x)
$$
for every $g\in L_*^2$. We then have the following lemma. 
\begin{lemma}
  Suppose Assumptions \ref{as:bounded_sxx}, \ref{as:dgp}, \ref{as:regularity} and \ref{as:complete} hold. If
$\sup_n\|f_n\|_{L_*^2}<\infty$, then $\|f_n\|_s\to0$
implies $f_n\rightharpoonup0$ in $L_*^2$.
\end{lemma}
\begin{proof}
Every subsequence is bounded in $L_*^2$ and thus contains a further
weakly convergent subsequence, say $f_{n_j}\rightharpoonup h$. As $s(x,x')$ is uniformly bounded and thus $s(x,\,\cdot\,)\in L_*^2$, we may use it as a test function. For each $x$, we have
\begin{align*}
    (\mathcal Sh)(x)
    &=\lim_{j\to\infty}\int f_{n_j}(x')s(x,x')\,dP_X^*(x')=\lim_{j\to\infty}(\mathcal Sf_{n_j})(x)=0,
\end{align*}
where the last equality follows from
$\|\mathcal Sf_{n_j}\|_\infty\lesssim\|f_{n_j}\|_s\to0$.
Hence $\|h\|_s^2=\int h\,(\mathcal{S}h)\,dP_X^*=0$,
and Lemma \ref{lem:norm_s} gives $h=0$, since $L_*^2\subset L_*^1$.
Every subsequence thus has a further subsequence
converging weakly to 0, proving $f_n\rightharpoonup0$ in $L_*^2$. 
\end{proof}

Under the setting of Theorem \ref{th:consist} including Assumptions \ref{as:regularity} and \ref{as:complete}, we have $K_n^X=u_n^2+v_n-1-\log v_n\geq u_n^2$, so $\sup_n\|u_n\|_{L_*^2}^2
    \leq\sup_n K_n<\infty$ $P^*$-a.s. by Theorem \ref{th:amart_convergence}. Together with
$\|u_n\|_s\to0$, this gives $u_n\rightharpoonup0$ in $L_*^2$
$P^*$-a.s.
For $w_n$ however, the existing bounds give
$\sup_n\|w_n\|_{L_*^1}<\infty$ $P^*$-a.s., but neither
uniform $L_*^2$ boundedness nor uniform integrability has
been established. Thus, weak convergence of $w_n$ in $L_*^2$,
or even in $L_*^1$, would require further analysis. Finally, even if weak convergence for both $u_n$ and $w_n$ was shown, this would not directly translate into convergence statements about $\mu_n$ and $\sigma_n^2$, since squaring or taking reciprocals generally need not preserve weak convergence.

\section{Additional Methodological Details}

\subsection{Similarity functions}\label{app:sxx}
The mixture construction in (\ref{eq:dxx}) also accommodates discrete covariates by taking $k(x;\psi)$ to be a probability mass function. We give two examples below.
\begin{example}
For $x\in\{1,\ldots,K\}$, let $k(x;\psi)=\psi_x$ and $\Psi\sim\textnormal{Dirichlet}(\phi\mu_1,\ldots,\phi\mu_K)$, where $\phi>0$, $\mu_k>0$ and $\sum_k\mu_k=1$. The Dirichlet moments give
\begin{align*}
    s(x,x')=1-\rho_x+\rho_x\frac{\mathbf{1}(x=x')}{\mu_x},\qquad \rho_x=\frac{1}{\phi+1}\cdot
\end{align*}
The components $\mu_k$ can be set to empirical means, with smoothing if any categories have zero counts. Larger $\rho_x$ thus places more weight on matching categories, especially those that are rare, and we may tune $\rho_x$ prequentially.
\end{example}
\begin{example}
For counts $x\in\mathbb N_0$, let $k(x;\psi)=e^{-\psi}\psi^x/x!$ and $\Psi\sim\operatorname{Gamma}(\phi\mu,\phi)$, with mean $\mu>0$ and rate $\phi>0$. Conjugacy gives
\begin{align*}
    s(x,x')
=
\frac{\Gamma(\phi\mu)\,\Gamma(\phi\mu+x+x')}
{\Gamma(\phi\mu+x)\Gamma(\phi\mu+x')}
\frac{(\phi+1)^{2\phi\mu+x+x'}}
{\phi^{\phi\mu}(\phi+2)^{\phi\mu+x+x'}}\cdot
\end{align*}
We may set $\mu$ to the empirical mean and tune $\phi$ prequentially.
\end{example}
Both kernel families (and the Gaussian example) are complete following \citet[Theorem 4.3.1]{Lehmann2006}. Finally, product kernels with independent mixing parameters give a product of coordinate-wise similarities. This allows continuous, categorical and count covariates to be combined in the same update.

\subsection{General transformations}\label{app:gen}
As discussed in the main paper, one may be interested in other transforms of the Gaussian random variable to obtain different families for $p_i(y \mid x)$, which then induce different moment updates. Our framework is quite general, as Proposition \ref{prop:gen_update} allows for general transformations. We now briefly discuss two choices beyond the symmetric SU family introduced in the main paper.

\subsubsection{Skewed residuals}
For skewed residuals, use the full SU distribution of \citet{johnson1949systems} which generalizes the case in the main paper. Let $\lambda=(s,\delta)$ with $s\in\mathbb R$ and $\delta>0$, and define
\begin{align*}
    g_{s,\delta}(\varepsilon)
    =\frac{\sinh\left(\frac{\varepsilon-s}{\delta}\right)+\sqrt\omega\sinh(\frac{s}{\delta})}{D_{s,\delta}}, \quad 
    D_{s,\delta}^2
    =\frac{\omega-1}{2}\left\{2\omega\sinh^2\left(\frac{s}{\delta}\right)+\omega+1\right\}
\end{align*}
and $\omega=e^{1/\delta^2}$ as before.
The parameter $s$ introduces skewness, while $\delta$ controls tail weight, and $s=0$ recovers (\ref{eq:SU_fam}). We highlight that while the SU distribution cannot attain all admissible skewness-kurtosis pairs, it can attain the full skewness range as long as $\omega$ grows appropriately. In practice, we may prequentially tune $(s, \delta)$ jointly. The terms $R_{s,\delta}^{(1)}$ and $R_{s,\delta}^{(2)}$ also have similar closed forms:
\begin{align*}
R_{s,\delta}^{(1)}(\varepsilon_i)
&=
\frac{
\omega^{(1-\rho^2)/2}
\sinh\!\left(\frac{\rho\varepsilon_i-s}{\delta}\right)
+\sqrt{\omega}\sinh\left(\frac{s}{\delta}\right)}
{D_{s,\delta}},\\
R_{s,\delta}^{(2)}(\varepsilon_i)
&=
\left[R_{s,\delta}^{(1)}(\varepsilon_i)\right]^2
+
\frac{(\omega^{1-\rho^2}-1)
\left[\omega^{1-\rho^2}\cosh\!\left(\frac{2(\rho\varepsilon_i-s)}{\delta}\right)+1\right]}
{2D_{s,\delta}^2}
\end{align*}
where $\varepsilon_i =s+\delta\sinh^{-1}\left\{D_{s,\delta}\,Z_i-\sqrt\omega\sinh\left(\frac{s}{\delta}\right)\right\}.$

\subsubsection{Platykurtic residuals}
Consider the signed power transform \citep{box1964analysis}
\begin{align*}
    g_a(\varepsilon)=\frac{\operatorname{sign}(\varepsilon)|\varepsilon|^a}{C_a},
    \qquad C_a^2=\frac{2^a\Gamma(a+1/2)}{\sqrt\pi},\qquad a > 0.
\end{align*}
This is again standardized, with skewness 0 and kurtosis equal to \\
$E[g_a(\varepsilon)^4]={\sqrt\pi\,\Gamma(2a+1/2)}/{\Gamma(a+1/2)^2}$.
The range $0<a<1$ gives platykurtic residuals, where $a=1$ returns the Gaussian case. 
The terms $R_{a}^{(1)}$ and $R_{a}^{(2)}$ also have the forms:
\begin{align*}
R_a^{(1)}(\varepsilon_i)
&=
\frac{C^2_{(a+1)/2}}{C_a}\rho\varepsilon_i(1-\rho^2)^{(a-1)/2}
\,\,{}_1F_1\left(
\frac{1-a}{2};\frac32;
-\frac{\rho^2\varepsilon_i^2}{2(1-\rho^2)}
\right),\\
R_a^{(2)}(\varepsilon_i)
&=
(1-\rho^2)^a
\,\,{}_1F_1\left(
-a;\frac12;
-\frac{\rho^2\varepsilon_i^2}{2(1-\rho^2)}
\right),
\end{align*}
where ${}_1F_1$ denotes the confluent hypergeometric
function and
$\varepsilon_i=\operatorname{sign}(Z_i)(C_a|Z_i|)^{1/a}$.

\subsubsection{Multi-output}\label{app:gen_mo}
For the multi-output SU extension, we consider the family of $p_i(y \mid x)$ induced by
\begin{align*}
    Y=\mu(x)+L(x)\,\mathbf{g}_\delta(\varepsilon),\qquad
    \varepsilon\sim\mathcal N(0,I_q),
\end{align*}
where $\mathbf{g}_\delta$ applies $g_{\delta}$ as defined in (\ref{eq:g_delt}) component-wise to the vector $\varepsilon$ and $L(x)$ is a Cholesky factor. The standardized SU components are independent, so $E[Y\mid x]=\mu(x)$ and $\operatorname{Var}(Y\mid x)=L(x)L(x)^\top$. 

We now outline the multivariate moment updates by extending Proposition \ref{prop:SU_update}. Let $Z_i$ be as in (\ref{eq:mv_Zi}) and set $\varepsilon_i^{(j)}=g_\delta^{-1}(Z_i^{(j)})$. The triangular form implies that conditioning on $Y^{(1:j-1)}$ is equivalent to conditioning on $\varepsilon^{(1:j-1)}$.  As a result, the conditional CDFs in (\ref{eq:condit_copula_mv}) are $u_{i-1}^{(j)} = \Phi(\varepsilon^{(j)})$ and $v_{i-1}^{(j)} = \Phi(\varepsilon_i^{(j)})$, so the copula-weighted component is the distribution of the random vector
$$
\widetilde Y_i
=\mu_{i-1}
+L_{i-1} \, \mathbf{g}_\delta\!\left(
\sqrt{1-\rho^2}\,W+\rho\varepsilon_i
\right),
\qquad W\sim\mathcal N(0,I_q).
$$
Using the univariate terms in (\ref{eq:R_delt}), write
\begin{align*}
     R_i&=\left(R_\delta^{(1)}(\varepsilon_i^{(j)})\right)_{j=1}^q,\quad    V_i=\operatorname{Diag}\left\{\left(
    R_\delta^{(2)}(\varepsilon_i^{(j)})-R_\delta^{(1)}(\varepsilon_i^{(j)})^2\right)_{j=1}^q\right\}.
\end{align*}
Suppressing dependence on $x$, one can then show that the moment updates are
\begin{align*}
    \mu_i&=\mu_{i-1}+\beta_i\, L_{i-1} R_i,\quad \Sigma_i
=L_{i-1}\left[
I_q+\beta_i\left\{
(1-\beta_i) R_i R_i^\top
-(I_q-V_i)
\right\}
\right]L_{i-1}^\top.
\end{align*}
The term $V_i$ arises from computing the expectation of the outer product $\mathbf{g}_\delta\!\left(\varepsilon_i'
\right) \mathbf{g}_\delta\!\left(
\varepsilon_i'
\right)^\top$ where $\varepsilon_i' = \sqrt{1-\rho^2}\,W+\rho\varepsilon_i$ under $W \sim \mathcal{N}(0,I_q)$. As a sanity check, we recover the Gaussian update as $\delta \to \infty$ since $R_i \to \rho Z_i$, and $V_i \to (1-\rho^2) I$. 

Under order-compatible missingness, where \(O_i=\{1,\ldots,r\}\), the remaining latent variables satisfy $\varepsilon_i^{(r+1:q)}
\mid Y_i^{(1:r)},X_i
\sim N(0,I_{q-r})$ under the predictive model. Averaging the raw moment updates as in (\ref{eq:mk_missing}) then amounts to
computing, for each $j > r$,
$E\!\left[R_i^{(j)}\mid Y_i^{(1:r)},X_i\right]=0,$
and
$E\!\left[V_i^{(j,j)}+(R_i^{(j)})^2
\mid Y_i^{(1:r)},X_i\right]=1$ in the above updates. This will result in closed form moment updates that are similar to the Gaussian case. For arbitrary missingness patterns, this simplification need not hold, and conditional integration under the SU family is required which may not be tractable. A potential approximation is to apply a similar projection by $Q_i$ of the residuals as the Gaussian case, which we leave for future work.

\subsection{Implementation details}\label{app:imp}

We begin by discussing a few implementation details that apply to all the recursive moment updates.
As discussed in the main paper, for prediction, we average over permutations on the raw moment scale, that is we average $\mu_n$ and $\Sigma_n+\mu_n\mu_n^\top$ before recovering the covariance. Predictive resampling then carries forward from the averaged moment functions. To avoid calculating $s(x,X_i)$ repeatedly for $i \geq n+1$, we use the same covariate sequence $X_{n+1:N} \iid P_X$ for each predictive resampling trajectory; see \cite{sun2026design} for further discussion of this approach.

Next, we discuss hyperparameter optimization. For the Gaussian-copula similarity, we use a separate bandwidth for each covariate with $s(x,x')=\prod_{j=1}^p
    c_{\rho_{x}^{(j)}}\{\Phi(x^{(j)}),\Phi(x'^{(j)})\}$, where $\rho_{x}^{(j)}=0$ removes the contribution of covariate $j$. For the outcome, we use a single bandwidth $\rho$ shared across coordinates in the multi-output case. The bandwidths ($\rho,\rho_x)$ are selected jointly by maximizing the prequential log-likelihood averaged over permutations, and for the symmetric SU model,  we also optimize $\delta$ jointly with $(\rho,\rho_x)$. When there are missing outcomes, each summand in the prequential log-likelihood is the marginal predictive density of the observed coordinates instead. For optimization, we use L-BFGS-B with gradients obtained by automatic differentiation.

    We conclude with a discussion on algorithmic implementation. In an ideal setting, evaluating the prequential log-likelihood only requires $n(n-1)/2$ moment updates, but this would entail dynamically-sized arrays which are not supported by JAX's scan operation. For larger datasets, a compromise is to evaluate the prequential objective in blocks, and discard the earlier blocks whose prequential contributions have already been evaluated, avoiding unnecessary subsequent updates. For the multi-output model, we store the mean components and the $q(q+1)/2$ lower-triangular entries of $L_i(x)$ as arrays during updating, and only reconstruct $\Sigma_i(x)$ afterwards. The Gaussian covariance update (\ref{eq:var_gauss_mv}) involves computing the Cholesky factorization of a term $I + c zz^\top$ then scaling, allowing $L_{i-1}$ to be updated to $L_i$ in $O(q^2)$ operations per observation and $x$-value, e.g. \citet[Codes 2.9.6 and 2.9.8]{hiptmair2016numerical}. This cost also holds under order-compatible missingness for (\ref{eq:cov_gauss_missing}), since $Q_i$ is diagonal so again we just have a rank-one Cholesky update. For arbitrary missingness patterns, $Q_i$ is generally dense, and standard projection and factorization routines can incur a worst-case $O(q^3)$ cost, although low-rank methods reduce this when few outputs are observed or missing. We leave the investigation of more efficient algorithms or approximations in the large $q$ setting for future work. 

 \section{Additional experimental details}
 
 \subsection{Simulation details}
In this section, we include additional details on the simulations.
 For the DDP baselines, we follow the notation of \cite{Antoniano2014} for the hyperparameters. For the heavy-tailed settings, we use the same prior hyperparameters as the simulation study of \cite{Antoniano2014}, with:
\begin{align*}
 \beta_0=(5/2,5/8)^\top,\quad C^{-1}=\operatorname{Diag}(4,1/4),
 \quad \alpha_1 = 1,\quad \alpha_2 = 1,\\
 \mu_0=0,\quad c=1/8,\quad a=1, \quad b=1.
\end{align*}
For the multivariate setting, we use similar settings to \citet[Example 3]{wade2025bayesian}, but extend it to 5 covariates:
\begin{align*}
 \beta_0=\boldsymbol{0}_6,\quad
 C^{-1}=\frac{\widehat{\operatorname{Var}}(Y)}{0.0405}\cdot 
              \operatorname{Diag}(10,1,1,1,1,1),\quad
 \alpha_1 = 2,\quad \alpha_2 = 0.0405,\\
 \mu_0=\boldsymbol{0}_5, \quad c=(1/36)\cdot\boldsymbol{1}_5, \quad  a=2 \cdot  \boldsymbol{1}_5, \quad b = 0.0375\cdot \boldsymbol{1}_5.
\end{align*}
The Matlab implementation of \cite{wade2025bayesian} can be found here: \url{https://github.com/sarawade/BayesXMix/tree/main/NormalizedWeights}.
For evaluating the MSE of the DDP approach, we use the posterior mean for estimating $\mu_*$. For $\sigma_*^2$, we use the posterior median as the variance posterior is heavy-tailed and the posterior mean performs poorly. As the DDP runs are slow and must be carried out on a cluster, the run-times are for a single run on the reported local machine.

For the single-output GP in the simulations, we use a radial basis function (RBF) kernel with one length scale per dimension, and optimize the marginal likelihood to select the length scale as well as the noise variance and kernel amplitude, using a standard \texttt{sklearn} implementation. Here, the posterior mean is used as the point estimate for $\mu_*$, and the optimized noise variance is used for estimating $\sigma_*^2$. 

For hyperparameter tuning in the recursive moment updates, we use $10$ permutations for all examples with $n = 200$, and $2$ permutations for the larger examples with $n = 600$ and $1000$. For actual fitting and prediction, $10$ permutations are used in all cases.
For predictive resampling, we set $P_X$ to be uniform on the known compact $\mathcal{X}$, i.e. $[-7,7]$ for the univariate covariate simulation and $[-2,2]^5$ for the multivariate covariate simulation. Predictive resampling uses $N = n + 4000$ and $B = 1000$ throughout. We naturally use the posterior means of $\mu_*$ and $\sigma_*^2$ as the point estimates for calculating the MSE. Note that while the posterior mean of $\mu_*$ equals the fitted mean $\mu_n$ due to the martingale, this is not necessarily the case for $\sigma_*^2$. For test prediction, we use the fitted predictive density $p_n(y \mid x)$, which can be computed without predictive resampling. 

 \subsection{Weather dataset details}\label{app:weather}

\begin{figure}[!h]
    \centering
    \includegraphics[width=\linewidth]{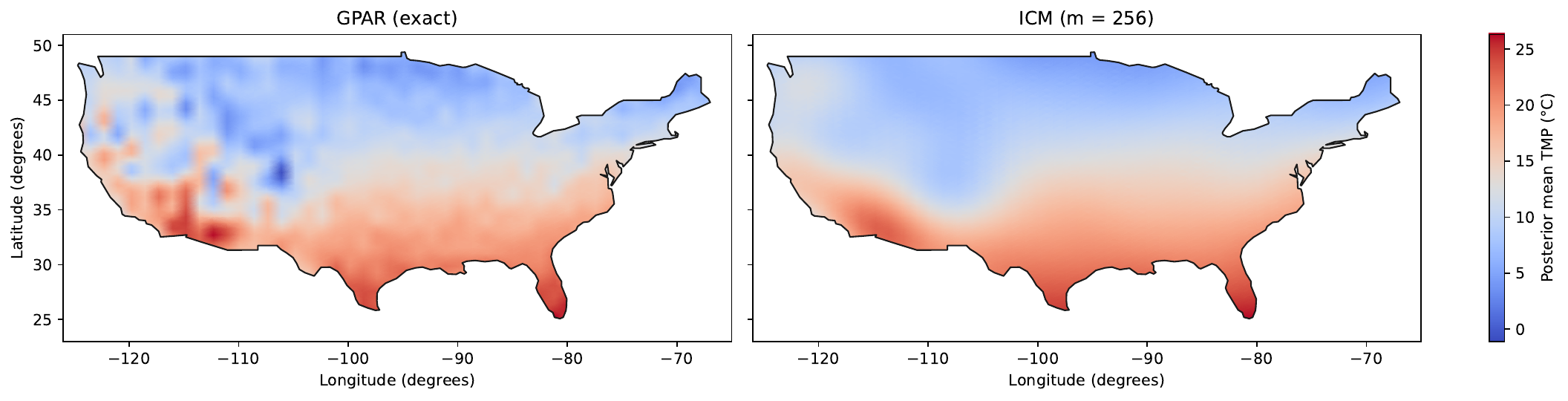}
    
    \caption{Posterior mean of the mean function $\mu(x)$ for TMP in the year 2021 for exact GPAR (left) and ICM with $m = 256$ (right).}
    \label{fig:weather_tmp_gp}
\end{figure}

We begin by outlining the key data preprocessing steps to obtain our curated dataset from the WEATHER-5K dataset of \citet{han2026benchmarking}. We extracted the 1033 weather stations from the 48 contiguous US states by matching the weather station codes obtained from the Integrated Surface Database of the National Oceanic and Atmospheric Administration \citep{smith2011integrated}. For each station and year, we averaged all TMP/DEW/SLP hourly measurements from the WEATHER-5K dataset to obtain the full dataset. To produce Figures \ref{fig:weather_corr}, \ref{fig:weather_tmp} and \ref{fig:weather_tmp_gp}, country outlines were obtained from the Natural Earth website \citep{kelso2010introducing}.

For GPAR, we use the package implementation with both linear and nonlinear dependencies on preceding outputs. Since GPAR is order-dependent, we use the ordering $(\mathrm{TMP},\mathrm{SLP},\mathrm{DEW})$, so prediction of DEW given the other outputs uses the final GP. We use 200 Monte Carlo samples to evaluate the predictive mean function, which is shown in Figure \ref{fig:weather_tmp_gp} (left) for TMP in the year 2021. We attribute the apparent roughness to Monte Carlo variation, and highlight that it is non-trivial to produce the GPAR equivalent of Figure \ref{fig:weather_corr} as moments of the joint output distribution generally lack closed forms.

For ICM, we use an RBF kernel with a separate length scale for each input dimension and a full-rank coregionalization matrix. Each output has its own Gaussian observation noise variance. Following the construction in the \texttt{GPflow} documentation, we augment the inputs with an output index and retain only observed responses, using \texttt{SwitchedLikelihood} for the output-specific likelihoods. For inference, we use the \texttt{SVGP} class for sparse variational inference which optimizes the variational parameters, hyperparameters and continuous coordinates of the \(m\) inducing inputs. The posterior mean of TMP for the ICM with $m = 256$ is shown in Figure \ref{fig:weather_tmp_gp} (right), which is quite similar to the RM approach. 

Finally, we outline the evaluation metrics in Table \ref{tab:weather_results}. Let $\mathcal D_{\mathrm{obs}}$ denote the training data, $p_n$ denote each method's predictive density, and \(\mathcal I_{\mathrm{test}}\) and \(\mathcal I_{\mathrm{DEW}}\) denote the indices of the held-out full data points and DEW measurements respectively, with sizes \(n_{\mathrm{test}}\) and \(n_{\mathrm{DEW}}\). For the fully held-out data points, we compute
\begin{align*}
    \frac{1}{n_{\mathrm{test}}}
    \sum_{i\in\mathcal I_{\mathrm{test}}}
    \log  p_n(Y_i\mid X_i,\mathcal D_{\mathrm{obs}}).
\end{align*}
For predicting DEW, writing $Y^{(1:2)}=(\mathrm{TMP},\mathrm{SLP})$ and $Y^{(3)}=\mathrm{DEW}$, we compute
\begin{align*}
    \frac{1}{n_{\mathrm{DEW}}}
    \sum_{i\in\mathcal I_{\mathrm{DEW}}}
    \log  p_n\left(
        Y_i^{(3)}\mid Y_i^{(1:2)}, X_i,\mathcal D_{\mathrm{obs}}
    \right).
\end{align*}
The methods incorporate the observed outputs differently for predicting DEW.
For GPAR, $(X_i, Y_i^{(1:2)})$ enter as inputs to the final layer GP. Similarly, for RM (Multi-output), we apply the Gaussian conditional formulas to the joint $p_n(y\mid x)$, conditioning $Y_i^{(3)}$ on $Y_i^{(1:2)}$. For ICM, 
conditional on the latent functions, the observation noises are independent. As $Y_i^{(1:2)}$ is already included in $\mathcal D_{\mathrm{obs}}$, their information enters through the posterior and no additional conditioning on $Y_i^{(1:2)}$ is needed. Finally, RM (Independent) predicts $Y_i^{(3)}$ independently given $X_i$.

\end{appendices}

\end{document}